\documentclass[a4paper,11pt]{article}

\pdfoutput=1

\usepackage{jheppub}

\usepackage[T1]{fontenc}

\usepackage{amsmath,amssymb}
\usepackage{graphicx}

\usepackage{booktabs}
\usepackage{tabularx}
\usepackage{array}

\usepackage{xcolor}
\usepackage{pifont}

\usepackage{microtype}

\usepackage{placeins}
\usepackage{float}

\usepackage[
    colorlinks=true,
    citecolor=blue,
    linkcolor=blue,
    urlcolor=blue
]{hyperref}

\newcolumntype{Y}{>{\raggedright\arraybackslash}X}

\newcommand{\YesMark}{\ding{51}}
\newcommand{\NoMark}{\ding{55}}

\numberwithin{equation}{section}

\title{\boldmath
The Triadic Texture: Neutrino Predictions, Viable Vacuum,
and Phenomenological Constraints
}

\author[a]{Sagar Tirtha Goswami}
\author[a]{Pralay Chakraborty}
\author[a,1]{Subhankar Roy,\note{Corresponding author.}}

\affiliation[a]{Gauhati University, India}

\emailAdd{sagartirtha@gauhati.ac.in}
\emailAdd{pralaychakraborty8@gmail.com}
\emailAdd{subhankar@gauhati.ac.in}

\abstract{
A minimal and predictive Majorana neutrino mass texture is proposed. The texture correlates
the atmospheric mixing angle and Dirac CP phase and constrains the absolute
neutrino masses, Majorana phases and the effective Majorana mass $m_{\beta\beta}$. A correlated oscillation
likelihood admits both normal and inverted orderings, with the normal-ordering
solution giving the lower minimum in the present fit, and the predicted mass
scale is confronted with recent DESI limits. The texture is realised in a
hybrid Type-I seesaw plus Weinberg-like dimension-six framework with
$A_4\otimes Z_{10}\otimes Z_7\otimes Z_5\otimes Z_3$ flavour symmetry. Both
the Altarelli--Feruglio (A--F) and Ma--Rajasekaran (M--R) realisations reproduce the
texture at the retained order. Using the complete renormalisable scalar
potential, the A--F alignment is found to contain a physical CP-even zero mode
at tree level, while the M--R alignment admits a finite locally stable and bounded
region together with sufficient conditions for a global minimum.
Seven physical CP-odd zero modes are identified and lifted by controlled
dimension-two soft terms without shifting the desired real M--R vacuum or the
leading tree-level neutrino texture. Charged lepton flavour violation and
non-unitarity provide complementary constraints. The conventional Type-I
unflavoured and flavoured leptogenesis contributions vanish for the retained
Yukawa structure, and the residual discrete vacuum degeneracy and associated
domain-wall problem are discussed.
}

\begin{document}

\maketitle

\flushbottom

\section{Introduction}
\label{sec:introduction}

The neutrino ($\nu$) is a peculiar particle in the Standard Model (SM) of Particle Physics. The SM asserts that it is massless owing to the absence of its right handed (RH) counterpart in nature. However, the evidence of neutrino oscillation \cite{r1,r2,r3,r4,r5,r6,r7,r8,r9,r10,r11,r12} questions this very assertion of SM because neutrinos require small but finite masses for the phenomenon of oscillation \cite{r13}. Furthermore, the fact that the three flavour states of neutrinos ($\nu_l$, $l=e,\mu,\tau$) are actually linear superpositions of three mass eigenstates ($\nu_i$, $i=1,2,3$) with definite masses, which is supported by experimental data \cite{r14,r15}, is another point that the SM is completely silent on. Besides these two aspects directly connected with neutrinos, the SM also cannot explain some other physical facts such as the observed baryon asymmetry of the universe (BAU) \cite{r16}, dark matter \cite{r17,r18} etc. In spite of the astonishing accuracies of many of its predictions, the insufficiencies of the SM make a strong case for exploring Physics beyond it (BSM).

As the failure to find RH neutrinos persists, one of the common BSM steps is to consider the neutrino as a Majorana particle thanks to its electrical neutrality. Regarding the smallness of their masses, there are many mechanisms that can address this point such as seesaw mechanisms \cite{r19,r20,r22,r23}, radiative mass generation mechanisms \cite{r24,r25,r26,r27} etc. On the question of mixing, the extent to which the mass eigenstates mix as shown by oscillation data can be explained through the application of various discrete symmetry groups \cite{r28,r29}. Models built with neutrino mass generation mechanisms along with discrete symmetry groups aim to explain, and in favourable cases predict, nine physical parameters related to neutrinos, viz., the absolute values of the three neutrino mass eigenvalues ($m_1$, $m_2$, $m_3$), three mixing angles ($\theta_{12}$, $\theta_{13}$, $\theta_{23}$), and three CP violating phases ($\delta$, $\alpha$, $\beta$). Among these, neutrino oscillation experiments are sensitive to the three mixing angles, the two independent mass-squared differences $\Delta m^2_{21}$ and $\Delta m^2_{31}$, and the Dirac CP-violating phase $\delta$. By contrast, the absolute neutrino mass scale and the two Majorana CP-violating phases $\alpha$ and $\beta$ remain inaccessible to oscillation experiments. The absolute neutrino mass scale can be probed through beta-decay endpoint experiments, cosmological observations, and neutrinoless double beta decay, while the Majorana phases may only be indirectly constrained through lepton number violating observables such as the effective Majorana mass \cite{r18,r30,r31}. Although the observable parameters are now being measured with increasing precision, several important questions remain unresolved, including the octant of $\theta_{23}$, the neutrino mass ordering, the precise value of $\delta$, and the possible values of the Majorana phases.

The information about all the physical parameters is encoded in the neutrino mass matrix, $M_\nu$, which originates from the Yukawa Lagrangian. It is a $3\times3$ complex symmetric matrix having 12 real parameters in general. By applying suitable conditions on the elements of $M_\nu$, some of the physical parameters can be predicted. This part is usually done by the discrete symmetry groups present in the models. Many constrained $M_\nu$'s, commonly referred to as textures, are widely studied in the literature. These include texture zeroes \cite{r32}, vanishing minors \cite{r33}, hybrid textures \cite{r34}, $\mu-\tau$ symmetric textures \cite{r35,r36,r37}, among many others. While some of these phenomenological ideas are now excluded by current experimental data (for example, exact $\mu-\tau$ symmetry), others remain viable in one form or another \cite{r38,r39}. Nevertheless, the continued proposition of new textures is important from both theoretical and experimental standpoints. Theoretically, they often arise from underlying flavour symmetries or UV complete models, opening new avenues for BSM Physics. Experimentally, until the determination of a unique $M_\nu$ is possible, the different textures offer distinct and testable predictions for the next generation of precision experiments.

Recently, an interesting texture with the correlation $(M_\nu)_{22}=-2(M_\nu)_{13}$ was proposed in \cite{r40}, and subsequently studied in \cite{r41,r42}. The texture has five independent complex parameters. While this texture leads to useful constraints on several observable parameters, it accommodates both the normal and inverted orderings (NO and IO, respectively) of neutrino mass eigenvalues. Moreover, the presence of a single correlation condition naturally raises the question of whether an additional correlation can further constrain the texture and thereby sharpen its phenomenological predictions.

Beyond the texture-level analysis, it is equally important to ask whether the texture can be realised within a consistent flavour model. A given flavour symmetry may reproduce the desired mass matrix through suitable basis choices and vacuum expectation value (VEV) alignments; however, this does not automatically guarantee the viability of the full model \cite{r43}. The corresponding scalar sector must also support the chosen vacuum configuration. This requires examining the scalar spectrum, the boundedness of the potential from below, and the global minimality of the vacuum. In addition, the appearance of massless scalar states, if present, should be understood, since such states may point to accidental symmetries or flat directions hidden in the scalar potential.

The phenomenological consequences of the resulting model provide another important perspective on its flavour structure. Even when the scalar sector is viable, the same flavour structure can restrict the allowed charged lepton flavour violating (CLFV) channels \cite{r44,r45}, induce non-unitarity of the effective lepton mixing matrix \cite{r46,r47}, affect the generation of BAU and so on. Thus, CLFV, non-unitarity, and BAU provide complementary ways of probing the model and constraining its parameter space.

In the present work, we propose a more constrained texture by supplementing the correlation $(M_\nu)_{22}=-2(M_\nu)_{13}$ with an additional relation, $(M_\nu)_{12}=(M_\nu)_{13}$. This new texture, which is referred to as the triadic texture, yields several predictions for observable neutrino parameters. A complete solution of the texture constraints, with a consistent treatment of the physical branches and the current correlated oscillation likelihoods, shows that the texture admits both NO and IO solutions. However, the two orderings occupy distinctly different regions of the allowed parameter space, particularly in the $\theta_{23}$--$\delta$ plane, with the NO solution providing the better fit in the present analysis. The texture also constrains the neutrino masses, the Majorana phases, and the effective Majorana mass, and leads to a strong correlation between the atmospheric mixing angle and the absolute neutrino mass scale. This makes the recent cosmological limits on the sum of neutrino masses, including those from DESI \cite{r88}, an important test of the texture.

Having established the texture itself, we construct an explicit flavour model based on the $A_4$ symmetry that realises the texture in two different bases with two different sets of VEV alignments: one in the Ma--Rajasekaran (M--R) basis and another in the Altarelli--Feruglio (A--F) basis of $A_4$ \cite{r37,r48,r49,r50}. The subsequent analysis shows that the physical viability of the vacuum depends sensitively on the chosen VEV configuration. It also reveals how the same underlying flavour structure can lead to distinctive scalar-sector features and phenomenological signatures. In the present analysis, we study the complete renormalisable scalar potential consistent with the flavour symmetry and the chosen real vacuum branch. We find that the A--F alignment contains a physical CP-even zero mode in the tree-level Hessian, whereas the M--R vacuum is viable in a finite region of parameter space, where the physical CP-even spectrum can be strictly positive, the massive CP-odd sector is locally stable, and the quartic potential is bounded from below. A set of sufficient conditions can also be given under which the chosen M--R vacuum is the global minimum. The scalar sector contains seven physical CP-odd zero modes in addition to the electroweak Goldstone boson, and their origin is identified.

The phenomenological implications of the model are analysed using the complete flavour structure. The scalar and heavy-neutrino sectors lead to characteristic CLFV processes and induce non-unitarity of the effective PMNS matrix. We also examine baryogenesis through leptogenesis. For the leading RH Majorana structure that defines the triadic texture, the CP-odd invariants relevant for the conventional unflavoured and flavoured Type-I leptogenesis asymmetries vanish at leading order. Possible decay channels involving physical flavon excitations originate from the higher-dimensional operators and depend on the scalar spectrum; they are therefore kept distinct from the conventional two-body leptogenesis source.

The plan of the paper is as follows. In Section~\ref{sec:phenomenology}, we discuss the phenomenological implications of the proposed triadic texture in a model-independent way, including its numerical predictions, their response to controlled shifts of the oscillation likelihood, and the implications of the absolute neutrino mass scale and neutrinoless double beta decay. In Section~\ref{sec:symmetry_realisation}, we present its flavour-symmetry realisation and discuss two distinct vacuum-alignment constructions. A detailed scalar-potential analysis is performed in Section~\ref{sec:vacuum_scalar}, where we examine the scalar spectra, vacuum stability, boundedness from below, the global character of the vacuum, and the physical CP-odd zero modes, thereby identifying the viable realisation of the texture. In Section~\ref{sec:phenomenological_constraints}, we study the resulting constraints and predictions of the model, including charged lepton flavour violation, non-unitarity of the effective lepton mixing matrix, and baryogenesis through the conventional Type-I channel. In Section~\ref{sec:soft_breaking}, we introduce a controlled set of dimension-two soft-breaking terms that lift the physical CP-odd zero modes and discuss their consequences for the scalar and flavour sectors. Section~\ref{sec:domain_walls} briefly addresses the residual discrete vacuum degeneracy and the associated domain-wall issue. Finally, Section~\ref{sec:summary_discussion} contains the summary and discussion.

\section{Phenomenology}
\label{sec:phenomenology}

In this section, we study the phenomenology of our proposed triadic texture consistent with neutrino oscillation experiments. Based on the predictions of the texture, we will also study the absolute neutrino mass scale, the Majorana phases, and the effective Majorana neutrino mass pertinent to neutrinoless double beta decay. It is important to note that the discussion in this section will be solely on the texture itself in a model-independent way.

\subsection{The Triadic Texture}
\label{subsec:triadic_texture}

Considering the neutrinos as Majorana particles, our proposed triadic texture is given in the following:
\begin{equation}
M_\nu=
\begin{pmatrix}
 a & b & b\\
 b & -2b & t\\
 b & t & d
\end{pmatrix},
\qquad b>0,
\label{eq:triadic_texture}
\end{equation}
where the common parameter $b$ is taken to be real and strictly positive throughout the analysis, while $a$, $t$, and $d$ are in general complex. As noted in Section~\ref{sec:introduction}, the texture has two correlations among its elements, namely $(M_\nu)_{12}=(M_\nu)_{13}$ and $(M_\nu)_{22}=-2(M_\nu)_{13}$. These conditions reduce the arbitrariness of the texture and make it more predictive.

To obtain information about its predictions, we diagonalise $M_\nu$ with the Pontecorvo--Maki--Nakagawa--Sakata (PMNS) matrix \cite{r51,r52}. For the present analysis, we use the phase convention
\begin{equation}
U_{\rm PMNS}=V P_M,
\qquad
P_M=\operatorname{diag}(e^{i\alpha},e^{i\beta},1),
\label{eq:triadic_upmns_pm}
\end{equation}
where
\begin{equation}
V=
\begin{pmatrix}
 c_{12}c_{13} & -s_{12}c_{13} & -s_{13}e^{-i\delta_{\rm CP}}\\
 s_{12}c_{23}+c_{12}s_{23}s_{13}e^{i\delta_{\rm CP}} & c_{12}c_{23}-s_{12}s_{23}s_{13}e^{i\delta_{\rm CP}} & s_{23}c_{13}\\
 -s_{12}s_{23}+c_{12}c_{23}s_{13}e^{i\delta_{\rm CP}} & -c_{12}s_{23}-s_{12}c_{23}s_{13}e^{i\delta_{\rm CP}} & c_{23}c_{13}
\end{pmatrix},
\label{eq:triadic_vmatrix}
\end{equation}
with $s_{ij}=\sin\theta_{ij}$ and $c_{ij}=\cos\theta_{ij}$. The unphysical phases associated with the charged lepton fields have already been absorbed through appropriate field redefinitions, while the two Majorana phases $\alpha$ and $\beta$ are kept explicitly in $P_M$.

With this convention, the complex symmetric mass matrix is diagonalised through the transpose congruence
\begin{equation}
U_{\rm PMNS}^{T}M_\nu U_{\rm PMNS}=\operatorname{diag}(m_1,m_2,m_3),
\qquad m_i>0.
\label{eq:triadic_congruence}
\end{equation}
The labels 1, 2, and 3 are fixed by the columns of $U_{\rm PMNS}$ and are not reordered after the calculation. Thus, the mass ordering follows from the labelled spectrum: $m_1<m_2<m_3$ corresponds to normal ordering (NO), whereas $m_3<m_1<m_2$ corresponds to inverted ordering (IO).

It is useful to separate the overall dimensional scale from the shape of the texture by defining
\begin{equation}
A\equiv\frac{a}{b},
\qquad
T\equiv\frac{t}{b},
\qquad
D\equiv\frac{d}{b}.
\label{eq:triadic_ATD}
\end{equation}
Equation~\eqref{eq:triadic_texture} can then be written as
\begin{equation}
M_\nu=b
\begin{pmatrix}
 A & 1 & 1\\
 1 & -2 & T\\
 1 & T & D
\end{pmatrix}.
\label{eq:triadic_scalefree}
\end{equation}
Thus, $A$, $T$, and $D$ describe the dimensionless shape of the texture, while $b$ supplies the common mass scale. In the next subsection, we show that the three vanishing off-diagonal conditions determine $A$, $T$, and $D$ in terms of the mixing parameters, while the observed solar mass-squared difference fixes the remaining positive scale $b$.

\subsection{Analytic Consequences of the Texture}
\label{subsec:triadic_analytic}

Having fixed the convention, it is useful to see how many parameters of the texture are actually independent. The three off-diagonal elements of $U_{\rm PMNS}^{T}M_\nu U_{\rm PMNS}$ must vanish. Using the scale-free parameters introduced above, the three off-diagonal conditions can be written as
\begin{align}
 \cos 2\theta_{23}\,T-\frac{1}{2}\sin 2\theta_{23}\,D &=q_1,\label{eq:triadic_lin1}\\
 -e^{-2i\delta_{\rm CP}}A+\sin 2\theta_{23}\,T+\cos^2\theta_{23}\,D &=q_2,\label{eq:triadic_lin2}\\
 -A-\sin 2\theta_{23}\,T+\sin^2\theta_{23}\,D &=q_3,\label{eq:triadic_lin3}
\end{align}
where
\begin{align}
q_1&=\sin 2\theta_{23}+\tan\theta_{13}\bigl(\cos\theta_{23}-\sin\theta_{23}\bigr)e^{-i\delta_{\rm CP}},\\
q_2&=1-\cos 2\theta_{23}-2\cot 2\theta_{13}\bigl(\cos\theta_{23}+\sin\theta_{23}\bigr)e^{-i\delta_{\rm CP}},\\
q_3&=1+\cos 2\theta_{23}+\tan\theta_{13}\bigl(\cos\theta_{23}+\sin\theta_{23}\bigr)e^{i\delta_{\rm CP}}
-\frac{2\bigl(\cos\theta_{23}-\sin\theta_{23}\bigr)}{\cos\theta_{13}}\cot 2\theta_{12}.
\end{align}
For later convenience, we define
\begin{equation}
\mathcal{D}_{23}\equiv\cos^2\theta_{23}+e^{-2i\delta_{\rm CP}}\sin^2\theta_{23}.
\label{eq:triadic_denominator}
\end{equation}
For $\mathcal{D}_{23}\neq0$, the above equations determine the dimensionless texture parameters uniquely as
\begin{align}
A&=-\frac{\sin 2\theta_{23}\,q_1+\sin^2\theta_{23}\,q_2+\cos^2\theta_{23}\,q_3}{\mathcal{D}_{23}},\label{eq:triadic_A_solution}\\
T&=\frac{\bigl(\cos^2\theta_{23}-e^{-2i\delta_{\rm CP}}\sin^2\theta_{23}\bigr)q_1+\frac{1}{2}\sin 2\theta_{23}\bigl(q_2-e^{-2i\delta_{\rm CP}}q_3\bigr)}{\mathcal{D}_{23}},\label{eq:triadic_T_solution}\\
D&=\frac{-\sin 2\theta_{23}\bigl(1+e^{-2i\delta_{\rm CP}}\bigr)q_1+\cos 2\theta_{23}\bigl(q_2-e^{-2i\delta_{\rm CP}}q_3\bigr)}{\mathcal{D}_{23}}.\label{eq:triadic_D_solution}
\end{align}
Thus, $A$, $T$, and $D$ are not independent scan parameters. Once the three mixing angles and $\delta_{\rm CP}$ are specified, the dimensionless shape of the texture is fixed analytically.

After imposing the off-diagonal conditions, the three labelled diagonal quantities can be expressed as
\begin{align}
F_1 &= A+\tan\theta_{13}\bigl(\cos\theta_{23}+\sin\theta_{23}\bigr)e^{i\delta_{\rm CP}}
+\frac{\cos\theta_{23}-\sin\theta_{23}}{\cos\theta_{13}}\tan\theta_{12},\label{eq:triadic_F1}\\
F_2 &= A+\tan\theta_{13}\bigl(\cos\theta_{23}+\sin\theta_{23}\bigr)e^{i\delta_{\rm CP}}
-\frac{\cos\theta_{23}-\sin\theta_{23}}{\cos\theta_{13}}\cot\theta_{12},\label{eq:triadic_F2}\\
F_3 &= Ae^{-2i\delta_{\rm CP}}-\cot\theta_{13}\bigl(\cos\theta_{23}+\sin\theta_{23}\bigr)e^{-i\delta_{\rm CP}}.\label{eq:triadic_F3}
\end{align}
Defining
\begin{equation}
\rho_i\equiv|F_i|,
\label{eq:triadic_rhoi}
\end{equation}
the physical masses are simply
\begin{equation}
m_i=b\rho_i.
\label{eq:triadic_masses_b}
\end{equation}
The Majorana phases are also fixed by the same reconstruction. In the convention of Eq.~\eqref{eq:triadic_upmns_pm},
\begin{equation}
\alpha=\frac{1}{2}\operatorname{Arg}\!\left(\frac{F_3}{F_1}\right),
\qquad
\beta=\frac{1}{2}\operatorname{Arg}\!\left(\frac{F_3}{F_2}\right),
\label{eq:triadic_majorana}
\end{equation}
with both phases understood modulo $\pi$.

The measured solar mass-squared difference removes the remaining overall scale. Since
\begin{equation}
\Delta m^2_{21}=b^2\bigl(\rho_2^2-\rho_1^2\bigr),
\end{equation}
a physical solution requires
\begin{equation}
D_{21}\equiv\rho_2^2-\rho_1^2>0,
\label{eq:triadic_D21}
\end{equation}
and the unique positive scale is
\begin{equation}
b^2=\frac{\Delta m^2_{21}}{D_{21}}.
\label{eq:triadic_bscale}
\end{equation}
Consequently,
\begin{equation}
m_i^2=\Delta m^2_{21}\frac{\rho_i^2}{\rho_2^2-\rho_1^2},
\label{eq:triadic_mi_analytic}
\end{equation}
and the signed atmospheric splittings follow as
\begin{align}
\Delta m^2_{31}&=\Delta m^2_{21}\frac{\rho_3^2-\rho_1^2}{\rho_2^2-\rho_1^2},\label{eq:triadic_dm31}\\
\Delta m^2_{32}&=\Delta m^2_{21}\frac{\rho_3^2-\rho_2^2}{\rho_2^2-\rho_1^2}.\label{eq:triadic_dm32}
\end{align}
The mass ordering is therefore an output of the labelled spectrum rather than an input imposed after diagonalisation. In particular, $\rho_1<\rho_2<\rho_3$ corresponds to NO, while $\rho_3<\rho_1<\rho_2$ corresponds to IO. The sum of neutrino masses is similarly fixed by
\begin{equation}
\Sigma_\nu=\sqrt{\Delta m^2_{21}}\,\frac{\rho_1+\rho_2+\rho_3}{\sqrt{\rho_2^2-\rho_1^2}}.
\label{eq:triadic_sum_analytic}
\end{equation}
The beta-decay effective mass follows as
\begin{equation}
m_\beta^2=b^2\!\left(c_{12}^2c_{13}^2\rho_1^2+s_{12}^2c_{13}^2\rho_2^2+s_{13}^2\rho_3^2\right),
\end{equation}
while the texture gives the particularly simple identity
\begin{equation}
m_{\beta\beta}=|(M_\nu)_{ee}|=|a|=b|A|.
\label{eq:triadic_mbb_identity}
\end{equation}
This identity has a direct physical meaning.  The parameter $a$ is not a neutrino mass eigenvalue, but its modulus is itself the effective Majorana mass measured by neutrinoless double beta decay.  Thus $|a|$ is not merely an internal texture parameter: it can be confronted directly with $0\nu\beta\beta$ searches, while the phase of $a$ remains part of the complex texture reconstruction.

A particularly simple consequence follows from Eqs.~\eqref{eq:triadic_F1} and \eqref{eq:triadic_F2}. Their difference is
\begin{equation}
F_2-F_1=\frac{2\sqrt{2}\,\sin\!\left(\theta_{23}-\frac{\pi}{4}\right)}{\cos\theta_{13}\,\sin 2\theta_{12}}.
\label{eq:triadic_Fdifference}
\end{equation}
At exact maximal atmospheric mixing, $\theta_{23}=45^\circ$, one therefore obtains $F_2=F_1$ and hence $\rho_2=\rho_1$. The measured solar mass-squared difference would then vanish. Since $\Delta m^2_{21}$ is experimentally nonzero, every finite consistent solution of the triadic texture must satisfy
\begin{equation}
\theta_{23}\neq45^\circ.
\label{eq:triadic_nonmaximal}
\end{equation}
This result is independent of the neutrino mass ordering.

The behaviour close to maximal mixing also explains an important feature of the numerical solutions. Defining
\begin{equation}
P\equiv A+\tan\theta_{13}\bigl(\cos\theta_{23}+\sin\theta_{23}\bigr)e^{i\delta_{\rm CP}},
\end{equation}
one obtains
{\small
\begin{equation}
D_{21}=\frac{4\sqrt{2}\,\sin\!\left(\theta_{23}-\frac{\pi}{4}\right)}{\cos\theta_{13}\,\sin 2\theta_{12}}\operatorname{Re}P
+\frac{8\sin^2\!\left(\theta_{23}-\frac{\pi}{4}\right)\cos 2\theta_{12}}{\cos^2\theta_{13}\,\sin^2 2\theta_{12}}.
\label{eq:triadic_D21_expanded}
\end{equation}
}
Hence $D_{21}\to0$ as $\theta_{23}\to45^\circ$. Along a physical approach with $D_{21}>0$, reproducing the measured solar splitting therefore requires a larger value of $b$ through Eq.~\eqref{eq:triadic_bscale}, which in turn raises the neutrino mass scale and generally increases $\Sigma_\nu$. This establishes an analytic connection between the atmospheric angle and cosmological limits on the absolute neutrino mass scale. Equation~\eqref{eq:triadic_sum_analytic} also turns a cosmological upper limit $\Sigma_\nu<L$ into a direct restriction on the angular solution,
\begin{equation}
\frac{(\rho_1+\rho_2+\rho_3)^2}{\rho_2^2-\rho_1^2}<\frac{L^2}{\Delta m^2_{21}}.
\label{eq:triadic_cosmo_angular}
\end{equation}
The analytic discussion therefore already exposes several characteristic features of the texture before any statistical scan is performed.  Exact maximal atmospheric mixing is excluded by the observed solar splitting; the mass ordering follows from the labelled $\rho_i$; the absolute mass scale and $\Sigma_\nu$ are tied to the same angular solution through one common scale $b$; and $|a|$ is directly identified with the observable $m_{\beta\beta}$.  These relations explain why the numerical analysis below is strongly constrained rather than being a fit of independent mass-matrix entries.  We now quantify these consequences using the full oscillation likelihood and the corresponding phenomenological constraints.

\subsection{Numerical Analysis and Phenomenological Constraints}
\label{subsec:triadic_numerical}

The numerical analysis is performed in the five-dimensional input space
\begin{equation}
\left\{\sin^2\theta_{12},\,\sin^2\theta_{13},\,\sin^2\theta_{23},\,\delta_{\rm CP},\,\Delta m^2_{21}\right\}.
\label{eq:scan_inputs_2c}
\end{equation}
The ranges are taken from the union of the corresponding NuFIT~6.1 $3\sigma$ intervals for NO and IO \cite{r14,r15}. We use the union because the ordering is not imposed while generating the input points. It is identified only after the labelled neutrino spectrum is obtained.

\begin{table}[!htbp]
\centering
\small
\caption{Input ranges used in the numerical scan. They correspond to the union of the NuFIT~6.1 $3\sigma$ ranges for NO and IO.}
\label{tab:scan_ranges_2c}
\begin{tabularx}{\textwidth}{Y>{\centering\arraybackslash}p{0.38\textwidth}}
\toprule
Observable & Scan range \\
\midrule
$\sin^2\theta_{12}$ & $0.2893-0.3295$ \\
$\sin^2\theta_{13}$ & $0.02070-0.02433$ \\
$\sin^2\theta_{23}$ & $0.432-0.590$ \\
$\delta_{\rm CP}\;(^\circ)$ & $114-405$ \\
$\Delta m^2_{21}\;(\mathrm{eV}^2)$ & $(7.236-7.823)\times10^{-5}$ \\
\bottomrule
\end{tabularx}
\end{table}

For the production run we use a five-dimensional Latin-hypercube sample \cite{r98} with
\begin{equation}
N=2^{22}=4\,194\,304
\label{eq:lhs_size_2c}
\end{equation}
points, with a fixed random seed so that the scan can be reproduced. At every point, $A$, $T$ and $D$ are reconstructed from Eqs.~\eqref{eq:triadic_A_solution}--\eqref{eq:triadic_D_solution}. The quantities $F_i$ and $\rho_i$ are then calculated and the labels $1,2,3$ are kept fixed throughout. The solar mass-squared difference fixes the positive common scale $b$, while the atmospheric splitting and the mass ordering are predictions of the labelled spectrum. Table~\ref{tab:input_output_2c} summarises which quantities are inputs and which are reconstructed or predicted.

\raggedbottom
\begin{table}[H]
\centering
\small
\caption{Input--output structure used in the numerical analysis.}
\label{tab:input_output_2c}
\begingroup
\renewcommand{\arraystretch}{1.16}
\begin{tabularx}{\textwidth}{>{\bfseries\raggedright\arraybackslash}p{0.24\textwidth}Y}
\toprule
Role & Quantities \\
\midrule
Scan inputs & $\sin^2\theta_{12}$, $\sin^2\theta_{13}$, $\sin^2\theta_{23}$, $\delta_{\rm CP}$, $\Delta m^2_{21}$ \\
Analytically reconstructed & $A,T,D$, $F_{1,2,3}$ and $\rho_{1,2,3}$ \\
Common scale & $b>0$, fixed consistently from the labelled spectrum \\
Predicted oscillation information & mass ordering and $|\Delta m^2_{3\ell}|$ \\
Physical outputs & $m_1,m_2,m_3$, $\Sigma_\nu$, $m_\beta$, $m_{\beta\beta}$, $\alpha$, $\beta$, $a$, $t$, $d$ \\
Statistical outputs & $\chi^2_{\min}$, profile $\Delta\chi^2$ and confidence intervals \\
\bottomrule
\end{tabularx}
\endgroup
\end{table}

Points close to the singular surface of the analytic solution are checked separately before the closed-form expressions are used. After the texture and oscillation requirements are imposed, but before any cosmological mass-sum classification, we obtain $18\,205$ solutions for NO and $6\,943$ solutions for IO. Thus, the texture allows both orderings in the convention used here. The corresponding sampled ranges are approximately
\begingroup
\setlength{\abovedisplayskip}{5pt}
\setlength{\belowdisplayskip}{5pt}
\begin{align}
0.43202&\lesssim\sin^2\theta_{23}\lesssim0.55274 && (\mathrm{NO}),\\
0.47313&\lesssim\sin^2\theta_{23}\lesssim0.58996 && (\mathrm{IO}),
\end{align}
\endgroup
with $264.34^\circ\lesssim\delta_{\rm CP}\lesssim404.999^\circ$ for NO and $202.00^\circ\lesssim\delta_{\rm CP}\lesssim274.93^\circ$ for IO. For NO we retain the extended phase interval because the allowed region is continuous through $360^\circ$.

\begin{figure}[!htbp]
\centering
\includegraphics[width=0.54\textwidth]{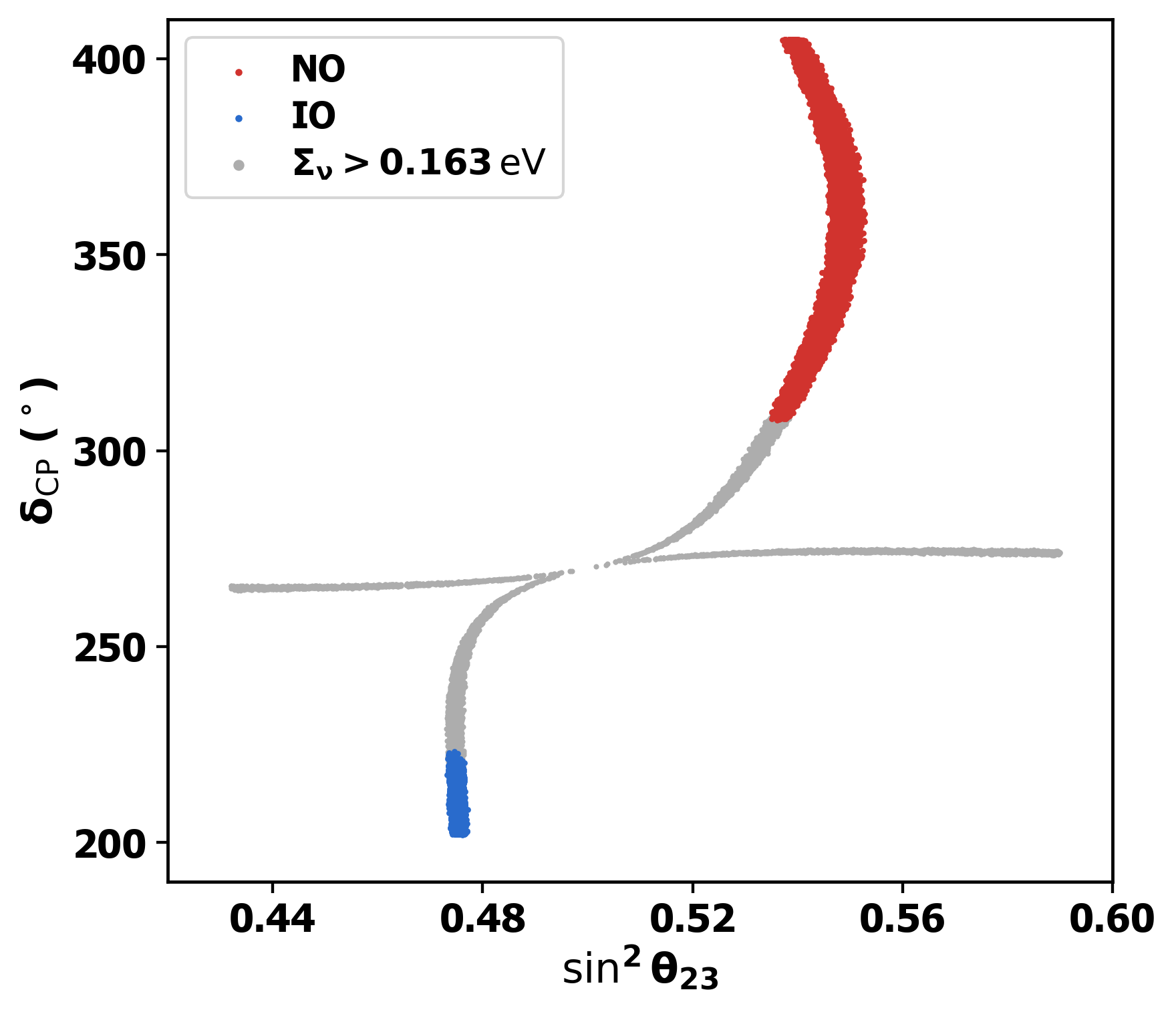}
\caption{Unweighted texture- and oscillation-consistent parameter space in the $\sin^2\theta_{23}$--$\delta_{\rm CP}$ plane. Red and blue points satisfy the $w_0w_a$CDM benchmark $\Sigma_\nu<0.163\,\mathrm{eV}$ in NO and IO, respectively; grey points lie above this benchmark. The sampled viable ranges are $0.43202\lesssim\sin^2\theta_{23}\lesssim0.55274$ for NO and $0.47313\lesssim\sin^2\theta_{23}\lesssim0.58996$ for IO. The NO phase is kept in an extended interval so that the region remains continuous across $360^\circ$.}
\label{fig:unweighted_2c}
\end{figure}

For the profiled results below we use the broader DESI benchmark $\Sigma_\nu<0.163\,\mathrm{eV}$ \cite{r88}; the cosmological interpretation is discussed in Subsection~\ref{subsec:absolute_mass}. This leaves $14\,724$ NO points and $1\,950$ IO points for the profiled analysis. Figure~\ref{fig:unweighted_2c} shows the corresponding unweighted texture- and oscillation-consistent model space. Red and blue points satisfy this benchmark for NO and IO, respectively, whereas the grey points lie above it. This scatter plot represents viable model points and should not be interpreted as a $\Delta\chi^2$ confidence region.

It is useful to go beyond these sampled ranges and compare the solutions with the correlated oscillation likelihood. For this purpose we use the ordering-dependent NuFIT~6.1 likelihood information \cite{r14,r15}. The solar and reactor angles are taken from the available two-dimensional likelihood in $(\sin^2\theta_{13},\sin^2\theta_{12})$, the solar mass-squared difference from its one-dimensional profile, and the atmospheric contribution from the three-dimensional likelihood in
\begin{equation}
\left(\sin^2\theta_{23},\,\Delta m^2_{3\ell},\,\delta_{\rm CP}\right),
\end{equation}
where $\ell=1$ for NO and $\ell=2$ for IO. The ordering contribution is also retained. Thus,
\begin{equation}
\chi^2_{\rm tot}=\chi^2_{13,12}+\chi^2_{21}+\chi^2_{23,3\ell,\delta}+\chi^2_{\rm ord}.
\label{eq:chi2_total_2c}
\end{equation}
The oscillation observables are not treated as independent Gaussian measurements; the likelihood prescription used throughout is the correlated combination in Eq.~\eqref{eq:chi2_total_2c}.

The best-fit oscillation parameters are listed in Table~\ref{tab:bestfit_2c}. The minimum values are
\begin{equation}
\chi^2_{\min}=6.4425\quad(\mathrm{NO}),\qquad
\chi^2_{\min}=9.6902\quad(\mathrm{IO}).
\end{equation}
The difference is therefore about $3.25$ in favour of NO for the likelihood used here. We do not interpret this number as a Gaussian significance for the ordering. The IO value also contains the NuFIT ordering offset $\chi^2_{\rm ord}=1.4859$.

\begin{table}[H]
\centering
\small
\caption{Oscillation quantities at the best-fit points of the triadic texture. The Dirac phase is quoted in the interval $[0,360^\circ)$.}
\label{tab:bestfit_2c}
\begin{tabularx}{\textwidth}{Y>{\centering\arraybackslash}p{0.24\textwidth}>{\centering\arraybackslash}p{0.24\textwidth}}
\toprule
Quantity & NO & IO \\
\midrule
$\chi^2_{\min}$ & $6.4425$ & $9.6902$ \\
$\sin^2\theta_{12}$ & $0.3080$ & $0.3080$ \\
$\sin^2\theta_{13}$ & $0.02250$ & $0.02260$ \\
$\sin^2\theta_{23}$ & $0.54943$ & $0.47475$ \\
$\theta_{23}\;(^\circ)$ & $47.84$ & $43.55$ \\
$\delta_{\rm CP}\;(^\circ)$ & $4.895$ & $220.615$ \\
$\Delta m^2_{21}\;(\mathrm{eV}^2)$ & $7.5336\times10^{-5}$ & $7.5336\times10^{-5}$ \\
$|\Delta m^2_{3\ell}|\;(\mathrm{eV}^2)$ & $2.5000\times10^{-3}$ & $2.4850\times10^{-3}$ \\
\bottomrule
\end{tabularx}
\end{table}

\begin{figure}[!htbp]
\centering
\includegraphics[width=0.76\textwidth]{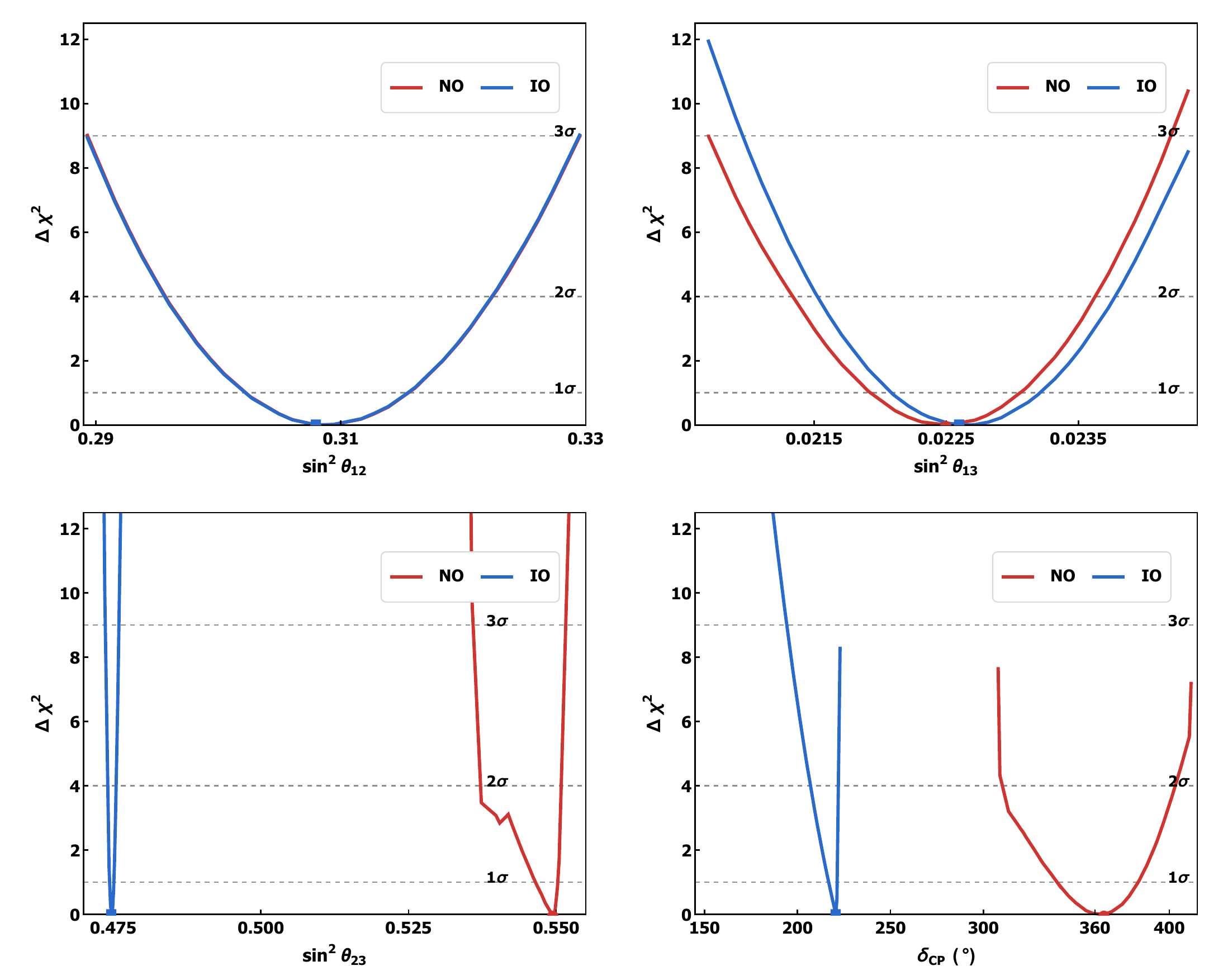}
\caption{One-dimensional profile likelihoods for $\sin^2\theta_{12}$, $\sin^2\theta_{13}$, $\sin^2\theta_{23}$ and $\delta_{\rm CP}$. NO and IO are shown separately, and the horizontal dashed lines indicate $\Delta\chi^2=1,4,9$. The best-fit atmospheric values are $\sin^2\theta_{23}=0.54943$ (NO) and $0.47475$ (IO), with $\delta_{\rm CP}=4.895^\circ$ and $220.615^\circ$, respectively. For the NO CP profile the phase is unwrapped across $360^\circ$ to display the continuous minimum correctly.}
\label{fig:1d_osc_2c}
\end{figure}

The one-dimensional profiles of the four oscillation parameters are shown in Fig.~\ref{fig:1d_osc_2c}. The solar and reactor angles stay close to their experimental preferred values. The more interesting behaviour appears in $\theta_{23}$ and $\delta_{\rm CP}$, where the two orderings clearly separate. NO is pushed to the upper octant, whereas IO is confined to a narrow lower-octant region. The CP phase also separates clearly between the two orderings.

We have also constructed separate one-dimensional profiles for the two mass-squared splittings. Fixing either $\Delta m^2_{21}$ or $|\Delta m^2_{3\ell}|$ fixes the corresponding common-scale combination, after which the remaining continuous variables are profiled. Both orderings select essentially the same solar minimum,
\begin{equation}
\Delta m^2_{21}=7.5336\times10^{-5}\,\mathrm{eV}^2,
\end{equation}
whereas the atmospheric minima are
\begin{equation}
|\Delta m^2_{3\ell}|=2.5000\times10^{-3}\,\mathrm{eV}^2\;(\mathrm{NO}),\qquad
|\Delta m^2_{3\ell}|=2.4850\times10^{-3}\,\mathrm{eV}^2\;(\mathrm{IO}).
\end{equation}
The upper side of the IO atmospheric profile is already influenced by the $\Sigma_\nu=0.163\,\mathrm{eV}$ boundary.

\begin{figure}[!htbp]
\centering
\includegraphics[width=0.44\textwidth]{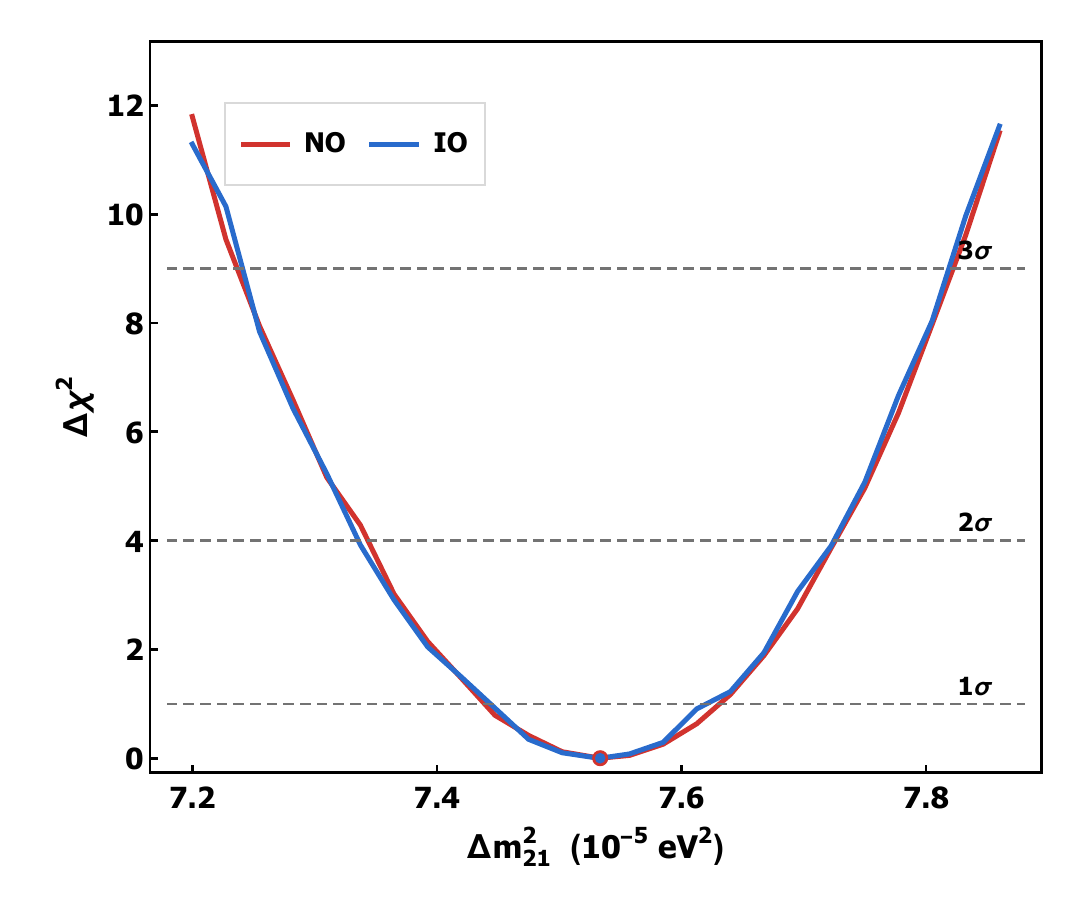}
\hspace{0.035\textwidth}
\includegraphics[width=0.44\textwidth]{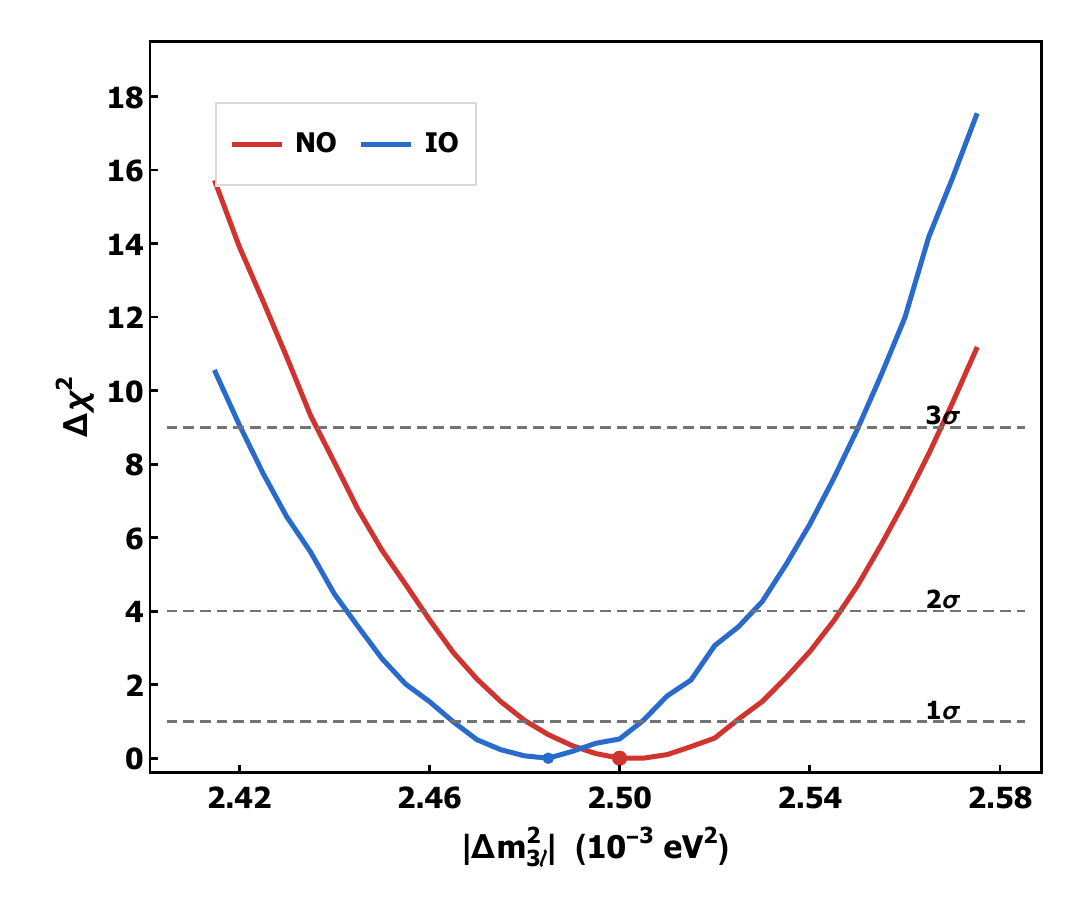}
\caption{One-dimensional profile likelihoods for the mass-squared splittings. Left: $\Delta m^2_{21}$. Right: $|\Delta m^2_{3\ell}|$, with $\ell=1$ for NO and $\ell=2$ for IO. Red and blue curves denote NO and IO, respectively. The dashed horizontal lines indicate the nominal one-parameter reference levels $\Delta\chi^2=1,4,9$.}
\label{fig:masssplit_2c}
\end{figure}

The corresponding intervals, obtained by linear interpolation between neighbouring points of the profiles, are given in Table~\ref{tab:masssplit_intervals_2c}.

\begin{table}[!htbp]
\centering
\scriptsize
\caption{One-dimensional profile intervals for the mass-squared splittings. The solar splitting is quoted in units of $10^{-5}\,\mathrm{eV}^2$ and the atmospheric splitting in units of $10^{-3}\,\mathrm{eV}^2$.}
\label{tab:masssplit_intervals_2c}
\begin{tabularx}{\textwidth}{>{\raggedright\arraybackslash}p{0.11\textwidth}Y>{\centering\arraybackslash}p{0.12\textwidth}>{\centering\arraybackslash}p{0.16\textwidth}>{\centering\arraybackslash}p{0.16\textwidth}>{\centering\arraybackslash}p{0.16\textwidth}}
\toprule
Ordering & Observable & Best fit & $1\sigma$ & $2\sigma$ & $3\sigma$ \\
\midrule
NO & $\Delta m^2_{21}$ & $7.5336$ & $[7.439,7.631]$ & $[7.344,7.726]$ & $[7.237,7.822]$ \\
IO & $\Delta m^2_{21}$ & $7.5336$ & $[7.444,7.621]$ & $[7.336,7.725]$ & $[7.241,7.819]$ \\
\addlinespace
NO & $|\Delta m^2_{3\ell}|$ & $2.5000$ & $[2.480,2.524]$ & $[2.459,2.546]$ & $[2.436,2.568]$ \\
IO & $|\Delta m^2_{3\ell}|$ & $2.4850$ & $[2.465,2.505]$ & $[2.443,2.528]$ & $[2.420,2.550]$ \\
\bottomrule
\end{tabularx}
\end{table}

\begin{figure}[!htbp]
\centering
\includegraphics[width=0.80\textwidth]{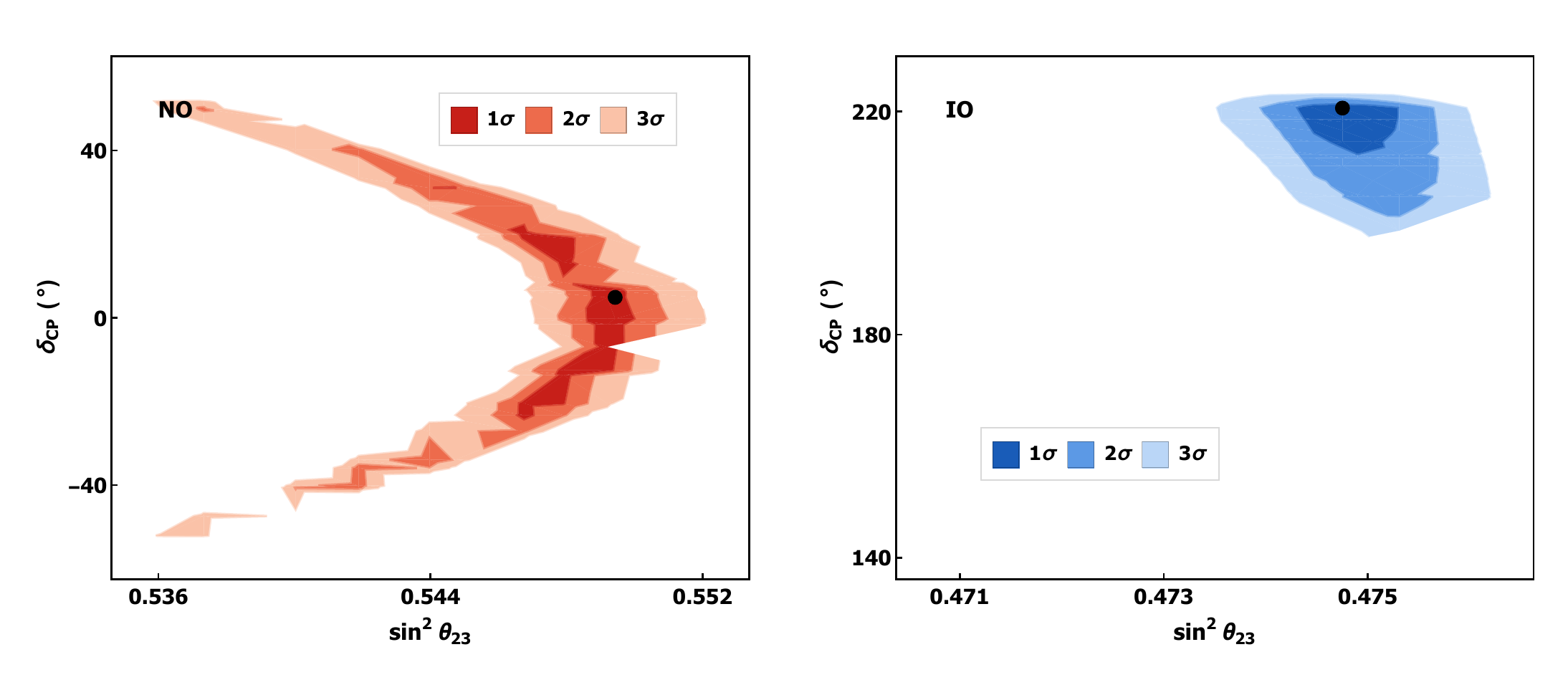}
\caption{Profiled regions in the $\sin^2\theta_{23}$--$\delta_{\rm CP}$ plane. NO is shown on the left and IO on the right. The contours correspond to the nominal two-parameter reference levels $\Delta\chi^2=2.30$, $6.18$, and $11.83$. The best-fit markers are at $(0.54943,4.895^\circ)$ for NO and $(0.47475,220.615^\circ)$ for IO.}
\label{fig:2d_2c}
\end{figure}

The difference between the two orderings becomes more transparent in the plane of $\sin^2\theta_{23}$ and $\delta_{\rm CP}$. The corresponding profiled regions are shown in Fig.~\ref{fig:2d_2c}. For two profiled parameters we display
\begin{equation}
\Delta\chi^2=2.30,\;6.18,\;11.83
\end{equation}
as the nominal $1\sigma$, $2\sigma$ and $3\sigma$ likelihood-ratio reference levels. They should not be understood as a separate coverage analysis. NO is concentrated near the upper-octant solution and the CP-phase boundary, whereas IO occupies a narrow lower-octant strip around $\delta_{\rm CP}\simeq220^\circ$.

\paragraph{Sensitivity to translated likelihood surfaces.}
The prediction of a rather restricted atmospheric region naturally raises another question: how much will the result change if a future global-fit best-fit point moves from its present position? We address this by translating the atmospheric likelihood surface by $\pm1\sigma$ and $\pm2\sigma$ separately along $\sin^2\theta_{23}$ and $\delta_{\rm CP}$. For every translated surface the texture conditions are kept unchanged and the remaining continuous parameters are minimised again. We define
\begin{equation}
\Delta\chi^2_{\min}=\chi^2_{\min,\mathrm{translated}}-\chi^2_{\min,\mathrm{present}}.
\end{equation}

\begin{figure}[!htbp]
\centering
\includegraphics[width=0.98\textwidth]{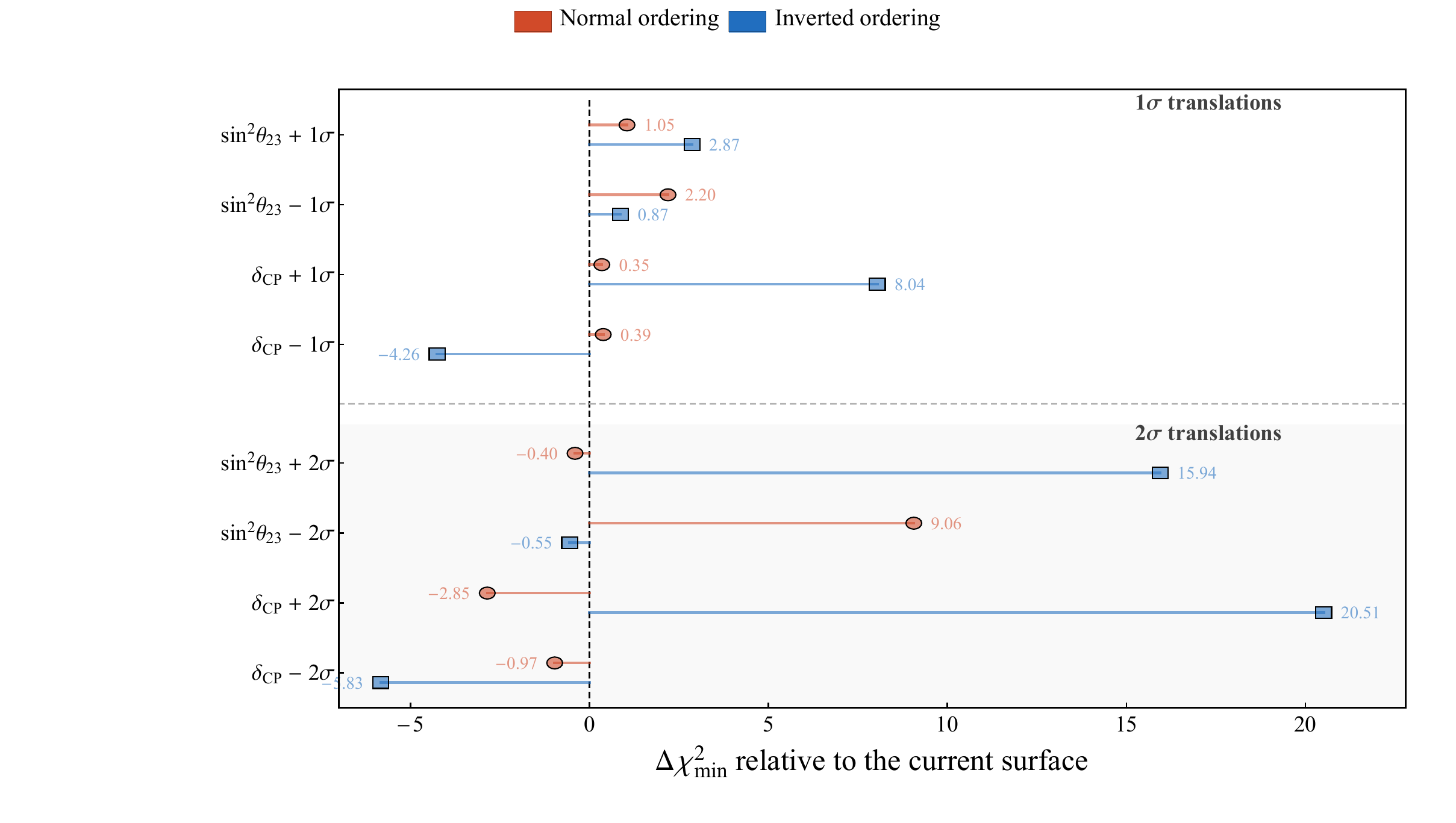}
\caption{Likelihood-translation sensitivity test. The plotted quantity is $\Delta\chi^2_{\min}$ after translating the oscillation likelihood surface by $\pm1\sigma$ and $\pm2\sigma$ in $\sin^2\theta_{23}$ and $\delta_{\rm CP}$. Red circles and blue squares denote NO and IO, respectively. Positive values correspond to a higher translated minimum, while negative values indicate a lower translated minimum. This test is distinct from the profile confidence regions.}
\label{fig:sensitivity_2c}
\end{figure}
A constrained minimum is obtained for every translation considered. The complete set of shifts is displayed directly in Fig.~\ref{fig:sensitivity_2c}. The NO solution changes only moderately under the $1\sigma$ translations, the largest increase being $\Delta\chi^2_{\min}=2.20$. The IO solution is more sensitive, particularly to displacement of $\delta_{\rm CP}$. For example, the $+1\sigma$ CP translation gives $8.04$, whereas the $-1\sigma$ translation gives $-4.26$. The asymmetry becomes stronger at $2\sigma$. We use this exercise only as a sensitivity diagnostic: it shows how the constrained minimum responds to controlled changes of the oscillation likelihood, rather than defining another confidence region.

\FloatBarrier

\subsection{Absolute Neutrino Mass Scale, Majorana Phases and Neutrinoless Double Beta Decay}
\label{subsec:absolute_mass}

The analytic relations derived in Subsection~\ref{subsec:triadic_analytic} show that the absolute neutrino mass scale is not independent of the mixing solution. It is therefore useful to examine the cosmological information and the mass observables together. We keep two recent DESI DR2 benchmark limits distinct,
\begin{equation}
\Sigma_\nu<0.0642\,\mathrm{eV}\qquad(\Lambda\mathrm{CDM}),
\end{equation}
and
\begin{equation}
\Sigma_\nu<0.163\,\mathrm{eV}\qquad(w_0w_a\mathrm{CDM}),
\end{equation}
which correspond to different cosmological assumptions \cite{r88}. These published limits are used here as external benchmark classifications; we do not recompute the cosmological likelihood for the model-specific non-degenerate spectrum. None of the texture- and oscillation-consistent points satisfies the tighter $\Lambda$CDM benchmark. Under the broader $w_0w_a$CDM benchmark, $14\,724$ NO points and $1\,950$ IO points remain.

\begin{figure}[H]
\centering
\includegraphics[width=0.52\textwidth]{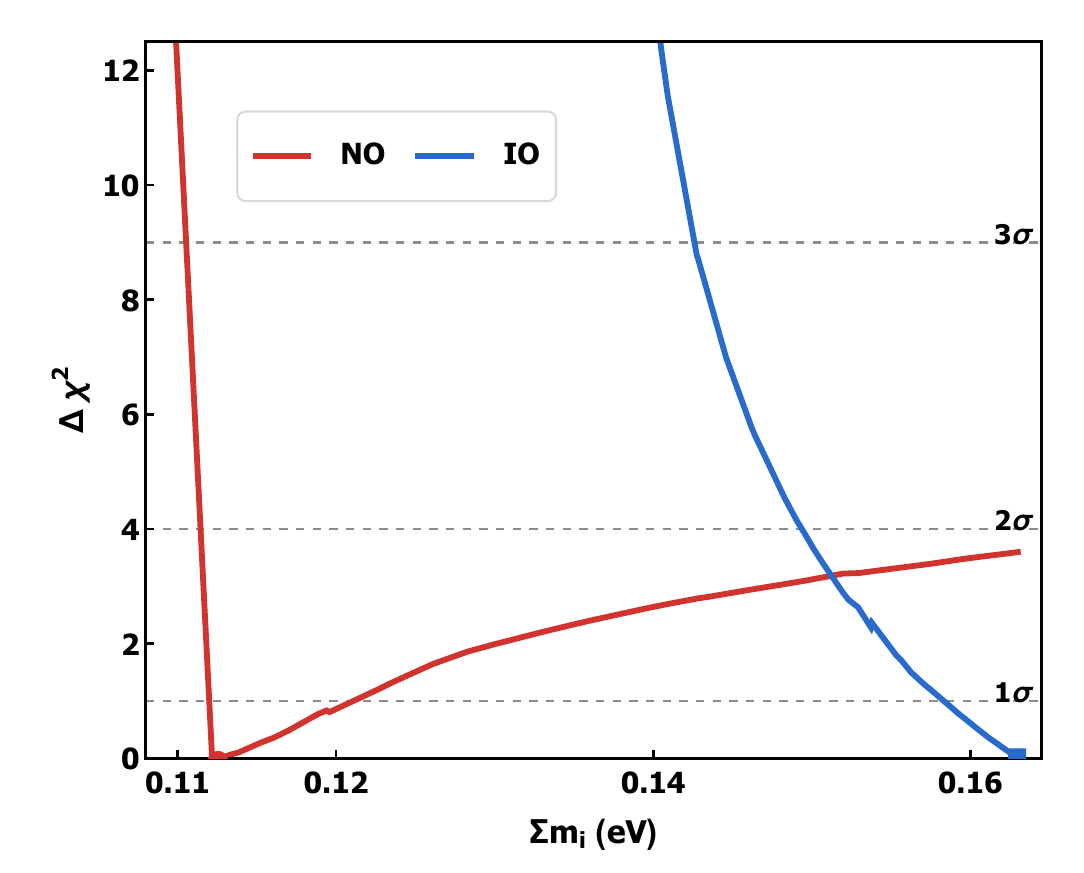}
\caption{One-dimensional profile likelihood for the neutrino mass sum $\Sigma_\nu$ in NO and IO. The dashed horizontal lines denote $\Delta\chi^2=1,4,9$. The IO minimum lies at the $w_0w_a$CDM benchmark boundary, while the NO minimum occurs at a smaller mass sum.}
\label{fig:sigma_profile_2d}
\end{figure}

Figure~\ref{fig:sigma_profile_2d} shows the one-dimensional profile of the neutrino mass sum. The NO best fit occurs at $\Sigma_\nu=0.11259\,\mathrm{eV}$, while the IO minimum is driven to $\Sigma_\nu\simeq0.163\,\mathrm{eV}$, essentially at the adopted benchmark boundary.

The principal mass and phase predictions at the two global minima are collected in Table~\ref{tab:mass_phase_bestfit_2d}. The Majorana phases are outputs of the analytic reconstruction rather than scan inputs.

\begin{table}[H]
\centering
\small
\caption{Absolute-mass observables and Majorana phases at the global best-fit points.}
\label{tab:mass_phase_bestfit_2d}
\begin{tabular*}{\textwidth}{@{\extracolsep{\fill}} lcc}
\toprule
Quantity & NO & IO \\
\midrule
$m_1\;(\mathrm{eV})$ & $0.02717$ & $0.06214$ \\
$m_2\;(\mathrm{eV})$ & $0.02852$ & $0.06275$ \\
$m_3\;(\mathrm{eV})$ & $0.05690$ & $0.03811$ \\
$\Sigma_\nu\;(\mathrm{eV})$ & $0.11259$ & $0.16300$ \\
$m_\beta\;(\mathrm{eV})$ & $0.02858$ & $0.06189$ \\
$m_{\beta\beta}\;(\mathrm{eV})$ & $0.02571$ & $0.06089$ \\
$\alpha\;(^{\circ})$ & $-89.24$ & $-86.83$ \\
$\beta\;(^{\circ})$ & $-89.30$ & $-86.79$ \\
\bottomrule
\end{tabular*}
\end{table}

\begin{figure}[H]
\centering
\includegraphics[width=0.54\textwidth]{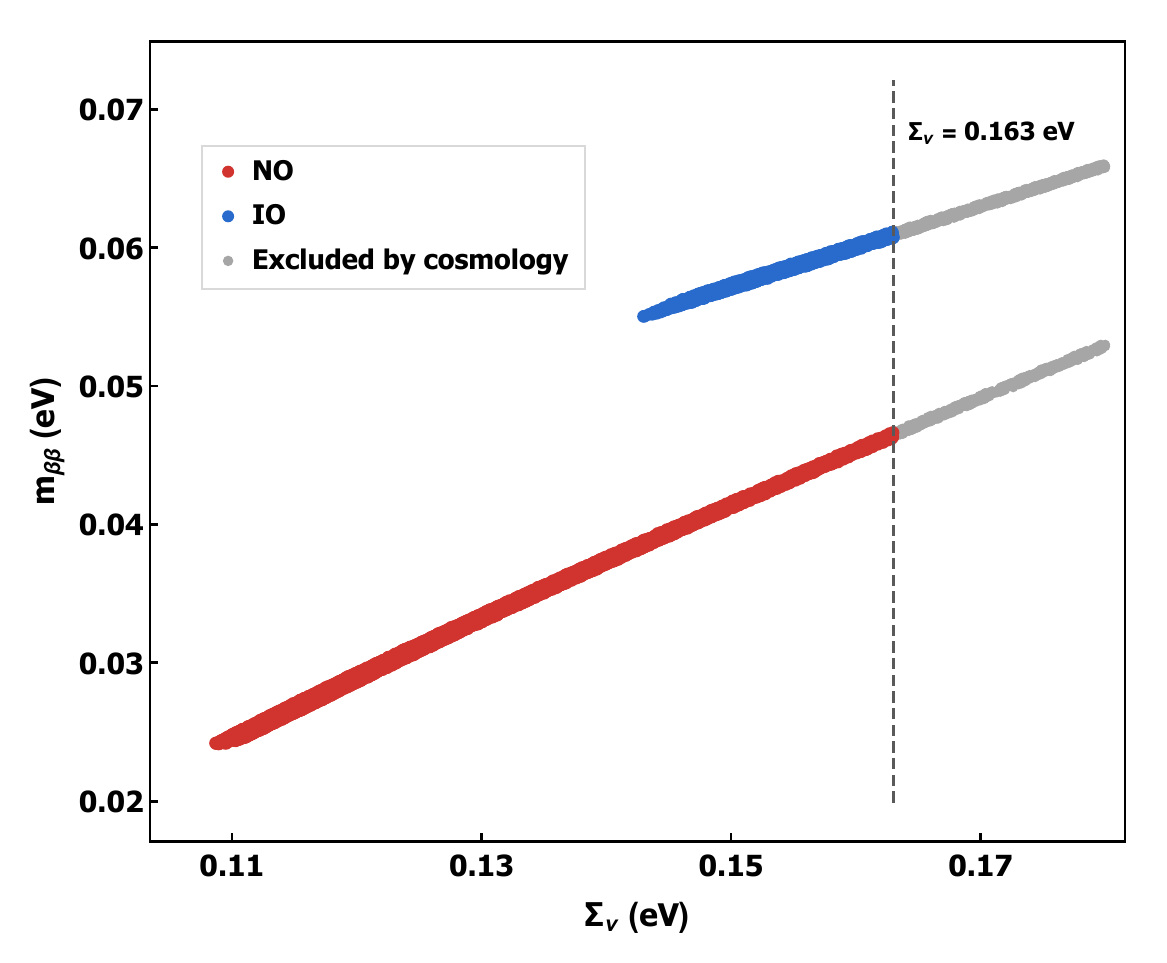}
\caption{Correlation between the effective Majorana mass $m_{\beta\beta}$ and the total neutrino mass $\Sigma_\nu$. Red and blue branches denote NO and IO points satisfying the $w_0w_a$CDM benchmark $\Sigma_\nu<0.163\,\mathrm{eV}$. Grey points show the near-boundary continuation above $0.163\,\mathrm{eV}$ within the focused plotting window $0.105\leq\Sigma_\nu\leq0.180\,\mathrm{eV}$. The vertical dashed line marks $\Sigma_\nu=0.163\,\mathrm{eV}$.}
\label{fig:mbb_sum_2d}
\end{figure}

The texture gives a sharp relation between the effective Majorana mass and the total neutrino mass. For the points satisfying the $w_0w_a$CDM benchmark, we have
\begin{align}
0.10869&\lesssim\Sigma_\nu\lesssim0.16298\;\mathrm{eV},
&0.02413&\lesssim m_{\beta\beta}\lesssim0.04665\;\mathrm{eV}
&& (\mathrm{NO}),\\
0.14298&\lesssim\Sigma_\nu\lesssim0.16300\;\mathrm{eV},
&0.05503&\lesssim m_{\beta\beta}\lesssim0.06113\;\mathrm{eV}
&& (\mathrm{IO}).
\end{align}
These branches are shown in Fig.~\ref{fig:mbb_sum_2d}. The coloured portions lie below $\Sigma_\nu=0.163\,\mathrm{eV}$; the grey portions show only the near-boundary continuation above this value in the focused plotting window, rather than the full high-mass rejected sample. The NO and IO branches remain well separated in $m_{\beta\beta}$ over the cosmologically relevant range.

The same reconstruction also fixes the texture parameters themselves. None of the quantities in Table~\ref{tab:texture_bestfit_2d} is independently scanned. Because $b$ is real and positive, the phases of $a$, $t$, and $d$ coincide with those of $A$, $T$, and $D$, respectively.

\begin{table}[!t]
\centering
\small
\caption{Best-fit reconstruction of the texture. The phases of $a$, $t$, and $d$ are not repeated because, for $b>0$, they are identical to those of $A$, $T$, and $D$, respectively.}
\label{tab:texture_bestfit_2d}
\begin{tabular*}{\textwidth}{@{\extracolsep{\fill}} lcc}
\toprule
Quantity & NO & IO \\
\midrule
$b\;(\mathrm{eV})$ & $0.008833$ & $0.007823$ \\
$|A|$ & $2.9104$ & $7.7840$ \\
$\arg A\;(^{\circ})$ & $-3.26$ & $-170.97$ \\
$|T|$ & $4.6805$ & $6.3525$ \\
$\arg T\;(^{\circ})$ & $178.38$ & $13.00$ \\
$|D|$ & $1.0495$ & $1.3727$ \\
$\arg D\;(^{\circ})$ & $-178.46$ & $174.26$ \\
$|a|\;(\mathrm{eV})$ & $0.025707$ & $0.060891$ \\
$|t|\;(\mathrm{eV})$ & $0.041342$ & $0.049693$ \\
$|d|\;(\mathrm{eV})$ & $0.009270$ & $0.010738$ \\
$\alpha\;(^{\circ})$ & $-89.24$ & $-86.83$ \\
$\beta\;(^{\circ})$ & $-89.30$ & $-86.79$ \\
\bottomrule
\end{tabular*}
\end{table}

For NO the best-fit texture has $|A|\simeq2.91$ and $|T|\simeq4.68$, with both $T$ and $D$ lying close to the negative real axis. The IO solution requires a substantially larger $|A|\simeq7.78$ and $|T|\simeq6.35$, with a qualitatively different phase pattern. In both orderings the predicted Majorana phases lie close to $-90^\circ$.

Finally, all $25\,148$ texture- and oscillation-consistent rows prior to cosmological classification pass the implemented consistency tests. These include the ordering assignment, the common-scale consistency, direct reconstruction/congruence checks and reproduction of the mass-squared splittings. The largest recorded numerical residual is $6.3\times10^{-13}$. The one-dimensional profile sets also cross the displayed nominal $1\sigma$, $2\sigma$, and $3\sigma$ reference levels for every observable shown above. These checks support the internal numerical consistency of the stored solution set and the profile results.

The discussion in this section has deliberately been model independent.  We have seen that the two texture correlations already lead to non-trivial consequences for the atmospheric angle, the ordering, the absolute neutrino mass scale, the Majorana phases and $m_{\beta\beta}$.  The next question is whether the same structure can arise from an explicit flavour construction without inserting the texture by hand.  We now turn to this question and show how the triadic form is obtained from an $A_4$-based effective theory in two commonly used representation bases.  The scalar-sector consistency of the corresponding vacuum alignments will be examined separately in the following section.

\FloatBarrier

\section{The Triadic Texture from a Symmetry Perspective}
\label{sec:symmetry_realisation}

In this section, we present a concrete model that realises the proposed triadic texture. We adopt a hybrid framework in which the light neutrino mass matrix receives contributions from both a Type-I seesaw mechanism and a Weinberg-like dimension-six operator. As the flavour symmetry, we choose the non-Abelian discrete group $A_4$, supplemented by the auxiliary Abelian symmetries $Z_{10}\otimes Z_7\otimes Z_5\otimes Z_3$. Apart from the SM fermions, we introduce three RH neutrino fields $\nu_{eR}$, $\nu_{\mu R}$ and $\nu_{\tau R}$. The scalar sector is also enriched with seven new scalars ($\Phi$, $\xi$, $\kappa$, $\chi$, $\eta$, $\rho$ and $\psi$) along with the SM Higgs, $H$.

\subsection{Model Setup}
\label{subsec:model_setup_eft}

The field assignments under
\begin{equation}
SU(2)_L\otimes U(1)_Y\otimes A_4\otimes Z_{10}\otimes Z_7\otimes Z_5\otimes Z_3
\label{eq:full_flavour_group}
\end{equation}
are given in Table~\ref{tab:field_assignments}. The relevant effective Yukawa Lagrangian is

\begin{table}[!htbp]
\centering
\scriptsize
\setlength{\tabcolsep}{3.2pt}
\caption{Field assignments under $A_4$, the electroweak gauge group and the additional cyclic symmetries.  The Abelian entries are additive charges except for the displayed $Z_3$ phases.}
\label{tab:field_assignments}
\resizebox{0.88\textwidth}{!}{%
\begin{tabular}{lccccccccccccccc}
\toprule
 & $D_L$ & $e_R$ & $\mu_R$ & $\tau_R$ & $H$ & $\nu_{eR}$ & $\nu_{\mu R}$ & $\nu_{\tau R}$ & $\Phi$ & $\xi$ & $\psi$ & $\chi$ & $\eta$ & $\rho$ & $\kappa$ \\
\midrule
$A_4$ & $3$ & $1$ & $1''$ & $1'$ & $1$ & $1''$ & $1$ & $1'$ & $3$ & $3$ & $1$ & $1$ & $1$ & $1''$ & $1$ \\
$SU(2)_L$ & $2$ & $1$ & $1$ & $1$ & $2$ & $1$ & $1$ & $1$ & $1$ & $1$ & $1$ & $1$ & $1$ & $1$ & $1$ \\
$U(1)_Y$ & $-1$ & $-2$ & $-2$ & $-2$ & $+1$ & $0$ & $0$ & $0$ & $0$ & $0$ & $0$ & $0$ & $0$ & $0$ & $0$ \\
$Z_{10}$ & $0$ & $0$ & $0$ & $0$ & $0$ & $0$ & $4$ & $3$ & $0$ & $0$ & $7$ & $6$ & $2$ & $0$ & $7$ \\
$Z_7$ & $-2$ & $4$ & $4$ & $4$ & $1$ & $6$ & $6$ & $6$ & $0$ & $5$ & $0$ & $0$ & $2$ & $2$ & $2$ \\
$Z_5$ & $0$ & $2$ & $2$ & $2$ & $0$ & $2$ & $0$ & $0$ & $3$ & $0$ & $2$ & $2$ & $0$ & $1$ & $3$ \\
$Z_3$ & $\omega^2$ & $\omega$ & $\omega$ & $\omega$ & $\omega$ & $1$ & $1$ & $1$ & $1$ & $1$ & $1$ & $1$ & $1$ & $1$ & $1$ \\
\bottomrule
\end{tabular}}
\end{table}

\begin{align}
-\mathcal L_Y={}&
\frac{y_{L1}}{\Lambda}(\overline{D_L}\Phi)_1 H e_R
+\frac{y_{L2}}{\Lambda}(\overline{D_L}\Phi)_{1'}H\mu_R
+\frac{y_{L3}}{\Lambda}(\overline{D_L}\Phi)_{1''}H\tau_R \notag\\
&+\frac{y_1}{\Lambda}(\overline{D_L}\Phi)_{1'}\widetilde H\nu_{eR}
+\frac{y_2}{\Lambda^2}(\overline{D_L}\Phi)_1\widetilde H\nu_{\mu R}\chi
+\frac{y_3}{\Lambda^2}(\overline{D_L}\Phi)_{1''}\widetilde H\nu_{\tau R}\psi \notag\\
&+\frac{y_c}{2}(\overline{\nu^c_{eR}}\nu_{eR})_{1'}\rho
+\frac{y_b}{2}\Big[(\overline{\nu^c_{eR}}\nu_{\tau R})_1+(\overline{\nu^c_{\tau R}}\nu_{eR})_1\Big]\kappa
+\frac{y_a}{2}(\overline{\nu^c_{\mu R}}\nu_{\mu R})_1\eta \notag\\
&+\frac{y_w}{\Lambda^2}\Big[(D_LD_L^c)_{3_s}\,\xi\Big]_1\widetilde H\widetilde H
+\mathrm{h.c.}
\label{eq:yukawa_effective}
\end{align}
The cyclic symmetries forbid unwanted lower dimensional operators and organise the effective expansion. Operators of dimension higher than those retained in Eq.~\eqref{eq:yukawa_effective} carry additional powers of the flavon ratios $|v_X|/\Lambda$. We therefore work with $|v_X|/\Lambda<1$, for which the effective-operator hierarchy is controlled.

The group $A_4$ can be represented in different bases \cite{r48}. In the A--F basis the generator $T$ is diagonal, whereas in the M--R basis $S$ is diagonal. The relevant representation matrices and product rules are summarised in Appendix~\ref{app:A4_rules}. We now show the realisation of the same leading texture in these two bases.

\subsection{The Triadic Texture in the Altarelli--Feruglio Basis}
\label{subsec:AF_realisation}

In the A--F basis, the VEV assignments of the $A_4$ triplet fields $\Phi$ and $\xi$ are chosen as
\begin{equation}
\langle\Phi\rangle_0=v_\Phi(1,0,0)^T,
\qquad
\langle\xi\rangle_0=v_\xi(0,-1,-1)^T,
\label{eq:AF_vevs}
\end{equation}
while the VEVs of the singlet fields are $\langle\kappa\rangle_0=v_\kappa$, $\langle\chi\rangle_0=v_\chi$, $\langle\eta\rangle_0=v_\eta$, $\langle\rho\rangle_0=v_\rho$, $\langle\psi\rangle_0=v_\psi$ and $\langle H\rangle_0=v_H$. With these choices, the charged-lepton mass matrix is diagonal and is given by
\begin{equation}
M_L=\frac{v_Hv_\Phi}{\Lambda}
\begin{pmatrix}
y_{L1}&0&0\\
0&y_{L2}&0\\
0&0&y_{L3}
\end{pmatrix}.
\label{eq:AF_ML}
\end{equation}
The Dirac and RH Majorana mass matrices are
\begin{equation}
M_D=
\begin{pmatrix}
0&\dfrac{y_2v_Hv_\Phi v_\chi}{\Lambda^2}&0\\[2mm]
\dfrac{y_1v_Hv_\Phi}{\Lambda}&0&0\\[2mm]
0&0&\dfrac{y_3v_Hv_\Phi v_\psi}{\Lambda^2}
\end{pmatrix},
\qquad
M_R=
\begin{pmatrix}
\dfrac{y_cv_\rho}{2\sqrt2}&0&\dfrac{y_bv_\kappa}{2\sqrt2}\\[2mm]
0&\dfrac{y_av_\eta}{2\sqrt2}&0\\[2mm]
\dfrac{y_bv_\kappa}{2\sqrt2}&0&0
\end{pmatrix}.
\label{eq:AF_MD_MR}
\end{equation}
The mass matrix $M_w$ coming from the Weinberg-like dim-6 operator is
\begin{equation}
M_w=\frac{y_wv_H^2v_\xi}{6\sqrt2\Lambda^2}
\begin{pmatrix}
0&1&1\\
1&-2&0\\
1&0&-2
\end{pmatrix}.
\label{eq:AF_Mw}
\end{equation}
As $M_L$ is diagonal, the final light neutrino mass matrix in the flavour basis (the basis where the charged-lepton mass matrix is diagonal) is given by
\begin{equation}
M_\nu=-M_D M_R^{-1}M_D^T+M_w.
\label{eq:AF_seesaw}
\end{equation}
The $M_\nu$ obtained from the model concurs with the triadic texture of Eq.~\eqref{eq:triadic_texture}, with the following relations,
\begin{align}
a&=-\frac{v_H^2v_\Phi^2v_\chi^2y_2^2}{4v_\eta y_a\Lambda^4},
&
 b&=\frac{v_H^2v_\xi y_w}{6\sqrt2\Lambda^2},\notag\\
t&=-\frac{v_H^2v_\Phi^2v_\psi y_1y_3}{2\sqrt2\,v_\kappa y_b\Lambda^3},
&
 d&=\frac{v_H^2v_\rho v_\Phi^2v_\psi^2y_3^2y_c}{4v_\kappa^2y_b^2\Lambda^4}
-\frac{v_H^2v_\xi y_w}{3\sqrt2\Lambda^2}.
\label{eq:AF_texture_parameters}
\end{align}
Thus, the A--F alignment reproduces the required triadic mass matrix at the level of the effective fermion operators. In the next subsection, we obtain the same texture in the M--R basis of $A_4$. Whether the two chosen alignments are supported by the scalar potential will be checked separately in Section~\ref{sec:vacuum_scalar}.

\subsection{The Triadic Texture in the Ma--Rajasekaran Basis}
\label{subsec:MR_realisation}

In the M--R basis, the VEV assignments of the $A_4$ triplet fields $\Phi$ and $\xi$ are chosen as
\begin{equation}
\langle\Phi\rangle_0=v_\Phi(1,1,1)^T,
\qquad
\langle\xi\rangle_0=v_\xi(1,-1,1)^T.
\label{eq:MR_vevs}
\end{equation}
The rest of the VEV assignments are the same as in the A--F basis. With these choices, $M_L$ is not diagonal, but has the special form
\begin{equation}
M_L=\frac{v_Hv_\Phi}{\Lambda}
\begin{pmatrix}
y_{L1}&y_{L2}&y_{L3}\\
y_{L1}&\omega y_{L2}&\omega^2y_{L3}\\
y_{L1}&\omega^2y_{L2}&\omega y_{L3}
\end{pmatrix},
\qquad
\omega=\frac{-1+i\sqrt3}{2}.
\label{eq:MR_ML}
\end{equation}
This $M_L$ can be diagonalised by singular value decomposition, $M_L^{\rm diag}=U_L^\dagger M_LU_R$. The two diagonalising matrices are
\begin{equation}
U_L=\frac1{\sqrt3}
\begin{pmatrix}
1&1&1\\
1&\omega&\omega^2\\
1&\omega^2&\omega
\end{pmatrix},
\qquad
U_R=\mathbf 1.
\label{eq:MR_UL_UR}
\end{equation}
The resulting diagonal matrix is $M_L^{\rm diag}=\sqrt3(v_Hv_\Phi/\Lambda)\operatorname{diag}(y_{L1},y_{L2},y_{L3})$.

The corresponding neutrino matrices are
\begin{equation}
M_D=v_\Phi
\begin{pmatrix}
\dfrac{y_1v_H}{2\Lambda}&\dfrac{y_2v_Hv_\chi}{2\sqrt2\Lambda^2}&\dfrac{y_3v_Hv_\psi}{2\sqrt2\Lambda^2}\\[2mm]
\dfrac{\omega y_1v_H}{2\Lambda}&\dfrac{y_2v_Hv_\chi}{2\sqrt2\Lambda^2}&\dfrac{\omega^2y_3v_Hv_\psi}{2\sqrt2\Lambda^2}\\[2mm]
\dfrac{\omega^2 y_1v_H}{2\Lambda}&\dfrac{y_2v_Hv_\chi}{2\sqrt2\Lambda^2}&\dfrac{\omega y_3v_Hv_\psi}{2\sqrt2\Lambda^2}
\end{pmatrix},
\label{eq:MR_MD}
\end{equation}
with the same $M_R$ as in Eq.~\eqref{eq:AF_MD_MR}, and
\begin{equation}
M_w=\frac{y_wv_\xi v_H^2}{2\sqrt2\Lambda^2}
\begin{pmatrix}
0&1&-1\\
1&0&1\\
-1&1&0
\end{pmatrix}.
\label{eq:MR_Mw}
\end{equation}

Unlike the A--F basis, $M_L$ is not diagonal in the M--R basis. The matrices in Eq.~\eqref{eq:MR_UL_UR} provide one valid diagonalisation, but the diagonal entries must still be assigned to the physical $e$, $\mu$ and $\tau$ states. To obtain the desired triadic texture of Eq.~\eqref{eq:triadic_texture} in the physical charged-lepton basis, we adopt an alternative but equally valid diagonalisation, obtained through the following permutations and phase redefinitions:
\begin{equation}
U'_L=U_L(R_1\leftrightarrow R_2),
\qquad
U'_R=\operatorname{diag}(1,\omega^2,\omega)\,U_R(R_2\leftrightarrow R_3),
\label{eq:PCLC_matrices}
\end{equation}
for which
\begin{equation}
(M_L^{\rm diag})'=U_L^{\prime\dagger}M_LU'_R
=\sqrt3\frac{v_Hv_\Phi}{\Lambda}\operatorname{diag}(y_{L1},y_{L3},y_{L2}).
\label{eq:PCLC_diag}
\end{equation}
This choice is the permuted charged lepton correction (PCLC) of Ref.~\cite{r67}. The coefficients $y_{L1}$, $y_{L2}$ and $y_{L3}$ are Lagrangian parameters in the flavour basis; they are not identified with the physical charged-lepton masses before diagonalisation and ordering. The physical $e$, $\mu$ and $\tau$ states are assigned after diagonalisation by matching the three mass eigenvalues to the observed charged-lepton spectrum. The allowed permutation in Eqs.~\eqref{eq:PCLC_matrices}--\eqref{eq:PCLC_diag} therefore changes the ordering convention, not the physical spectrum.

Once this left-handed charged-lepton ordering is fixed, the same $U'_L$ must be applied to the neutrino sector because the charged leptons and neutrinos belong to the same electroweak lepton doublets. Defining
\begin{equation}
M_\nu^{\rm sym}=-M_D M_R^{-1}M_D^T+M_w,
\end{equation}
the Majorana mass matrix in the physical charged-lepton basis is obtained by the congruence transformation
\begin{equation}
M_\nu=U_L^{\prime T}M_\nu^{\rm sym}U'_L.
\label{eq:MR_two_step}
\end{equation}
The resulting $M_\nu$ again has the form of Eq.~\eqref{eq:triadic_texture}. The corresponding relations between the texture parameters and the model parameters are
\begin{align}
a&=\frac{v_H^2\!\big(-9v_\Phi^2v_\chi^2y_2^2+2v_\eta v_\xi y_a y_w\Lambda^2\big)}{6\sqrt2\,v_\eta y_a\Lambda^4},
&
b&=\frac{v_H^2v_\xi y_w}{3\sqrt2\Lambda^2},\notag\\
t&=-\frac{v_H^2\!\big(18v_\Phi^2v_\psi y_1y_3+\sqrt2\,v_\kappa v_\xi y_b y_w\Lambda\big)}{12v_\kappa y_b\Lambda^3},
&
d&=\frac{v_H^2\!\big(18\omega v_\rho v_\Phi^2v_\psi^2y_3^2y_c-8v_\kappa^2v_\xi y_b^2y_w\Lambda^2\big)}{12\sqrt2\,v_\kappa^2y_b^2\Lambda^4}.
\label{eq:MR_texture_parameters}
\end{align}
The above relations can also be solved for the dependent couplings as
\begin{align}
y_a&=-\frac{3v_H^2v_\Phi^2v_\chi^2y_2^2}{2\sqrt2\,(a-b)v_\eta\Lambda^4},
&
y_w&=\frac{3\sqrt2\,b\Lambda^2}{v_H^2v_\xi},\notag\\
y_b&=-\frac{3v_H^2v_\Phi^2v_\psi y_1y_3}{(b+2t)v_\kappa\Lambda^3},
&
y_c&=\frac{6\omega^2\sqrt2\,(2b+d)v_H^2v_\Phi^2y_1^2}{(b+2t)^2v_\rho\Lambda^2}.
\label{eq:MR_dependent_couplings}
\end{align}

We have therefore obtained the same triadic texture from the two $A_4$ bases with two different triplet vacuum alignments. The $\Phi$ alignments are related by the usual change of $A_4$ basis, whereas the two $\xi$ alignments are not simply related in the same way. However, reproducing the required neutrino mass matrix is not enough to establish a viable model. The chosen VEV configuration must also arise as a stationary and stable vacuum of the scalar potential. We examine this point in the next section.

\section{Vacuum Structure and Viable $A_4$ Realisations}
\label{sec:vacuum_scalar}

The two $A_4$ realisations introduced in the flavour construction reproduce the
same triadic neutrino texture (Eq.\,(\ref{eq:triadic_texture})). This,
however, does not guarantee that the corresponding VEV configurations define
acceptable scalar vacua. The scalar test must start from one definite scalar
potential; only then is it meaningful to compare the stationary points and their
Hessians. We therefore first write the complete renormalisable scalar potential
used in the analysis. We then test the Altarelli--Feruglio (A--F) alignment and,
after that, the Ma--Rajasekaran (M--R) alignment. For the viable M--R branch we
subsequently study the complete neutral scalar spectrum, local stability,
boundedness from below and the global character of the vacuum.

\subsection{Complete renormalisable scalar potential and field parametrisation}
\label{subsec:full_scalar_potential}
The scalar sector contains the SM doublet $H$, the two $A_4$ triplets
$\Phi$ and $\xi$, and the singlet fields
$\psi,\chi,\eta,\rho,\kappa$. The renormalisable scalar potential can be
written as
\begin{equation}
V=V_2+V_4 ,
\label{eq:Vtotal_scalar}
\end{equation}
where the quadratic part is
\begin{equation}
V_2=
-\mu_H^2(H^\dagger H)
-\mu_\Phi^2(\Phi^\dagger\Phi)_1
-\mu_\xi^2(\xi^\dagger\xi)_1
-\sum_{s=\psi,\chi,\eta,\rho,\kappa}\mu_s^2|s|^2 .
\label{eq:V2_scalar}
\end{equation}
The imposed gauge and flavour symmetries do not allow any renormalisable
cubic scalar interaction.

For the quartic part, it is useful to keep the different classes of
interactions explicit. The self-interactions of $H$, $\Phi$ and $\xi$ are
\begin{align}
V_H^{(4)}={}&
\lambda_H(H^\dagger H)^2 ,
\label{eq:VH4}
\\
V_\Phi^{(4)}={}&
\lambda_{\Phi1}\big[(\Phi^\dagger\Phi)_1\big]^2
+\lambda_{\Phi2}
 (\Phi^\dagger\Phi)_{1'}(\Phi^\dagger\Phi)_{1''}
\notag\\
&+\lambda_{\Phi3}
 \big[(\Phi^\dagger\Phi)_{3_s}
      (\Phi^\dagger\Phi)_{3_s}\big]_1
+\lambda_{\Phi4}
 \big[(\Phi^\dagger\Phi)_{3_a}
      (\Phi^\dagger\Phi)_{3_a}\big]_1 ,
\label{eq:VPhi4}
\\
V_\xi^{(4)}={}&
\lambda_{\xi1}\big[(\xi^\dagger\xi)_1\big]^2
+\lambda_{\xi2}
 (\xi^\dagger\xi)_{1'}(\xi^\dagger\xi)_{1''}
\notag\\
&+\lambda_{\xi3}
 \big[(\xi^\dagger\xi)_{3_s}
      (\xi^\dagger\xi)_{3_s}\big]_1
+\lambda_{\xi4}
 \big[(\xi^\dagger\xi)_{3_a}
      (\xi^\dagger\xi)_{3_a}\big]_1 .
\label{eq:VXi4}
\end{align}
For the five $A_4$ singlet flavons we have
\begin{align}
V_S^{(4)}={}&
\sum_{s=\psi,\chi,\eta,\rho,\kappa}\lambda_s|s|^4
+\sum_{\substack{s<t\\
s,t=\psi,\chi,\eta,\rho,\kappa}}
\lambda_{st}|s|^2|t|^2 .
\label{eq:VS4}
\end{align}
The second term in Eq.~\eqref{eq:VS4} contains all ten pairwise
singlet--singlet quartic interactions.

The mixed interactions involving $H$, the triplets and the singlets are
\begin{align}
V_{\rm mix}^{(4)}={}&
\lambda_{H\Phi}(H^\dagger H)(\Phi^\dagger\Phi)_1
+\lambda_{H\xi}(H^\dagger H)(\xi^\dagger\xi)_1
\notag\\
&+(H^\dagger H)
 \sum_{s=\psi,\chi,\eta,\rho,\kappa}
 \lambda_{Hs}|s|^2
\notag\\
&+(\Phi^\dagger\Phi)_1
 \sum_{s=\psi,\chi,\eta,\rho,\kappa}
 \lambda_{\Phi s}|s|^2
\notag\\
&+(\xi^\dagger\xi)_1
 \sum_{s=\psi,\chi,\eta,\rho,\kappa}
 \lambda_{\xi s}|s|^2
+V_{\Phi\xi}^{(4)} .
\label{eq:Vmixed4}
\end{align}
The $\Phi$--$\xi$ part is
\begin{align}
V_{\Phi\xi}^{(4)}={}&
\lambda_{\Phi\xi1}J_1
+\lambda_{\Phi\xi2}J_2
+\lambda_{\Phi\xi4}J_4
+\lambda_{\Phi\xi5}J_5
+\lambda_{\Phi\xi8}J_8
\notag\\
&+\lambda_{\Phi\xi9}J_9
+\lambda_{\Phi\xi11}J_{11}
+\lambda_{\Phi\xi12}J_{12}
+\lambda_{\Phi\xi14}J_{14},
\label{eq:VPhiXi4}
\end{align}
where
\begin{equation}
\begin{aligned}
J_1={}&(\Phi^\dagger\Phi)_1(\xi^\dagger\xi)_1,
&\qquad
J_8={}&(\Phi^\dagger\xi)_1(\xi^\dagger\Phi)_1,
\\[1mm]
J_4={}&\big[(\Phi^\dagger\Phi)_{3_s}(\xi^\dagger\xi)_{3_s}\big]_1,
&
J_5={}&\big[(\Phi^\dagger\Phi)_{3_a}(\xi^\dagger\xi)_{3_a}\big]_1,
\\[1mm]
J_{11}={}&\big[(\Phi^\dagger\xi)_{3_s}(\xi^\dagger\Phi)_{3_s}\big]_1,
&
J_{12}={}&\big[(\Phi^\dagger\xi)_{3_a}(\xi^\dagger\Phi)_{3_a}\big]_1,
\\[1mm]
J_2={}&(\Phi^\dagger\Phi)_{1'}(\xi^\dagger\xi)_{1''}
 +(\Phi^\dagger\Phi)_{1''}(\xi^\dagger\xi)_{1'},
\\[1mm]
J_9={}&(\Phi^\dagger\xi)_{1'}(\xi^\dagger\Phi)_{1''}
 +(\Phi^\dagger\xi)_{1''}(\xi^\dagger\Phi)_{1'},
\\[1mm]
J_{14}={}&\big[(\Phi^\dagger\xi)_{3_s}(\xi^\dagger\Phi)_{3_a}\big]_1
 -\big[(\Phi^\dagger\xi)_{3_a}(\xi^\dagger\Phi)_{3_s}\big]_1 .
\end{aligned}
\label{eq:Jdefinitions}
\end{equation}
Some of the above $A_4$ contractions are related by exact recoupling
identities. We retain this notation because it is convenient for the
stationarity and scalar-mass analysis.

There is one further quartic interaction which is allowed by the stated
flavour symmetry and is also CP even. It can be written as
\begin{equation}
{\cal O}_\rho=
\big[(\Phi\Phi)_{3_s}\xi^\dagger\big]_{1''}\rho^\dagger ,
\qquad
V_{\rm add}^{(4)}
=\lambda_{\Phi\Phi\xi\rho}
\left({\cal O}_\rho+{\cal O}_\rho^\dagger\right) ,
\label{eq:Vadditional}
\end{equation}
where $\lambda_{\Phi\Phi\xi\rho}$ is real. Thus the complete CP-conserving potential has to be considered together with Eq.~\eqref{eq:Vadditional}. At this stage, $\lambda_{\Phi\Phi\xi\rho}$ is kept general. Its value is constrained only after imposing the stationarity conditions for the A--F and M--R vacuum alignments.

For studying the vacuum, the scalar fields are expanded as
\begin{align}
H&=
\begin{pmatrix}
G^+\\[1mm]
\dfrac{1}{\sqrt2}(v_H+h+i\Omega)
\end{pmatrix},
&
\Phi_j&=
\frac{1}{\sqrt2}
(v_{\Phi j}+\phi_{Rj}+i\phi_{Ij}),
\notag\\
\xi_j&=
\frac{1}{\sqrt2}
(v_{\xi j}+\xi_{Rj}+i\xi_{Ij}),
&
X&=
\frac{1}{\sqrt2}
(v_X+X_R+iX_I),
\quad
X=\psi,\chi,\eta,\rho,\kappa .
\label{eq:scalar_field_expansion}
\end{align}
Since all the additional flavons are singlets under $SU(2)_L\otimes U(1)_Y$, electroweak
symmetry breaking is entirely due to $H$. The three Standard-Model Goldstone states are therefore
identified directly from the Higgs doublet as
\begin{equation}
G^+\longrightarrow W_L^+,
\qquad
G^-\longrightarrow W_L^-,
\qquad
G^0\equiv\Omega\longrightarrow Z_L^0.
\label{eq:SM_goldstones_main}
\end{equation}
They provide the longitudinal degrees of freedom of $W^+$, $W^-$ and $Z^0$ after electroweak
symmetry breaking and are not part of the physical scalar spectrum. Since the flavons are
electroweak singlets, none of their phase directions can replace or mix with these gauge
Goldstone modes. In particular, there is no additional physical charged scalar in this
field content: $G^\pm$ are the only charged scalar components and are removed from the
physical spectrum by the Higgs mechanism.

The potential in Eqs.~\eqref{eq:Vtotal_scalar}--\eqref{eq:Vadditional}, together with the field
parametrisation in Eq.~\eqref{eq:scalar_field_expansion}, is used for both vacuum alignments.
We now test the A--F and M--R VEV configurations using their tadpole and Hessian conditions.

\subsection{Viability of the VEVs in the Altarelli--Feruglio Basis}
\label{subsec:AF_viability}

The stability test must be performed with the scalar theory already fixed. Accordingly, the Hessian used below is obtained from the complete potential of Subsection~\ref{subsec:full_scalar_potential}, after imposing the stationary-point conditions appropriate to the A--F alignment.

For the A--F realisation the triplet VEVs are
\begin{equation}
\langle\Phi\rangle_{\rm AF}=v_\Phi(1,0,0),
\qquad
\langle\xi\rangle_{\rm AF}=v_\xi(0,-1,-1).
\label{eq:AFvacuum4}
\end{equation}
The A--F tadpoles are evaluated with the same complete potential. We denote a component tadpole by $T_X\equiv(\partial V/\partial X)|_0$. For nonzero VEVs the residual component equations require
\begin{equation}
\lambda_{\xi2}=\frac{4}{9}\lambda_{\xi3},
\qquad
\lambda_{\Phi\xi11}=-\frac32\lambda_{\Phi\xi14}
+\frac92\lambda_{\Phi\xi2}-\lambda_{\Phi\xi4}
+\frac92\lambda_{\Phi\xi9},
\qquad
\lambda_{\Phi\Phi\xi\rho}=0.
\label{eq:AF_residual_tadpoles}
\end{equation}
For example, $T_{\xi_{R1}}\propto -(9\lambda_{\xi2}-4\lambda_{\xi3})v_\xi^3$, while the two independent $\phi_{R2,3}$ equations first fix the mixed-quartic combination in Eq.~\eqref{eq:AF_residual_tadpoles} and then force $\lambda_{\Phi\Phi\xi\rho}=0$. The remaining independent tadpoles determine the quadratic parameters. With $\mathcal S=\{\psi,\chi,\eta,\rho,\kappa\}$, they are
\begin{align}
\mu_H^2={}&\frac12\left[2\lambda_Hv_H^2+\lambda_{H\Phi}v_\Phi^2+2\lambda_{H\xi}v_\xi^2+\sum_{s\in\mathcal S}\lambda_{Hs}v_s^2\right],\notag\\
\mu_\Phi^2={}&\frac1{36}\Big[18\lambda_{H\Phi}v_H^2+(36\lambda_{\Phi1}+16\lambda_{\Phi3})v_\Phi^2+18\sum_{s\in\mathcal S}\lambda_{\Phi s}v_s^2\notag\\
&+v_\xi^2(36\lambda_{\Phi\xi1}+4\lambda_{\Phi\xi11}-9\lambda_{\Phi\xi12}+12\lambda_{\Phi\xi14}-8\lambda_{\Phi\xi4}+36\lambda_{\Phi\xi9})\Big],\notag\\
\mu_\xi^2={}&\frac1{72}\Big[36\lambda_{H\xi}v_H^2+(144\lambda_{\xi1}+36\lambda_{\xi2}+48\lambda_{\xi3})v_\xi^2+36\sum_{s\in\mathcal S}\lambda_{\xi s}v_s^2\notag\\
&+v_\Phi^2(36\lambda_{\Phi\xi1}+4\lambda_{\Phi\xi11}-9\lambda_{\Phi\xi12}+12\lambda_{\Phi\xi14}-8\lambda_{\Phi\xi4}+36\lambda_{\Phi\xi9})\Big],\notag
\end{align}
\begin{align}
\mu_s^2={}&\frac12\left[\lambda_{Hs}v_H^2+\lambda_{\Phi s}v_\Phi^2+2\lambda_{\xi s}v_\xi^2+2\lambda_sv_s^2+\sum_{\substack{t\in\mathcal S\\t\ne s}}\lambda_{st}v_t^2\right],\quad s\in\mathcal S.
\label{eq:AF_mu2_solutions}
\end{align}
Substitution of Eqs.~\eqref{eq:AF_residual_tadpoles}--\eqref{eq:AF_mu2_solutions} into all real and imaginary first derivatives makes every A--F tadpole vanish. The decisive question is then whether this stationary configuration is locally stable in the physical scalar directions.

\subsubsection{CP-even sector}
\label{subsubsec:AF_CPeven}

In the neutral CP-even interaction basis
\begin{equation}
\mathcal B_{\rm even}^{\rm AF}=
(h,\phi_{R1},\phi_{R2},\phi_{R3},\xi_{R1},\xi_{R2},\xi_{R3},
\psi_R,\chi_R,\eta_R,\rho_R,\kappa_R),
\label{eq:AF_even_basis}
\end{equation}
the Hessian obtained after imposing the A--F tadpole conditions is rank
deficient by one,
\begin{equation}
\operatorname{rank}\!\left[(M_{\rm even}^2)_{\rm AF}\right]=11.
\label{eq:AF_rank11}
\end{equation}
The null direction can be exhibited explicitly as
\begin{equation}
(M_{\rm even}^2)_{\rm AF}
\begin{pmatrix}
0&0&0&0&0&-1/\sqrt2&1/\sqrt2&0&0&0&0&0
\end{pmatrix}^{T}=0,
\label{eq:AF_nullvector}
\end{equation}
which corresponds to the physical CP-even state
\begin{equation}
S_{\rm AF}^{(0)}=\frac{1}{\sqrt2}
(\xi_{R3}-\xi_{R2}),
\qquad
m_{S_{\rm AF}^{(0)}}^2=0.
\label{eq:AF_zero_state}
\end{equation}
This state is CP even and therefore cannot be one of the electroweak
Goldstone modes associated with $W^\pm$ or $Z^0$. The A--F stationary
configuration therefore contains a physical CP-even zero mode at tree level.
We accordingly turn to the alternative M--R realisation, for which the physical
CP-even spectrum can instead be strictly positive in a finite region of
parameter space. The explicit A--F Hessian and its null eigenvector are given in
Appendix~\ref{app:AF_CPeven_flat}.

\subsection{Viability of the VEVs in the Ma--Rajasekaran Basis}
\label{subsec:MR_viability}

\subsubsection{Stationarity of the M--R vacuum}
\label{subsec:scalar_potential_stationarity}

We take the scalar potential to be CP conserving and work with the real
vacuum branch, so that all the VEVs are real. It is important that a real
vacuum must be stationary not only along the real scalar directions but
also along the imaginary ones. Therefore
\begin{equation}
\left.
\frac{\partial V}{\partial\varphi_{Ra}}
\right|_0=0,
\qquad
\left.
\frac{\partial V}{\partial\varphi_{Ia}}
\right|_0=0
\label{eq:full_stationarity_condition}
\end{equation}
have to hold simultaneously.

For the M--R realisation the triplet VEVs are
\begin{equation}
\langle\Phi\rangle=v_\Phi(1,1,1),
\qquad
\langle\xi\rangle=v_\xi(1,-1,1),
\label{eq:MRvacuum4}
\end{equation}
while $v_H,v_\psi,v_\chi,v_\eta,v_\rho$ and $v_\kappa$ are also
taken to be real and nonzero.

Eight independent real tadpole equations may be chosen as
\begin{equation}
T_h=T_{\phi_{R1}}=T_{\xi_{R2}}
=T_{\psi_R}=T_{\chi_R}=T_{\eta_R}
=T_{\rho_R}=T_{\kappa_R}=0 .
\label{eq:independent_real_tadpoles}
\end{equation}
These equations determine the eight quadratic parameters
\begin{equation}
\mu_H^2,\quad
\mu_\Phi^2,\quad
\mu_\xi^2,\quad
\mu_\psi^2,\quad
\mu_\chi^2,\quad
\mu_\eta^2,\quad
\mu_\rho^2,\quad
\mu_\kappa^2
\label{eq:quadratic_parameters_eliminated}
\end{equation}
in terms of the VEVs and quartic couplings. Writing again $\mathcal S=\{\psi,\chi,\eta,\rho,\kappa\}$, the solutions are
\begin{align}
\mu_H^2={}&\frac12\left[2\lambda_Hv_H^2+3\lambda_{H\Phi}v_\Phi^2+3\lambda_{H\xi}v_\xi^2+\sum_{s\in\mathcal S}\lambda_{Hs}v_s^2\right],\notag\\
\mu_\Phi^2={}&\frac12\Big[\lambda_{H\Phi}v_H^2+2(3\lambda_{\Phi1}+4\lambda_{\Phi3})v_\Phi^2
+(3\lambda_{\Phi\xi1}+4\lambda_{\Phi\xi11}+4\lambda_{\Phi\xi4}+3\lambda_{\Phi\xi8})v_\xi^2
+\sum_{s\in\mathcal S}\lambda_{\Phi s}v_s^2\Big],\notag\\
\mu_\xi^2={}&\frac12\Big[\lambda_{H\xi}v_H^2+2(3\lambda_{\xi1}+4\lambda_{\xi3})v_\xi^2
+(3\lambda_{\Phi\xi1}+4\lambda_{\Phi\xi11}+4\lambda_{\Phi\xi4}+3\lambda_{\Phi\xi8})v_\Phi^2
+\sum_{s\in\mathcal S}\lambda_{\xi s}v_s^2\Big],\notag\\
\mu_s^2={}&\frac12\left[\lambda_{Hs}v_H^2+3\lambda_{\Phi s}v_\Phi^2+3\lambda_{\xi s}v_\xi^2+2\lambda_sv_s^2+\sum_{\substack{t\in\mathcal S\\t\ne s}}\lambda_{st}v_t^2\right],\quad s\in\mathcal S.
\label{eq:MR_mu2_solutions_main}
\end{align}
These expressions are written after using the residual alignment relation derived below in Eq.~\eqref{eq:MR_tadpole_relation}; the component check and the neutral Hessian construction are collected in Appendix~\ref{app:complete_scalar_diagonalisation}.

After these quadratic parameters are eliminated, the remaining real
component tadpoles contain the common combination
\begin{equation}
{\cal C}_{\rm MR}
=
\lambda_{\Phi\xi11}
+\lambda_{\Phi\xi12}
+2\lambda_{\Phi\xi4}
+\lambda_{\Phi\xi8}
-\lambda_{\Phi\xi9}.
\label{eq:CMR_tadpole}
\end{equation}
For example,
\begin{equation}
T_{\phi_{R2}}\propto
-v_\xi^2v_\Phi\,{\cal C}_{\rm MR},
\qquad
T_{\xi_{R1}},\,T_{\xi_{R3}}
\propto
v_\xi v_\Phi^2\,{\cal C}_{\rm MR}.
\label{eq:residual_real_tadpoles}
\end{equation}
Hence, for $v_\Phi,v_\xi\neq0$, stationarity of the M--R
alignment requires
\begin{equation}
\lambda_{\Phi\xi12}
=
-\lambda_{\Phi\xi11}
-2\lambda_{\Phi\xi4}
-\lambda_{\Phi\xi8}
+\lambda_{\Phi\xi9}.
\label{eq:MR_tadpole_relation}
\end{equation}
This is a genuine relation among the mixed quartic couplings which
follows directly from the chosen vacuum alignment.

The additional interaction in Eq.~\eqref{eq:Vadditional} has to be
checked separately. Although it is allowed by the flavour symmetry and
is CP even for real $\lambda_{\Phi\Phi\xi\rho}$, it produces
nonvanishing imaginary-direction tadpoles at the real M--R
configuration. Two representative terms are
\begin{equation}
\left.
\frac{\partial V_{\rm add}^{(4)}}{\partial\phi_{I1}}
\right|_0
\propto
-\sqrt3\,
\lambda_{\Phi\Phi\xi\rho}
v_\Phi v_\xi v_\rho ,
\qquad
\left.
\frac{\partial V_{\rm add}^{(4)}}{\partial\rho_I}
\right|_0
\propto
+\sqrt3\,
\lambda_{\Phi\Phi\xi\rho}
v_\Phi^2v_\xi .
\label{eq:additional_odd_tadpoles}
\end{equation}
Since the VEVs appearing in these expressions are nonzero, the same
real M--R vacuum is stationary only when
\begin{equation}
\lambda_{\Phi\Phi\xi\rho}=0 .
\label{eq:extra_quartic_stationarity}
\end{equation}
It is worth stressing the origin of Eq.~\eqref{eq:extra_quartic_stationarity}.
The interaction in Eq.~\eqref{eq:Vadditional} is neither forbidden by
the flavour symmetry nor removed by CP invariance of the potential.
Its coefficient vanishes because the chosen real M--R configuration
otherwise fails the imaginary-direction stationarity conditions.

Thus the minimisation conditions have a simple structure. The eight
independent real tadpoles determine the eight quadratic mass
parameters, while the residual component equations give
Eqs.~\eqref{eq:MR_tadpole_relation} and
\eqref{eq:extra_quartic_stationarity}. No further quartic coupling is
required to vanish by stationarity. The remaining quartic couplings
will instead be constrained by the requirement that the stationary
configuration has an acceptable scalar spectrum and that the potential
is stable. We turn to these questions next.

Equations~\eqref{eq:MR_tadpole_relation} and \eqref{eq:extra_quartic_stationarity} follow directly from the stationarity of the chosen M--R vacuum and are not simplifying assumptions on the quartic couplings. The remaining quartic couplings are kept general, subject to the stability conditions discussed below. Their radiative stability is not studied here.

\label{subsec:scalar_local_stability}

Once the stationary point is fixed, the next requirement is that the physical scalar
fluctuations around it should not contain any negative mass-squared direction. Since the
vacuum is real, the neutral scalar fluctuations separate into CP-even and CP-odd sectors.
We discuss them separately. In order to keep the connection with the physical scalar states
explicit, we also display the vacuum-adapted transformations and the corresponding
diagonalising matrices. The longer component expressions are collected in Appendix~\ref{app:complete_scalar_diagonalisation}.

\subsubsection{CP-even scalar sector}
\label{subsubsec:CPeven_local}

We collect the twelve neutral CP-even fields in the basis
\begin{equation}
\mathcal B_{\rm even}=
(h,\phi_{R1},\phi_{R2},\phi_{R3},\xi_{R1},\xi_{R2},\xi_{R3},
\psi_R,\chi_R,\eta_R,\rho_R,\kappa_R).
\label{eq:CPeven_basis_main}
\end{equation}
The two triplet VEVs in Eq.~\eqref{eq:MRvacuum4} single out one direction parallel to each
VEV and two orthogonal directions. We define the corresponding triplet transformations by
\begin{equation}
\begin{pmatrix}\phi_{R1}\\ \phi_{R2}\\ \phi_{R3}\end{pmatrix}
=U_\Phi
\begin{pmatrix}\Phi_{T1}\\ \Phi_{T2}\\ \Phi_R\end{pmatrix},
\qquad
\begin{pmatrix}\xi_{R1}\\ \xi_{R2}\\ \xi_{R3}\end{pmatrix}
=U_\xi
\begin{pmatrix}\xi_{T1}\\ \xi_{T2}\\ \xi_R\end{pmatrix},
\label{eq:triplet_adapted_fields_main}
\end{equation}
where
\begin{equation}
U_\Phi=
\begin{pmatrix}
-1/\sqrt2&-1/\sqrt6&1/\sqrt3\\
0&2/\sqrt6&1/\sqrt3\\
1/\sqrt2&-1/\sqrt6&1/\sqrt3
\end{pmatrix},
\qquad
U_\xi=
\begin{pmatrix}
-1/\sqrt2&1/\sqrt6&1/\sqrt3\\
0&2/\sqrt6&-1/\sqrt3\\
1/\sqrt2&1/\sqrt6&1/\sqrt3
\end{pmatrix}.
\label{eq:UPhi_Uxi_main}
\end{equation}
Both matrices are orthogonal. Their third columns are precisely the directions of the two
triplet VEVs, while the first two columns span the corresponding transverse subspaces. With
\begin{equation}
U_{\rm trip}=\operatorname{diag}(1,U_\Phi,U_\xi,I_5),
\label{eq:Utrip_main}
\end{equation}
and a permutation $P_{8+4}$ which only places the radial fields before the transverse ones,
the separation of the original twelve-dimensional Hessian is the orthogonal congruence
\begin{equation}
P_{8+4}^{T}U_{\rm trip}^{T}M_{\rm even}^2U_{\rm trip}P_{8+4}
=M_R^2\oplus M_T^2,
\label{eq:even_8plus4_main}
\end{equation}
where $M_R^2$ is an $8\times8$ matrix acting on the ordered radial basis
\begin{equation}
\mathbf r=(h,\Phi_R,\xi_R,\psi_R,\chi_R,\eta_R,\rho_R,\kappa_R)^T,
\label{eq:radial_basis_definition_main}
\end{equation}
and $M_T^2$ is a $4\times4$ matrix acting on
$\mathbf t=(\Phi_{T1},\Phi_{T2},\xi_{T1},\xi_{T2})^T$. Here $\Phi_R$ and $\xi_R$ are the directions parallel to the corresponding VEVs, while the $T1,T2$ directions are orthogonal to them. This separation is useful because the transverse part can be diagonalised analytically, whereas the radial part keeps the mixing among $H$, the radial triplet modes and the five singlet flavons. We define the radial mass eigenstates by $\mathbf H_R=(H_{R1},\ldots,H_{R8})^T$ and $\mathbf r=O_R\mathbf H_R$, where $O_R$ is orthogonal and satisfies
\begin{equation}
O_R^T M_R^2 O_R
=\operatorname{diag}(m_{R1}^2,\ldots,m_{R8}^2).
\label{eq:OR_radial_main}
\end{equation}
The eight radial mass-squared eigenvalues are defined by
\begin{equation}
\chi_R(\mu)\equiv\det(M_R^2-\mu I_8),
\qquad
\chi_R(m_{Ri}^2)=0,\qquad i=1,\ldots,8,
\label{eq:radial_characteristic_main}
\end{equation}
and the columns of $O_R$ are the corresponding normalised eigenvectors. Splitting the radial
basis as $(h,\Phi_R,\xi_R)\oplus(\psi_R,\chi_R,\eta_R,\rho_R,\kappa_R)$, we write
\begin{equation}
M_R^2=\begin{pmatrix}
M_{3R}^2&X_R\\
X_R^T&M_S^2
\end{pmatrix},
\label{eq:MR_radial_blocks_raw_main}
\end{equation}
with
\begin{equation}
\resizebox{0.98\textwidth}{!}{$\displaystyle
M_{3R}^2=\begin{pmatrix}
2\lambda_Hv_H^2&\sqrt3\lambda_{H\Phi}v_Hv_\Phi&\sqrt3\lambda_{H\xi}v_Hv_\xi\\
\sqrt3\lambda_{H\Phi}v_Hv_\Phi&2(3\lambda_{\Phi1}+4\lambda_{\Phi3})v_\Phi^2&
(3\lambda_{\Phi\xi1}+4\lambda_{\Phi\xi11}+4\lambda_{\Phi\xi4}+3\lambda_{\Phi\xi8})v_\Phi v_\xi\\
\sqrt3\lambda_{H\xi}v_Hv_\xi&
(3\lambda_{\Phi\xi1}+4\lambda_{\Phi\xi11}+4\lambda_{\Phi\xi4}+3\lambda_{\Phi\xi8})v_\Phi v_\xi&
2(3\lambda_{\xi1}+4\lambda_{\xi3})v_\xi^2
\end{pmatrix}.$}
\label{eq:MR_3radial_raw_main}
\end{equation}
For $s,t\in\{\psi,\chi,\eta,\rho,\kappa\}$ the remaining entries are
\begin{equation}
(X_R)_{1s}=\lambda_{Hs}v_Hv_s,\qquad
(X_R)_{2s}=\sqrt3\lambda_{\Phi s}v_\Phi v_s,\qquad
(X_R)_{3s}=\sqrt3\lambda_{\xi s}v_\xi v_s,
\label{eq:MR_XR_raw_main}
\end{equation}
\begin{equation}
(M_S^2)_{ss}=2\lambda_s v_s^2,\qquad
(M_S^2)_{st}=\lambda_{st}v_sv_t\quad(s\ne t).
\label{eq:MR_MS_raw_main}
\end{equation}
Equations~\eqref{eq:MR_radial_blocks_raw_main}--\eqref{eq:MR_MS_raw_main} give every
entry of the eight-dimensional radial mass matrix directly in terms of the quartic couplings and
VEVs. Appendix~\ref{app:complete_scalar_diagonalisation} displays the same matrix in a compact
square form, together with the complete component form of the diagonalising transformation.

For later compact formulae we define
\begin{equation}
A_\Phi=3\lambda_{\Phi1}+4\lambda_{\Phi3},
\qquad
A_\xi=3\lambda_{\xi1}+4\lambda_{\xi3},
\label{eq:Aphi_Axi_main}
\end{equation}
and
\begin{align}
C_R={}&3\lambda_{\Phi\xi1}+4\lambda_{\Phi\xi11}
+4\lambda_{\Phi\xi4}+3\lambda_{\Phi\xi8},
\notag\\
C_T={}&-2\lambda_{\Phi\xi11}+3\lambda_{\Phi\xi2}
-2\lambda_{\Phi\xi4}+3\lambda_{\Phi\xi9}.
\label{eq:CR_CT_main}
\end{align}
The radial matrix can be written in the form $M_R^2=D_v K_R D_v$, where $D_v$ contains
the appropriate VEV factors. Writing
\begin{equation}
K_R=
\begin{pmatrix}
K_3&P\\
P^T&K_S
\end{pmatrix},
\qquad
K_3=
\begin{pmatrix}
2\lambda_H&\sqrt3\lambda_{H\Phi}&\sqrt3\lambda_{H\xi}\\
\sqrt3\lambda_{H\Phi}&2A_\Phi&C_R\\
\sqrt3\lambda_{H\xi}&C_R&2A_\xi
\end{pmatrix},
\label{eq:radial_KR_main}
\end{equation}
the exact positivity test for this sector is
\begin{equation}
K_3>0,
\qquad
K_S-P^T K_3^{-1}P>0.
\label{eq:radial_Schur_main}
\end{equation}
The first of these conditions gives explicitly
\begin{equation}
\begin{aligned}
&\lambda_H>0,\\[-1mm]
&4\lambda_HA_\Phi-3\lambda_{H\Phi}^2>0,\\[-1mm]
&4\lambda_HA_\Phi A_\xi-3A_\Phi\lambda_{H\xi}^2-3A_\xi\lambda_{H\Phi}^2
-\lambda_HC_R^2+3C_R\lambda_{H\Phi}\lambda_{H\xi}>0.
\end{aligned}
\label{eq:K3_conditions_main}
\end{equation}
The second condition in Eq.~\eqref{eq:radial_Schur_main} contains the remaining mixing of the
radial triplet modes with the singlet flavons. It is important that these mixed quartic couplings
need not vanish. They are allowed as long as the complete Schur complement remains positive.
The observed Higgs boson has to be identified among the eight eigenstates of this full radial
block. In the mass-basis definition above, if $H_{Rh}$ denotes the state identified with
the observed Higgs, then
\begin{equation}
H_{Rh}=\sum_{a=1}^{8}(O_R)_{ah}\,r_a,
\qquad
m_{H_{Rh}}\simeq125~{\rm GeV}.
\label{eq:physical_Higgs_state_main}
\end{equation}
It must also be dominantly aligned with the electroweak-doublet direction,
$|(O_R)_{1h}|^2$ being close to unity as required by the observed SM-like Higgs couplings.
Thus the physical Higgs is identified by both its mass and its doublet fraction. In the present
mixed scalar sector its mass is an eigenvalue of $M_R^2$ and is not, in general, equal to a
single diagonal entry.

The transverse block can be written more explicitly. Defining
\begin{equation}
B_\Phi=3\lambda_{\Phi2}-4\lambda_{\Phi3},
\qquad
B_\xi=3\lambda_{\xi2}-4\lambda_{\xi3},
\label{eq:Bphi_Bxi_main}
\end{equation}
and
\begin{equation}
a=B_\Phi v_\Phi^2,\qquad
b=B_\xi v_\xi^2,\qquad
c=C_Tv_\Phi v_\xi,\qquad
d=2\sqrt3\lambda_{\Phi\xi14}v_\Phi v_\xi,
\label{eq:abcd_even_main}
\end{equation}
we obtain
\begin{equation}
M_T^2=
\begin{pmatrix}
a&0&c&-d\\
0&a&-d&-c\\
c&-d&b&0\\
-d&-c&0&b
\end{pmatrix}.
\label{eq:MT_even_main}
\end{equation}
Its two mass-squared eigenvalues are
\begin{equation}
m_{T\pm}^2=\frac12\left[a+b\pm\sqrt{(a-b)^2+4(c^2+d^2)}\right],
\qquad \text{each with multiplicity two}.
\label{eq:transverse_even_masses_main}
\end{equation}
The transverse diagonalising matrix can also be written
analytically. Defining
\begin{equation}
s_T=\sqrt{c^2+d^2},\qquad
\cos2\gamma=\frac{c}{s_T},\quad
\sin2\gamma=-\frac{d}{s_T},
\label{eq:gamma_even_main}
\end{equation}
and
\begin{equation}
D_T=\sqrt{(a-b)^2+4s_T^2},\qquad
\cos2\beta=\frac{a-b}{D_T},\quad
\sin2\beta=\frac{2s_T}{D_T},
\label{eq:beta_even_main}
\end{equation}
we use $c_\beta=\cos\beta$, $s_\beta=\sin\beta$, $c_\gamma=\cos\gamma$ and
$s_\gamma=\sin\gamma$ to write
\begin{equation}
O_T=
\begin{pmatrix}
c_\beta c_\gamma&-c_\beta s_\gamma&-s_\beta c_\gamma&-s_\beta s_\gamma\\
c_\beta s_\gamma& c_\beta c_\gamma&-s_\beta s_\gamma& s_\beta c_\gamma\\
s_\beta c_\gamma& s_\beta s_\gamma& c_\beta c_\gamma&-c_\beta s_\gamma\\
s_\beta s_\gamma&-s_\beta c_\gamma& c_\beta s_\gamma& c_\beta c_\gamma
\end{pmatrix},
\label{eq:OT_even_main}
\end{equation}
so that
\begin{equation}
O_T^T M_T^2 O_T
=\operatorname{diag}(m_{T+}^2,m_{T+}^2,m_{T-}^2,m_{T-}^2).
\label{eq:OT_even_diag_main}
\end{equation}
Combining the vacuum-adapted transformation, the radial diagonalisation and the transverse
diagonalisation gives the complete CP-even transformation in the original component basis,
\begin{equation}
U_{\rm even}\equiv O_{\rm even}^{(12)}
=U_{\rm trip}P_{8+4}\operatorname{diag}(O_R,O_T).
\label{eq:Oeven_complete_main}
\end{equation}
The columns of $U_{\rm even}$ are therefore the twelve CP-even mass eigenstates written
in the original interaction basis. The complete CP-even mass spectrum is
\begin{equation}
\operatorname{Spec}(M_{\rm even}^2)=
\left\{
 m_{R1}^2,\ldots,m_{R8}^2,
 m_{T+}^2,m_{T+}^2,m_{T-}^2,m_{T-}^2
\right\}.
\label{eq:CPeven_complete_spectrum_main}
\end{equation}
Here the four transverse masses are the explicit analytic expressions in
Eq.~\eqref{eq:transverse_even_masses_main}, whereas the eight radial masses are the roots of
Eq.~\eqref{eq:radial_characteristic_main}. Appendix~\ref{app:complete_scalar_diagonalisation} gives the complete CP-even transformation in an exact factorised form, with every basis, index and constituent matrix defined explicitly.

Both transverse masses are positive when
\begin{equation}
B_\Phi>0,\qquad
B_\xi>0,\qquad
C_T^2+12\lambda_{\Phi\xi14}^2<B_\Phi B_\xi.
\label{eq:even_transverse_positive_main}
\end{equation}
Therefore the mixed $\Phi$--$\xi$ terms can be present without spoiling the CP-even vacuum
stability. Their magnitudes are instead restricted by the inequality in
Eq.~\eqref{eq:even_transverse_positive_main}.

\subsubsection{CP-odd scalar sector}
\label{subsubsec:CPodd_local}

The neutral CP-odd interaction basis is
\begin{equation}
\mathcal B_{\rm odd}=
(\Omega,\phi_{I1},\phi_{I2},\phi_{I3},\xi_{I1},\xi_{I2},\xi_{I3},
\psi_I,\chi_I,\eta_I,\rho_I,\kappa_I).
\label{eq:CPodd_basis_main}
\end{equation}
After using the stationarity conditions of Subsection~\ref{subsec:scalar_potential_stationarity},
the massive part of the CP-odd triplet sector again reduces to the transverse directions. The
same matrices $U_\Phi$ and $U_\xi$ in Eq.~\eqref{eq:UPhi_Uxi_main} act on the imaginary triplet fields. The subscript $0$ denotes the phase direction parallel to the corresponding VEV, while $T1$ and $T2$ denote the two mutually orthogonal transverse phase directions. Explicitly,
\begin{align}
\Phi_{IT1}&=\frac{-\phi_{I1}+\phi_{I3}}{\sqrt2},&
\Phi_{IT2}&=\frac{-\phi_{I1}+2\phi_{I2}-\phi_{I3}}{\sqrt6},&
\Phi_{I0}&=\frac{\phi_{I1}+\phi_{I2}+\phi_{I3}}{\sqrt3},
\notag\\
\xi_{IT1}&=\frac{-\xi_{I1}+\xi_{I3}}{\sqrt2},&
\xi_{IT2}&=\frac{\xi_{I1}+2\xi_{I2}+\xi_{I3}}{\sqrt6},&
\xi_{I0}&=\frac{\xi_{I1}-\xi_{I2}+\xi_{I3}}{\sqrt3}.
\label{eq:odd_adapted_fields_main}
\end{align}
We define
\begin{equation}
P_\Phi=-(\lambda_{\Phi3}+\lambda_{\Phi4}),
\qquad
P_\xi=-(\lambda_{\xi3}+\lambda_{\xi4}),
\label{eq:Pphi_Pxi_main}
\end{equation}
and
\begin{equation}
\Delta=\lambda_{\Phi\xi4}-\lambda_{\Phi\xi5}
+\lambda_{\Phi\xi8}-\lambda_{\Phi\xi9}.
\label{eq:Delta_odd_main}
\end{equation}
With
\begin{equation}
A=6P_\Phi v_\Phi^2,\qquad
B=6P_\xi v_\xi^2,\qquad
D=\Delta v_\Phi v_\xi,
\label{eq:ABD_odd_main}
\end{equation}
the four massive CP-odd states are obtained from two $2\times2$ matrices,
\begin{equation}
M_1^2=\begin{pmatrix}A&D\\D&B\end{pmatrix},
\qquad
M_2^2=\begin{pmatrix}A&3D\\3D&B\end{pmatrix}.
\label{eq:odd_two_blocks_main}
\end{equation}
Their mass-squared eigenvalues are
\begin{align}
m_{1\pm}^2&=\frac12\left[A+B\pm\sqrt{(A-B)^2+4D^2}\right],\notag\\
m_{2\pm}^2&=\frac12\left[A+B\pm\sqrt{(A-B)^2+36D^2}\right].
\label{eq:odd_four_masses_main}
\end{align}
The two blocks are diagonalised by different rotations because their off-diagonal entries differ.
The corresponding mixing angles are defined through
\begin{equation}
\tan 2\theta_1=\frac{2D}{A-B},\qquad
\tan 2\theta_2=\frac{6D}{A-B}.
\label{eq:odd_angles_main}
\end{equation}
The branches of $\theta_1$ and $\theta_2$ are chosen so that the rotations below diagonalise the two
blocks continuously and the eigenstates are ordered according to Eq.~\eqref{eq:odd_four_masses_main}.
For $A=B$, the corresponding limiting value is $|\theta_i|=\pi/4$, with the sign fixed by that of the
off-diagonal entry.
The four massive states may then be written directly as
\begin{align}
A_{1+}&=\cos\theta_1\,\Phi_{IT1}+\sin\theta_1\,\xi_{IT1},&
A_{1-}&=-\sin\theta_1\,\Phi_{IT1}+\cos\theta_1\,\xi_{IT1},
\notag\\
A_{2+}&=\cos\theta_2\,\Phi_{IT2}+\sin\theta_2\,\xi_{IT2},&
A_{2-}&=-\sin\theta_2\,\Phi_{IT2}+\cos\theta_2\,\xi_{IT2}.
\label{eq:odd_massive_states_main}
\end{align}
Writing $c_i=\cos\theta_i$ and $s_i=\sin\theta_i$, the transformation from the adapted
triplet basis to the ordered set $(A_{1+},A_{2+},\Phi_{I0},A_{1-},A_{2-},\xi_{I0})$ is
\begin{equation}
O_{\rm pair}=
\begin{pmatrix}
c_1&0&0&-s_1&0&0\\
0&c_2&0&0&-s_2&0\\
0&0&1&0&0&0\\
s_1&0&0&c_1&0&0\\
0&s_2&0&0&c_2&0\\
0&0&0&0&0&1
\end{pmatrix}.
\label{eq:Opair_odd_main}
\end{equation}
Thus the complete triplet-phase diagonaliser in the original component basis is
\begin{equation}
O_6=\operatorname{diag}(U_\Phi,U_\xi)O_{\rm pair},
\label{eq:O6_odd_main}
\end{equation}
and the complete neutral CP-odd transformation is
\begin{equation}
U_{\rm odd}\equiv O_{\rm odd}=\operatorname{diag}(1,O_6,I_5).
\label{eq:Oodd_complete_main}
\end{equation}
With the mass-basis ordering
$(\Omega,A_{1+},A_{2+},\Phi_{I0},A_{1-},A_{2-},\xi_{I0},
\psi_I,\chi_I,\eta_I,\rho_I,\kappa_I)$, this gives
\begin{equation}
U_{\rm odd}^T M_{\rm odd}^2 U_{\rm odd}
=\operatorname{diag}(0,m_{1+}^2,m_{2+}^2,0,m_{1-}^2,m_{2-}^2,0,0,0,0,0,0).
\label{eq:Oodd_diag_main}
\end{equation}
Thus all twelve neutral CP-odd mass-squared eigenvalues are
\begin{equation}
\operatorname{Spec}(M_{\rm odd}^2)=
\left\{m_{1+}^2,m_{2+}^2,m_{1-}^2,m_{2-}^2,
\underbrace{0,0,0,0,0,0,0,0}_{8\;\text{zero modes}}\right\}.
\label{eq:CPodd_complete_spectrum_main}
\end{equation}
The four nonzero masses are given explicitly in Eq.~\eqref{eq:odd_four_masses_main}. Appendix~\ref{app:complete_scalar_diagonalisation} gives the complete CP-odd transformation in block form and displays the non-trivial triplet-phase matrix explicitly.
The positivity of all four massive states requires
\begin{equation}
P_\Phi>0,\qquad
P_\xi>0,\qquad
\Delta^2<4P_\Phi P_\xi.
\label{eq:odd_positivity_main}
\end{equation}
Thus the CP-odd mixing controlled by $\Delta$ need not vanish. The last inequality directly
limits how large this mixing may be relative to the self-couplings of the two triplets.

\subsubsection{Accidental global $U(1)$ symmetries and CP-odd zero modes}
\label{subsubsec:accidental_U1_zero_modes}

The neutral CP-odd Hessian also contains eight exact zero eigenvalues. One of them is
$\Omega\equiv G^0$, the neutral Standard-Model Goldstone boson identified in
Eq.~\eqref{eq:SM_goldstones_main}, which is absorbed by $Z^0$. The charged pair $G^\pm$ belongs
to the charged sector and is absorbed by $W^\pm$, and hence it does not appear in the neutral
CP-odd Hessian. The counting of the zero directions is therefore
\begin{equation}
8\;\hbox{neutral CP-odd zero modes}
=1\;G^0+7\;\hbox{physical zero modes}.
\label{eq:CPodd_zero_count_main}
\end{equation}
For the two triplets, the zero directions parallel to the VEVs may be chosen as
\begin{equation}
\Phi_{I0}=\frac{\phi_{I1}+\phi_{I2}+\phi_{I3}}{\sqrt3},
\qquad
\xi_{I0}=\frac{\xi_{I1}-\xi_{I2}+\xi_{I3}}{\sqrt3}.
\label{eq:triplet_zero_modes_main}
\end{equation}
The remaining seven zero modes are therefore physical and may be represented by
\begin{equation}
\mathcal Z_{\rm phys}=
(\Phi_{I0},\xi_{I0},\psi_I,\chi_I,\eta_I,\rho_I,\kappa_I).
\label{eq:seven_physical_zero_modes_main}
\end{equation}
Their origin is not the same for all seven states. We denote a continuous rephasing charge by $q_F$. Invariance of the retained Yukawa and Majorana operators gives the independent constraints
\begin{equation}
\begin{alignedat}{2}
-q_D+q_\Phi+q_H+q_{e,\mu,\tau}&=0,\qquad&
-q_D+q_\Phi-q_H+q_{\nu_e}&=0,\\
-q_D+q_\Phi-q_H+q_{\nu_\mu}+q_\chi&=0,&
-q_D+q_\Phi-q_H+q_{\nu_\tau}+q_\psi&=0,\\
2q_{\nu_e}+q_\rho&=0,&
q_{\nu_e}+q_{\nu_\tau}+q_\kappa&=0,\\
2q_{\nu_\mu}+q_\eta&=0,&
-2q_D-2q_H+q_\xi&=0.
\end{alignedat}
\label{eq:U1_charge_constraints_main}
\end{equation}
The constraint matrix has a five-dimensional solution space. One direction is the gauged hypercharge assignment; quotienting by it leaves four independent accidental global $U(1)$ symmetries. The detailed charge basis and the action of the soft terms are given in Appendix~\ref{app:soft_scalar_details}.

Since the chosen vacuum breaks these four accidental symmetries spontaneously, four of the seven
physical massless states are genuine Goldstone bosons of accidental global symmetries. The
remaining three states are tree-level phase-flat directions: their vanishing masses follow from the
structure of the renormalisable potential, but they are not protected by three additional exact
continuous symmetries. Thus the seven physical zero modes consist of
\begin{equation}
\begin{aligned}
7\;\hbox{physical CP-odd zero modes}
&=4\;\hbox{accidental-symmetry Goldstones}\\
&\quad+3\;\hbox{phase-flat directions}.
\end{aligned}
\label{eq:physical_zero_origin_main}
\end{equation}
This distinction is important when the phenomenological consequences of the massless sector and
the effect of controlled symmetry-breaking terms are discussed later.

Equations~\eqref{eq:radial_Schur_main}, \eqref{eq:K3_conditions_main},
\eqref{eq:even_transverse_positive_main} and \eqref{eq:odd_positivity_main}, together with the
stationarity relations in Eqs.~\eqref{eq:MR_tadpole_relation} and
\eqref{eq:extra_quartic_stationarity}, give the conditions required at this stage for a stationary
M--R vacuum with a locally acceptable neutral scalar spectrum. These conditions, however, do
not yet establish stability of the potential at arbitrarily large field values. We therefore turn
next to the boundedness from below of the quartic potential. The global character of the selected
vacuum will be examined separately afterwards.

\subsubsection{Boundedness from below}
\label{subsec:BFB_main}

Local stability and boundedness from below are different requirements. The former concerns the
curvature of the potential around the chosen stationary configuration, whereas boundedness from
below (BFB) asks whether the quartic part remains positive in every large-field direction. For the
present multi-scalar potential it is therefore necessary to keep both the overall field magnitudes
and the internal directions of the two $A_4$ triplets under control \cite{r70,r71}.

We introduce the eight non-negative radial variables
\begin{equation}
\begin{split}
z_H&=H^\dagger H,\qquad z_\Phi=(\Phi^\dagger\Phi)_1,\qquad
z_\xi=(\xi^\dagger\xi)_1,\\
z_\psi&=|\psi|^2,\qquad z_\chi=|\chi|^2,\qquad z_\eta=|\eta|^2,
\qquad z_\rho=|\rho|^2,\qquad z_\kappa=|\kappa|^2,
\end{split}
\label{eq:BFB_radial_variables_main}
\end{equation}
and collect them as
\begin{equation}
{\bf z}=(z_H,z_\Phi,z_\xi,z_\psi,z_\chi,z_\eta,z_\rho,z_\kappa)^T,
\qquad z_i\ge0.
\label{eq:BFB_zvector_main}
\end{equation}
For the triplets we write $\Phi=\sqrt{z_\Phi}\,n_\Phi$ and
$\xi=\sqrt{z_\xi}\,n_\xi$, with $n_\Phi^\dagger n_\Phi=n_\xi^\dagger n_\xi=1$.
The quartic potential can then be separated into a radial part and an angular part.

For either triplet $X=\Phi,\xi$, we define
\begin{equation}
U_X=\frac{(X^\dagger X)_{1'}(X^\dagger X)_{1''}}{z_X^2},
\qquad
W_X=-\frac{[(X^\dagger X)_{3_a}(X^\dagger X)_{3_a}]_1}{z_X^2},
\label{eq:UX_WX_main}
\end{equation}
for which $U_X\ge0$ and $W_X\ge0$. With the combinations already introduced in
Eqs.~\eqref{eq:Aphi_Axi_main}, \eqref{eq:Bphi_Bxi_main} and
\eqref{eq:Pphi_Pxi_main}, the two triplet self-interactions take the form
\begin{equation}
V_X^{(4)}=z_X^2\left(\frac{A_X}{3}+\frac{B_X}{3}U_X+P_XW_X\right).
\label{eq:triplet_BFB_decomposition_main}
\end{equation}
For the real equal-modulus M--R directions in Eq.~\eqref{eq:MRvacuum4},
$U_\Phi=W_\Phi=U_\xi=W_\xi=0$. Hence the terms proportional to $B_X$ and $P_X$
measure the increase of the quartic energy away from these directions. Their positivity is already
required by the transverse CP-even and massive CP-odd sectors.

The mixed $\Phi$--$\xi$ quartic also admits a useful exact reduction. Using the $A_4$ recoupling
identities among the $J_i$ of Eq.~\eqref{eq:Jdefinitions}, together with the stationarity relation
in Eq.~\eqref{eq:MR_tadpole_relation}, it becomes
\begin{equation}
V_{\Phi\xi}^{(4)}=
\frac{C_R}{3}J_1+\frac{C_T}{3}J_2-\Delta J_5
+\lambda_{\Phi\xi14}J_{14},
\label{eq:mixed_quartic_reduced_main}
\end{equation}
where $J_1=z_\Phi z_\xi$, while $C_R$, $C_T$ and $\Delta$ are defined in
Eqs.~\eqref{eq:CR_CT_main} and \eqref{eq:Delta_odd_main}. Introducing
$j_i=J_i/(z_\Phi z_\xi)$ for the angular mixed invariants, the exact identities relevant here are
\begin{equation}
j_2^2+\frac34j_{14}^2=4U_\Phi U_\xi,
\qquad
|j_5|\le\sqrt{W_\Phi W_\xi}.
\label{eq:mixed_angular_identities_main}
\end{equation}
These relations make it possible to control the mixed angular terms for arbitrary relative
orientations and phases of the two triplets, not only for infinitesimal fluctuations about the
vacuum.

Indeed, the first identity and the Cauchy--Schwarz inequality give
\begin{equation}
\left|\frac{C_T}{3}j_2+\lambda_{\Phi\xi14}j_{14}\right|
\le \frac{2}{3}\sqrt{C_T^2+12\lambda_{\Phi\xi14}^2}\,
\sqrt{U_\Phi U_\xi},
\label{eq:CT_angular_bound_main}
\end{equation}
whereas the second identity gives
\begin{equation}
|\Delta j_5|\le |\Delta|\sqrt{W_\Phi W_\xi}.
\label{eq:Delta_angular_bound_main}
\end{equation}
Therefore the same inequalities obtained from the local transverse spectra,
\begin{equation}
C_T^2+12\lambda_{\Phi\xi14}^2<B_\Phi B_\xi,
\qquad
\Delta^2<4P_\Phi P_\xi,
\label{eq:BFB_angular_conditions_main}
\end{equation}
together with $B_\Phi,B_\xi,P_\Phi,P_\xi>0$, ensure that the positive triplet self-terms in
Eq.~\eqref{eq:triplet_BFB_decomposition_main} dominate the mixed angular contribution.
Consequently,
\begin{equation}
\Delta V_{\rm ang}\ge0
\label{eq:angular_nonnegative_main}
\end{equation}
for arbitrary triplet directions. This is stronger than the local Hessian statement: it controls the
triplet angular dependence over the full field space.

It remains to control the radial quartic form. Along the M--R angular directions it can be written as
\begin{equation}
V_{4,{\rm rad}}={\bf z}^TQ_{\rm MR}{\bf z},
\qquad
Q_{\rm MR}=\begin{pmatrix}Q_3&R\\R^T&Q_S\end{pmatrix},
\label{eq:QMR_BFB_main}
\end{equation}
where, without introducing any composite quartic parameter,
\begin{equation}
\resizebox{0.98\textwidth}{!}{$\displaystyle
Q_3=\begin{pmatrix}
\lambda_H&\lambda_{H\Phi}/2&\lambda_{H\xi}/2\\
\lambda_{H\Phi}/2&(3\lambda_{\Phi1}+4\lambda_{\Phi3})/3&
(3\lambda_{\Phi\xi1}+4\lambda_{\Phi\xi11}+4\lambda_{\Phi\xi4}+3\lambda_{\Phi\xi8})/6\\
\lambda_{H\xi}/2&
(3\lambda_{\Phi\xi1}+4\lambda_{\Phi\xi11}+4\lambda_{\Phi\xi4}+3\lambda_{\Phi\xi8})/6&
(3\lambda_{\xi1}+4\lambda_{\xi3})/3
\end{pmatrix}.$}
\label{eq:Q3_BFB_main}
\end{equation}
For $s,t\in\{\psi,\chi,\eta,\rho,\kappa\}$,
\begin{equation}
R_{1s}=\frac{\lambda_{Hs}}2,\qquad
R_{2s}=\frac{\lambda_{\Phi s}}2,\qquad
R_{3s}=\frac{\lambda_{\xi s}}2,
\qquad
(Q_S)_{ss}=\lambda_s,\qquad
(Q_S)_{st}=\frac{\lambda_{st}}2\;(s\ne t).
\label{eq:QS_R_BFB_main}
\end{equation}
These equations specify every entry of the radial quartic matrix directly in terms of the original
quartic couplings; Appendix~\ref{app:raw_radial_quartic} gives the same $8\times8$ matrix as explicit $3+5$ blocks, which is algebraically identical but considerably easier to read.
Because all components of ${\bf z}$ are non-negative, the exact radial language for BFB is
copositivity of $Q_{\rm MR}$. A convenient sufficient condition, which is somewhat stronger but
analytically transparent, is positive definiteness,
\begin{equation}
Q_{\rm MR}>0.
\label{eq:QMR_positive_main}
\end{equation}
Using the Schur complement, this condition is equivalent to
\begin{equation}
Q_3>0,
\qquad
Q_S-R^TQ_3^{-1}R>0.
\label{eq:QMR_Schur_BFB_main}
\end{equation}
Using $A_\Phi$, $A_\xi$ and $C_R$ defined in Eq.~\eqref{eq:CR_CT_main}, the positivity of $Q_3$ gives
\begin{equation}
\begin{aligned}
&\lambda_H>0,\\[-1mm]
&4\lambda_HA_\Phi-3\lambda_{H\Phi}^2>0,\\[-1mm]
&4\lambda_HA_\Phi A_\xi-3A_\Phi\lambda_{H\xi}^2-3A_\xi\lambda_{H\Phi}^2
-\lambda_HC_R^2+3C_R\lambda_{H\Phi}\lambda_{H\xi}>0.
\end{aligned}
\label{eq:Q3_BFB_conditions_main}
\end{equation}
The second condition in Eq.~\eqref{eq:QMR_Schur_BFB_main} constrains simultaneously the
five singlet self-couplings, the ten singlet cross-couplings, and their couplings to $H$, $\Phi$ and
$\xi$. These couplings therefore need not vanish; what is required is positivity of the complete
radial form.

It is useful to note the close relation with the CP-even radial analysis. The matrix
$Q_{\rm MR}$ and the radial CP-even Hessian are related by a non-singular rescaling with the
nonzero VEVs. Hence the same coupling combinations already encountered in
Eqs.~\eqref{eq:radial_Schur_main} and \eqref{eq:K3_conditions_main} appear here. In the present
subsection they are used together with the global angular bounds to test the quartic potential
in the large-field limit.

We therefore obtain a finite region of quartic-coupling space in which
\begin{equation}
V_4>0
\qquad\text{for every nonzero large-field direction},
\label{eq:BFB_final_main}
\end{equation}
so that the scalar potential is bounded from below. This result establishes the asymptotic
stability of the potential. It does not by itself prove that the chosen M--R stationary point is the
deepest finite-field configuration; the global character of the vacuum will be addressed in the
next subsection.

\subsubsection{Global character of the M--R vacuum}
\label{subsec:global_minimum_main}

The conditions derived above establish stationarity, local stability and boundedness from below.
A further question is whether another field configuration at a finite distance can have a lower
energy than the selected M--R vacuum. This has to be checked separately, since neither a positive
Hessian nor BFB alone is sufficient to establish a global minimum.

The angular analysis in the previous subsection already gives the first part of the argument. For
arbitrary triplet directions we may write
\begin{equation}
V({\bf z},n_\Phi,n_\xi)=V_{\rm MR}({\bf z})+\Delta V_{\rm ang}({\bf z},n_\Phi,n_\xi),
\qquad
\Delta V_{\rm ang}\ge0,
\label{eq:global_split_main}
\end{equation}
where $V_{\rm MR}({\bf z})$ denotes the potential evaluated with the same eight radial magnitudes
but with the M--R triplet directions. Thus, for fixed magnitudes, the M--R angular configuration
cannot be lowered by changing the triplet orientations or phases, provided the transverse
conditions in Eq.~\eqref{eq:BFB_angular_conditions_main} are satisfied.

It remains to compare different radial magnitudes. We denote the radial variables at the chosen stationary vacuum by
\begin{equation}
{\bf z}_0=(z_{H0},z_{\Phi0},z_{\xi0},z_{\psi0},z_{\chi0},z_{\eta0},z_{\rho0},z_{\kappa0})^T.\label{eq:z0_global_main}
\end{equation}
Along the M--R directions the potential
has the quadratic form in ${\bf z}$ introduced in Eq.~\eqref{eq:QMR_BFB_main}. After the tadpole
conditions are used, the difference from the vacuum energy can be written exactly as
\begin{equation}
V_{\rm MR}({\bf z})-V_{\rm MR}({\bf z}_0)
=({\bf z}-{\bf z}_0)^TQ_{\rm MR}({\bf z}-{\bf z}_0).
\label{eq:radial_completed_square_main}
\end{equation}
Therefore, if
\begin{equation}
Q_{\rm MR}>0,
\label{eq:global_Q_positive_main}
\end{equation}
the radial contribution is non-negative for every ${\bf z}$ and vanishes only at
${\bf z}={\bf z}_0$. The same positive-definiteness condition is already encoded in the complete
CP-even radial Hessian through the congruence relation discussed in Sec.~\ref{subsec:BFB_main}.
In practice it may be imposed through the Schur-complement conditions in
Eq.~\eqref{eq:QMR_Schur_BFB_main}.

Combining Eqs.~\eqref{eq:global_split_main} and \eqref{eq:radial_completed_square_main} gives
\begin{align}
&V({\bf z},n_\Phi,n_\xi)
-V({\bf z}_0,n_\Phi^{\rm MR},n_\xi^{\rm MR})
\notag\\
&\qquad=({\bf z}-{\bf z}_0)^TQ_{\rm MR}({\bf z}-{\bf z}_0)
+\Delta V_{\rm ang}\ge0.
\label{eq:global_energy_difference_main}
\end{align}
Hence the chosen M--R configuration is a global minimum in the finite region of coupling space
specified by
\begin{equation}
\begin{gathered}
Q_{\rm MR}>0,
\qquad B_\Phi>0,\quad B_\xi>0,\quad P_\Phi>0,\quad P_\xi>0,\\
C_T^2+12\lambda_{\Phi\xi14}^2<B_\Phi B_\xi,
\qquad
\Delta^2<4P_\Phi P_\xi.
\end{gathered}
\label{eq:global_conditions_summary_main}
\end{equation}

The conditions in Eq.~\eqref{eq:global_conditions_summary_main}, together with the stationarity
relations obtained in Sec.~\ref{subsec:scalar_potential_stationarity}, define a finite sufficient
region in which the selected M--R vacuum is globally viable. The radial requirement is kept in
its general positive-definite form $Q_{\rm MR}>0$, equivalently in the Schur-complement form
given above, while the CP-even and CP-odd transverse inequalities are exact. No additional
restriction on the individual quartic couplings is introduced.

The algebra leading to the angular identities in Eq.~\eqref{eq:mixed_angular_identities_main}
and to the radial completed-square relation in Eq.~\eqref{eq:radial_completed_square_main} is
given explicitly in Appendix~\ref{app:global_minimum_derivation}. The conditions above are
sufficient, but not necessary, for global minimality. Because the potential still contains the phase
and discrete degeneracies discussed above, the selected M--R configuration need not be unique in
the full complex field space. The result established here is that it is a global minimum within the
stated region of couplings.

We have therefore established, within one consistent set of finite inequalities, the stationarity,
local scalar stability, boundedness from below and global viability of the M--R vacuum. The
remaining issue in the exact scalar theory is the presence of the seven physical CP-odd zero modes.
We first proceed with the phenomenological analysis of the model established above. The lifting of
the physical zero modes by controlled dimension-two symmetry-breaking terms will be treated
separately after that analysis.

\FloatBarrier

% ================================================================
% SECTION 5 -- CURRENT DRAFT THROUGH THE NUMERICAL ANALYSIS IN SEC. 5.2
% ================================================================

\section[Phenomenological Constraints and Predictions of the Model]{Phenomenological Constraints and Predictions\texorpdfstring{\\}{ }of the Model}
\label{sec:phenomenological_constraints}

Our analysis includes CLFV, non-unitarity of the effective PMNS matrix, and baryogenesis through leptogenesis. We first use the present upper bounds on various CLFV processes as inputs to constrain the relevant combinations of the model parameters. The resulting parameter space is then further examined using the bounds on non-unitarity, which can provide complementary constraints. Finally, we study leptogenesis \cite{r72,r73}, one of the mechanisms to generate the observed baryon asymmetry of the Universe \cite{r18,r74}, in the light of the present model in the M--R basis of $A_4$.

\subsection{Charged Lepton Flavour Violation}
\label{subsec:CLFV}

In this section, we study various possible CLFV processes allowed by the proposed model. The augmented scalar sector containing CP-even and CP-odd scalars makes this analysis particularly interesting. The main sources of possible CLFV are the charged-lepton sector of the Yukawa Lagrangian in Eq.~\eqref{eq:yukawa_effective} and the charged-current gauge interaction enabled by the neutrino sector through the Type-I seesaw mechanism. While the former gives flavour-changing processes mediated by neutral scalars, the latter gives CLFV through $W^\pm$ exchange together with the heavy Majorana neutrinos. We discuss these two possibilities in turn below \cite{r44,r45}.

\subsubsection{Scalar Mediators}
\label{subsubsec:scalar_mediators}

We first consider the charged-lepton part of the Yukawa Lagrangian. For the M--R vacuum, $\langle\Phi\rangle=v_\Phi(1,1,1)$, the fluctuation along the VEV direction gives the diagonal charged-lepton mass matrix and hence does not generate a tree-level flavour-changing neutral current (FCNC) by itself. The two directions transverse to the VEV, on the other hand, carry non-diagonal charged-lepton Yukawa couplings. In the physical charged-lepton ordering already fixed in Section~\ref{sec:symmetry_realisation}, the corresponding mass-normalised flavour matrices may be written as
\begin{equation}
\Gamma_A=\frac{1}{\sqrt6\,v_\Phi}
\begin{pmatrix}
0&i m_\mu&-i m_\tau\\
-i m_e&0&i m_\tau\\
i m_e&-i m_\mu&0
\end{pmatrix},
\qquad
\Gamma_B=\frac{1}{\sqrt6\,v_\Phi}
\begin{pmatrix}
0&m_\mu&m_\tau\\
m_e&0&m_\tau\\
m_e&m_\mu&0
\end{pmatrix}.
\label{eq:GammaAB_main}
\end{equation}
Both matrices have vanishing diagonal entries but non-vanishing off-diagonal entries. Thus it is the transverse part of $\Phi$ that carries the neutral-scalar flavour-changing structure. The detailed derivation of Eq.~\eqref{eq:GammaAB_main}, including its consistency with the charged-lepton basis used earlier, is collected in Appendix~\ref{app:scalar_clfv_details}.

For the complete scalar potential of Section~\ref{sec:vacuum_scalar}, the physical amplitude is obtained by propagating the flavour-active interaction-basis directions through the full scalar mass matrix. The explicit-Yukawa and propagator descriptions are related through
\begin{equation}
\sum_n\frac{\Gamma_n^{(f)}\Gamma_n^{(i)}}{q^2-m_n^2}
=
\sum_{r,s}\Gamma_r^{(f)}
\left[(q^2I-M^2)^{-1}\right]_{rs}\Gamma_s^{(i)} ,
\label{eq:yukawa_propagator_bridge}
\end{equation}
where $r,s$ denote interaction-basis directions and $n$ labels the physical scalar states. Thus the propagator treatment does not replace the Yukawa analysis; the two are simply equivalent ways of writing the same amplitude.

For the CP-even transverse sector of Section~\ref{sec:vacuum_scalar}, we define
\begin{equation}
X_E=B_\Phi v_\Phi^2,\qquad
Y_E=B_\xi v_\xi^2,\qquad
Z_E^2=v_\Phi^2v_\xi^2\left(C_T^2+12\lambda_{\Phi\xi14}^2\right).
\label{eq:XEYEZE_main}
\end{equation}
The projection of the complete CP-even propagator on the two flavour-active $\Phi$ directions is exactly proportional to the identity,
\begin{equation}
\left[(M_E^2-q^2I)^{-1}\right]_{\Phi\Phi}=g_E(q^2)I_2,
\qquad
g_E(q^2)=\frac{Y_E-q^2}{(X_E-q^2)(Y_E-q^2)-Z_E^2}.
\label{eq:gE_maintext}
\end{equation}
Consequently, the two transverse flavour tensors in Eq.~\eqref{eq:GammaAB_main} receive the same CP-even spectral weight. This exact result of the complete scalar potential is the origin of the CP-even cancellations discussed below.

The CP-odd sector is different. With the quantities already introduced in Section~\ref{sec:vacuum_scalar},
\begin{equation}
X_O=6P_\Phi v_\Phi^2,\qquad
Y_O=6P_\xi v_\xi^2,\qquad
Z_O=\Delta v_\Phi v_\xi,
\label{eq:XOYOZO_main}
\end{equation}
the two massive transverse blocks give, in the contact limit,
\begin{equation}
g_{O1}=\frac{Y_O}{X_OY_O-Z_O^2},\qquad
g_{O2}=\frac{Y_O}{X_OY_O-9Z_O^2}.
\label{eq:gO12_main}
\end{equation}
It is useful to define
\begin{equation}
g_\Sigma=g_{O1}+g_{O2},\qquad
\delta g=g_{O1}-g_{O2},\qquad
F_\mp=2g_E\mp g_\Sigma,
\label{eq:gSigma_delta_Fpm_main}
\end{equation}
where $g_E\equiv g_E(0)$. In particular,
\begin{equation}
\delta g=-\frac{8Y_OZ_O^2}
{(X_OY_O-Z_O^2)(X_OY_O-9Z_O^2)}
={\cal O}(\Delta^2).
\label{eq:deltag_maintext}
\end{equation}
Thus the unequal CP-odd spectral weights reintroduce flavour-changing amplitudes that disappear when the two CP-odd structures become degenerate.

Keeping the final-state charges explicit, the scalar sector gives four mixed-flavour three-body $\tau$ modes,
\begin{align}
\tau^-&\to e^-\mu^-\mu^+,
&\tau^-&\to\mu^-e^-e^+,
\notag\\
\tau^-&\to e^-e^-\mu^+,
&\tau^-&\to\mu^-\mu^-e^+ .
\label{eq:scalar_four_channels_main}
\end{align}
The first two modes are controlled by $\delta g$: their CP-even pieces cancel exactly and their remaining CP-odd amplitudes begin at ${\cal O}(\Delta^2)$, so that the rates begin at ${\cal O}(\Delta^4)$. The two same-sign modes are generically non-zero and depend on the combinations $F_-$ and $F_+$ in Eq.~\eqref{eq:gSigma_delta_Fpm_main}. In particular, $\tau^-\to\mu^-\mu^-e^+$ contains the mass-enhanced coefficient $m_\mu m_\tau F_+$. The charge-resolved Wilson coefficients, identical-particle antisymmetrisation and contact-limit widths are given in Appendix~\ref{app:scalar_clfv_details}; the treatment follows the standard organisation of CLFV amplitudes \cite{r75,r76,r77,r78}. Current searches for LFV $\tau$ decays provide the experimental input for the corresponding bounds \cite{r79}.

The corresponding contact-limit branching ratios are, with $\Gamma_\tau$ denoting the measured total tau width,
\begin{align}
\mathrm{BR}(\tau^-\to e^-\mu^-\mu^+)
&=\frac{m_\tau^5(m_e^2+m_\mu^2)(m_\tau^2+m_\mu^2)}
{221184\pi^3\Gamma_\tau v_\Phi^4}\,|\delta g|^2,
\label{eq:BRtaueMuMu_scalar_main}\\[1mm]
\mathrm{BR}(\tau^-\to\mu^-e^-e^+)
&=\frac{m_\tau^5(m_e^2+m_\mu^2)(m_\tau^2+m_e^2)}
{221184\pi^3\Gamma_\tau v_\Phi^4}\,|\delta g|^2.
\label{eq:BRtaumuEE_scalar_main}
\end{align}
For the two same-sign modes one obtains
\begin{align}
\mathrm{BR}(\tau^-\to e^-e^-\mu^+)
={}&\frac{m_\tau^5}{442368\pi^3\Gamma_\tau v_\Phi^4}
\Big[(m_\mu^2m_\tau^2+m_e^4)|F_-|^2\notag\\
&\hspace{12mm}+2m_e^2(m_\tau^2+m_\mu^2)|F_+|^2\Big],
\label{eq:BRtaueeMu_scalar_main}\\[1mm]
\mathrm{BR}(\tau^-\to\mu^-\mu^-e^+)
={}&\frac{m_\tau^5}{442368\pi^3\Gamma_\tau v_\Phi^4}
\Big[(m_e^2m_\tau^2+m_\mu^4)|F_-|^2\notag\\
&\hspace{12mm}+2m_\mu^2(m_\tau^2+m_e^2)|F_+|^2\Big].
\label{eq:BRtaumumuE_scalar_main}
\end{align}
These expressions follow from the charge-resolved effective operators and the three-body phase-space normalisation; the detailed trace and identical-particle derivations are left to Appendix~\ref{app:scalar_clfv_details} \cite{r75,r76,r77,r78}. They are valid in the heavy-mediator contact regime, $m_S^2\gg q^2\lesssim m_\tau^2$. The experimental limits will be used only after the gauge/heavy neutral lepton (HNL) contribution has also been established, since some physical final states receive both mechanisms and must then be combined at the amplitude level \cite{r79}.

The neutral scalars also contribute to radiative decays $\ell_\alpha\to\ell_\beta\gamma$. Using the standard neutral-scalar dipole organisation of Ref.~\cite{r77}, the equal CP-even spectral weights in Eq.~\eqref{eq:gE_maintext} lead to an exact cancellation of the CP-even contribution. The two CP-odd transverse blocks are not spectrally identical when $\Delta\ne0$; consequently the CP-odd dipole amplitude is in general non-zero and begins at ${\cal O}(\Delta^2)$, while the corresponding rate begins at ${\cal O}(\Delta^4)$. The explicit loop-kernel translation is given in Appendix~\ref{app:scalar_clfv_details}.

At this point we do not convert every experimental three-body limit into an independent scalar bound. Some of the same final states also receive a heavy-neutrino contribution through the charged-current interaction, and the two mechanisms must be combined at the amplitude level whenever they overlap. We therefore first discuss the gauge-boson mediated contribution and return to the phenomenological constraints after both mechanisms have been established.

\subsubsection{Gauge Boson Mediator}
\label{subsubsec:gauge_mediator}

Apart from the scalar contribution discussed above, CLFV can also arise from the ordinary charged-current interaction once the active neutrinos mix with the heavy Majorana states. In the heavy-neutrino mass basis the relevant interaction is
\begin{equation}
-\mathcal L_{CC}\supset\frac{g}{\sqrt2}W^-_\mu\,
\overline\ell_\alpha\gamma^\mu P_L B_{\alpha I}N_I+\mathrm{h.c.},
\label{eq:LCC_HNL_main}
\end{equation}
where $B_{\alpha I}$ denotes the charged-current coupling of the charged lepton $\ell_\alpha$ to the heavy Majorana state $N_I$. At leading seesaw order, the active--heavy mixing that enters the charged current is
\begin{equation}
\Theta=\left[M_D M_R^{-1}\right]^*,
\label{eq:Theta_main}
\end{equation}
where the charged-lepton ordering is the one already fixed in Section~\ref{sec:symmetry_realisation}. From this point onward $M_D$ denotes the Dirac matrix in that physical charged-lepton ordering. The complete neutral-fermion matrix also contains the Weinberg-like light block $M_w$, so Eq.~\eqref{eq:Theta_main} is the leading seesaw expression rather than an exact all-orders identity.

Using the same physical flavour ordering, it is convenient to write
\begin{equation}
M_D=
\begin{pmatrix}
0&P&0\\
Q&0&0\\
0&0&R
\end{pmatrix},
\qquad
M_R=
\begin{pmatrix}
\mathcal M_1&0&\mathcal M_2\\
0&\mathcal M_3&0\\
\mathcal M_2&0&0
\end{pmatrix}.
\label{eq:MDc_MR_main}
\end{equation}
Here the three Dirac-sector entries are only short notations for the combinations already fixed by the Yukawa Lagrangian,
\begin{equation}
P=\frac{\sqrt3\,y_2v_Hv_\Phi v_\chi}{2\sqrt2\,\Lambda^2},\qquad
Q=\frac{\sqrt3\,\omega y_1v_Hv_\Phi}{2\Lambda},\qquad
R=\frac{\sqrt3\,\omega^2y_3v_Hv_\Phi v_\psi}{2\sqrt2\,\Lambda^2}.
\label{eq:PQR_def_main}
\end{equation}
Direct multiplication then gives
\begin{equation}
\Theta=\left[
\begin{pmatrix}
0&P/\mathcal M_3&0\\
0&0&Q/\mathcal M_2\\
R/\mathcal M_2&0&-\mathcal M_1R/\mathcal M_2^2
\end{pmatrix}
\right]^* .
\label{eq:Theta_explicit_main}
\end{equation}
It follows immediately that
\begin{equation}
(\Theta\Theta^\dagger)_{e\mu}=(\Theta\Theta^\dagger)_{e\tau}=0,
\qquad
(\Theta\Theta^\dagger)_{\mu\tau}\neq0
\quad\text{generically}.
\label{eq:ThetaTheta_support_main}
\end{equation}
The derivation of Eqs.~\eqref{eq:MDc_MR_main}--\eqref{eq:Theta_explicit_main}, including the inversion of $M_R$, is given in Appendix~\ref{app:gauge_clfv_details}. The same reconstruction also makes transparent how the lower-case texture parameters enter this sector. In particular,
\begin{equation}
\mathcal M_3=-\frac{P^2}{a-b},\qquad
\mathcal M_2=-\frac{2QR}{b+2t},\qquad
\mathcal M_1=\frac{4(2b+d)Q^2}{(b+2t)^2}.
\label{eq:Mi_texture_main}
\end{equation}
Thus the heavy-neutrino contribution probes the three combinations $a-b$, $b+2t$ and $2b+d$ without requiring numerical values of $a,b,t,d$ to be inserted at this stage.

The $(1,3)$ block of $M_R$ is diagonalised by a Takagi transformation. After the usual phase redefinitions of the displayed $\mathcal M_1$ and $\mathcal M_2$ parameters, define
\begin{equation}
\Delta_N=\sqrt{\mathcal M_1^2+4\mathcal M_2^2},\qquad
M_\pm=\frac{\Delta_N\pm\mathcal M_1}{2},\qquad
M_0=|\mathcal M_3|,
\label{eq:HNL_masses_main}
\end{equation}
and $\tan2\theta_N=2\mathcal M_2/\mathcal M_1$, the heavy-state charged-current couplings may be written as
\begin{equation}
B_{\ell N}=
\begin{pmatrix}
P^*/\mathcal M_3&0&0\\
0&Q^*c_N/M_+&iQ^*s_N/M_-\\
0&R^*s_N/M_+&-iR^*c_N/M_-
\end{pmatrix},
\label{eq:Bmatrix_main}
\end{equation}
where the columns are ordered as $(N_0,N_+,N_-)$ and $c_N=\cos\theta_N$, $s_N=\sin\theta_N$. Equation~\eqref{eq:Bmatrix_main} shows the leading flavour support directly:
\begin{equation}
N_0:\ e\ \text{only},\qquad
N_\pm:\ \mu,\tau\ \text{only}.
\label{eq:HNL_support_main}
\end{equation}
This flavour structure determines which radiative and three-body CLFV amplitudes can be generated at leading order.

For the radiative modes we use the gauge-invariant one-loop convention of Ref.~\cite{r89}. Defining $x_I=M_I^2/M_W^2$ and
\begin{equation}
G_\gamma^{\alpha\beta}=\sum_I B_{\alpha I}^*B_{\beta I}G_\gamma(x_I),
\label{eq:Ggamma_ab_main}
\end{equation}
the branching ratio is
\begin{equation}
\mathrm{BR}(\ell_\alpha\to\ell_\beta\gamma)=
\frac{\alpha_w^3s_w^2}{256\pi^2}
\frac{m_\alpha^4}{M_W^4}\frac{m_\alpha}{\Gamma_\alpha}
\left|G_\gamma^{\alpha\beta}\right|^2 .
\label{eq:BRrad_HNL_main}
\end{equation}
The support in Eq.~\eqref{eq:HNL_support_main} gives
\begin{equation}
G_\gamma^{\mu e}=G_\gamma^{\tau e}=0,
\qquad
G_\gamma^{\tau\mu}=z_+G_\gamma(x_+)+z_-G_\gamma(x_-),
\label{eq:Ggamma_support_main}
\end{equation}
with
\begin{equation}
z_+=Q^*R\frac{s_Nc_N}{M_+^2},\qquad
z_-=-Q^*R\frac{s_Nc_N}{M_-^2}.
\label{eq:zpm_main}
\end{equation}
Hence, among the standard radiative HNL channels, only $\tau\to\mu\gamma$ is non-zero at leading order in the present texture. The explicit form of $G_\gamma(x)$ and the common-heavy limit are collected in Appendix~\ref{app:gauge_clfv_details}.

The three-body decays require more care. In addition to the photon penguin they receive $Z$-penguin and box contributions; for Majorana neutrinos there is also a genuinely Majorana box contraction \cite{r89}. The leading contributions to the different channels are summarised in Table~\ref{tab:HNL_channel_support_main}.
\begin{table}[H]
\centering
\caption{Leading heavy-neutrino/$W$ contributions to the relevant CLFV channels. A check mark indicates a non-zero leading contribution and a cross indicates a vanishing one. ``Majorana box'' denotes a contribution that survives only through the Majorana box structure at the order considered.}
\label{tab:HNL_channel_support_main}
\renewcommand{\arraystretch}{1.10}
\begin{tabularx}{\textwidth}{Y>{\centering\arraybackslash}p{0.28\textwidth}}
\toprule
Process & HNL/$W$ contribution\\
\midrule
$\tau\to\mu\gamma$ & \YesMark\\
$\mu\to e\gamma$, $\tau\to e\gamma$ & \NoMark\\
$\tau^-\to\mu^-\mu^-\mu^+$ & \YesMark\\
$\tau^-\to\mu^-e^-e^+$ & \YesMark\\
$\tau^-\to e^-e^-\mu^+$ & Majorana box\\
$\tau^-\to e^-\mu^-\mu^+$ & \NoMark\\
$\tau^-\to\mu^-\mu^-e^+$ & \NoMark\\
$\tau\to3e$, $\mu\to3e$ & \NoMark\\
\bottomrule
\end{tabularx}
\end{table}
The last result is particularly useful: the scalar modes $\tau^-\to e^-\mu^-\mu^+$ and $\tau^-\to\mu^-\mu^-e^+$ do not receive a leading HNL/$W$ contribution, whereas $\tau^-\to\mu^-e^-e^+$ and $\tau^-\to e^-e^-\mu^+$ receive both mechanisms.

The corresponding HNL three-body branching ratios are written in terms of the
standard form factors $F_\gamma$, $F_Z$ and $F_{\rm Box}$, together with the
corresponding loop kernels defined in Appendix~\ref{app:gauge_clfv_details}. We define
\begin{equation}
\mathcal K_\tau=\frac{\alpha_w^4}{24576\pi^3}
\frac{m_\tau^4}{M_W^4}\frac{m_\tau}{\Gamma_\tau},
\quad
\mathcal P_{\tau\mu}=F_Z^{\tau\mu}-F_\gamma^{\tau\mu},
\label{eq:K_P_main}
\end{equation}
and
\begin{equation}
\mathcal Q_{3\mu}=F_Z^{\tau\mu}+\frac12F_{\rm Box}^{\tau\mu\mu\mu},
\qquad
\mathcal Q_{\mu ee}=F_Z^{\tau\mu}+F_{\rm Box}^{\tau\mu ee}.
\label{eq:Q_main}
\end{equation}
With these definitions, the three-body rates are \cite{r89}
\begin{align}
\mathrm{BR}(\tau\to3\mu)=\mathcal K_\tau\Big\{&
2|\mathcal Q_{3\mu}-2s_w^2\mathcal P_{\tau\mu}|^2
+4s_w^4|\mathcal P_{\tau\mu}|^2\notag\\
&+16s_w^2\Re[\mathcal Q_{3\mu}G_\gamma^{\tau\mu *}]
-48s_w^4\Re[\mathcal P_{\tau\mu}G_\gamma^{\tau\mu *}]\notag\\
&+32s_w^4|G_\gamma^{\tau\mu}|^2
\left(\ln\frac{m_\tau^2}{m_\mu^2}-\frac{11}{4}\right)\Big\},
\label{eq:BR3mu_HNL_main}
\end{align}
while
\begin{align}
\mathrm{BR}(\tau^-\to\mu^-e^-e^+)=\mathcal K_\tau\Big\{&
|\mathcal Q_{\mu ee}-2s_w^2\mathcal P_{\tau\mu}|^2
+4s_w^4|\mathcal P_{\tau\mu}|^2\notag\\
&+8s_w^2\Re[\mathcal Q_{\mu ee}G_\gamma^{\tau\mu *}]
-32s_w^4\Re[\mathcal P_{\tau\mu}G_\gamma^{\tau\mu *}]\notag\\
&+32s_w^4|G_\gamma^{\tau\mu}|^2
\left(\ln\frac{m_\tau^2}{m_e^2}-3\right)\Big\}.
\label{eq:BRmuee_HNL_main}
\end{align}
For the channel supported only by the genuine Majorana box at leading HNL order,
\begin{equation}
\mathrm{BR}(\tau^-\to e^-e^-\mu^+)=
\frac{\alpha_w^4}{49152\pi^3}
\frac{m_\tau^4}{M_W^4}\frac{m_\tau}{\Gamma_\tau}
\left|F_{\rm Box}^{\tau ee\mu}\right|^2.
\label{eq:BRMajBox_main}
\end{equation}
These expressions show why the three-body HNL modes cannot be reduced to a single universal coefficient multiplying $|(\Theta\Theta^\dagger)_{\tau\mu}|^2$: the photon penguin, $Z$ penguin and box terms carry different heavy-mass dependences and interfere with each other.

The scalar and HNL contributions have been derived separately above. For channels receiving both contributions, however, the amplitudes must be added before squaring,
\begin{equation}
\mathcal M_{\rm tot}=\mathcal M_{\rm scalar}+\mathcal M_{W/N},
\label{eq:Mtot_main}
\end{equation}
and not, in general, by simply adding two separately calculated branching ratios. At the order considered here this applies to $\tau\to\mu\gamma$, $\tau^-\to\mu^-e^-e^+$ and $\tau^-\to e^-e^-\mu^+$. The remaining charge-resolved channels in Table~\ref{tab:HNL_channel_support_main} provide clean support tests of one mechanism or the other. We use these results in the phenomenological constraint discussion below.

\subsubsection{Phenomenological constraints}
\label{subsubsec:clfv_constraints}

We now use the present experimental upper limits to illustrate what can be constrained without making an additional assumption on the scalar or heavy-neutrino spectrum. For the four charge-resolved three-lepton modes, the Heavy Flavor Averaging Group (HFLAV) end-2023 combinations give, at $90\%$ C.L. \cite{r90},
\begin{equation}
\begin{gathered}
{\rm BR}(\tau^-\to e^-\mu^-\mu^+)<2.0\times10^{-8},\qquad
{\rm BR}(\tau^-\to\mu^-e^-e^+)<1.3\times10^{-8},\\
{\rm BR}(\tau^-\to e^-e^-\mu^+)<1.1\times10^{-8},\qquad
{\rm BR}(\tau^-\to\mu^-\mu^-e^+)<1.4\times10^{-8}.
\end{gathered}
\label{eq:HFLAV_tau3l_limits_main}
\end{equation}
The first and the last of these are particularly useful here because they do not receive a leading HNL/$W$ contribution in Eq.~\eqref{eq:HNL_support_main}. The first one gives directly
\begin{equation}
\left|\frac{\delta g}{v_\Phi^2}\right|<7.05\times10^{-7}\ {\rm GeV}^{-4}.
\label{eq:deltag_direct_numeric_main}
\end{equation}
The clean same-sign mode $\tau^-\to\mu^-\mu^-e^+$ gives the stronger structural constraint. In the stable massive scalar region, $F_+>0$ and the propagator combinations obey $|F_-|<F_+$ and $|\delta g|<F_+$. Using Eq.~\eqref{eq:BRtaumumuE_scalar_main} therefore gives
\begin{equation}
0<\frac{F_+}{v_\Phi^2}<C_S,
\qquad
\frac{|F_-|}{v_\Phi^2}<C_S,
\qquad
\frac{|\delta g|}{v_\Phi^2}<C_S,
\label{eq:scalar_contact_master_main}
\end{equation}
with
\begin{equation}
C_S=5.91\times10^{-7}\ {\rm GeV}^{-4}.
\label{eq:CS_master_main}
\end{equation}
The first limit in Eq.~\eqref{eq:HFLAV_tau3l_limits_main} also gives the direct clean scalar bound in Eq.~\eqref{eq:deltag_direct_numeric_main}. A stronger constraint follows once the overlapping $\tau^-\to\mu^-e^-e^+$ channel is treated at the amplitude level. Here the massless-daughter approximation means that daughter charged-lepton masses are neglected in the three-body phase-space integration and external kinematics wherever they enter only through corrections of order $m_\ell^2/m_\tau^2$, while masses required by chirality flips and Yukawa/chiral coefficients are retained. This is adequate for the quoted $\tau$ limits because the omitted electron- and muon-mass effects are subleading at the precision of the present bounds; see, for example, the standard CLFV review in Ref.~\cite{r76}. In this approximation only the $m_e^2$-suppressed scalar chiral coefficient interferes with the left-handed HNL vector current, while the dominant scalar term remains positive definite. The resulting full-model bound is
\begin{equation}
\frac{|\delta g|}{v_\Phi^2}<5.6903\times10^{-7}\ {\rm GeV}^{-4}.
\label{eq:deltag_combined_numeric_main}
\end{equation}
For the same-sign channel $\tau^-\to e^-e^-\mu^+$ the only leading HNL contribution is the genuine Majorana box, and its interference with the scalar term vanishes in the same approximation. The experimental limit therefore also implies
\begin{equation}
|F_{\rm Box}^{\tau ee\mu}|<0.2601.
\label{eq:MajBox_numeric_bound_main}
\end{equation}
The numerical steps leading to Eqs.~\eqref{eq:deltag_direct_numeric_main}--\eqref{eq:MajBox_numeric_bound_main} are given in Appendix~\ref{app:numerical_constraints}. These results constrain propagator and loop combinations rather than providing universal lower limits on individual scalar or heavy-neutrino masses. For example,
\begin{align}
F_+={}&2\left(\frac{\cos^2\beta}{m_{T+}^2}+\frac{\sin^2\beta}{m_{T-}^2}\right)
+\left(\frac{\cos^2\theta_1}{m_{1+}^2}+\frac{\sin^2\theta_1}{m_{1-}^2}\right)\notag\\
&+\left(\frac{\cos^2\theta_2}{m_{2+}^2}+\frac{\sin^2\theta_2}{m_{2-}^2}\right),
\label{eq:Fplus_spectral_main}
\end{align}
so the bound acts jointly on the scalar masses, their mixing angles and $v_\Phi$.

The radiative modes provide complementary information. Since the leading HNL contribution to $\mu\to e\gamma$ vanishes for the support in Eq.~\eqref{eq:HNL_support_main}, the MEG~II limit
\begin{equation}
{\rm BR}(\mu\to e\gamma)<1.5\times10^{-13}\qquad(90\%\ {\rm C.L.})
\label{eq:MEGII_limit_main}
\end{equation}
\cite{r91} is, at the order considered, a clean probe of the scalar-loop contribution. However, the CP-odd dipole amplitude depends on a difference of spectral loop weights and can approach zero continuously; hence Eq.~\eqref{eq:MEGII_limit_main} cannot be converted into a model-independent lower bound on one scalar mass \cite{r77}. For the surviving HNL radiative mode, the HFLAV limit ${\rm BR}(\tau\to\mu\gamma)<2.7\times10^{-8}$ \cite{r90} gives
\begin{equation}
|G_\gamma^{\tau\mu}|\lesssim6.47\times10^{-3},
\label{eq:Ggamma_numeric_main}
\end{equation}
for the electroweak input convention specified in Appendix~\ref{app:numerical_constraints}. The three-body HNL modes contain photon-penguin, $Z$-penguin and box terms with nontrivial interference, and therefore no universal one-dimensional numerical bound on $P,Q$ or $R$ is inferred from them without specifying the heavy spectrum. For the overlapping scalar/HNL channels the amplitudes are combined according to Eq.~\eqref{eq:Mtot_main}.

\subsection{Non-unitarity of the Effective PMNS Matrix}
\label{subsec:nonunitarity}

In seesaw frameworks, the mixing of the active light neutrinos with heavy neutral fermions generally makes the effective $3\times3$ lepton mixing matrix slightly non-unitary \cite{r46,r47,r80,r81}. After the heavy states are integrated out, the charged-current mixing matrix may be written, to leading order in the active--heavy mixing, as
\begin{equation}
N\simeq (I-\widetilde\eta)U_{\rm PMNS},
\qquad
\widetilde\eta\simeq\frac12\Theta\Theta^\dagger .
\label{eq:eta_definition_main}
\end{equation}
Here $\widetilde\eta$ denotes the non-unitarity matrix and should not be confused with the scalar field $\eta$ introduced in Section~\ref{sec:symmetry_realisation}. The same active--heavy charged-current mixing matrix $\Theta$ has already appeared in the gauge-mediated CLFV discussion above, so no new basis transformation is required at this stage.

Using Eqs.~\eqref{eq:Theta_explicit_main} and \eqref{eq:Mi_texture_main}, this becomes
\begin{equation}
\Theta=-\left[
\begin{pmatrix}
0&\dfrac{a-b}{P}&0\\[2mm]
0&0&\dfrac{b+2t}{2R}\\[2mm]
\dfrac{b+2t}{2Q}&0&\dfrac{2b+d}{R}
\end{pmatrix}
\right]^*.
\label{eq:Theta_texture_main}
\end{equation}
Consequently,
\begin{equation}
\widetilde\eta=
\begin{pmatrix}
\dfrac{|a-b|^2}{2|P|^2}&0&0\\[3mm]
0&\dfrac{|b+2t|^2}{8|R|^2}&
\dfrac{(b+2t)^*(2b+d)}{4|R|^2}\\[3mm]
0&\dfrac{(b+2t)(2b+d)^*}{4|R|^2}&
\dfrac{|b+2t|^2}{8|Q|^2}+\dfrac{|2b+d|^2}{2|R|^2}
\end{pmatrix}.
\label{eq:eta_matrix_main}
\end{equation}
The detailed multiplication leading to Eq.~\eqref{eq:eta_matrix_main}, together with checks in the heavy-neutrino mass basis, is given in Appendix~\ref{app:nonunitarity_details}. Two features are immediate. First,
\begin{equation}
\widetilde\eta_{e\mu}=\widetilde\eta_{e\tau}=0
\label{eq:eta_ezeros_main}
\end{equation}
at the leading order retained in the model. Second, the $\mu\tau$ entry is correlated with the muon diagonal entry through
\begin{equation}
\frac{|\widetilde\eta_{\mu\tau}|}{\widetilde\eta_{\mu\mu}}
=\frac{2|2b+d|}{|b+2t|}.
\label{eq:eta_ratio_main}
\end{equation}
Thus the non-unitarity pattern is highly restricted by the texture rather than being a general Hermitian matrix.

For the numerical limits we use the general heavy-neutrino scenario of the current global analysis in Ref.~\cite{r47}. Since CLFV has been calculated independently in Subsection~\ref{subsec:CLFV}, we use the lepton-flavour-conserving information for the non-unitarity test so that direct CLFV limits are not counted a second time. In particular, positive semidefiniteness of $\widetilde\eta=\tfrac12\Theta\Theta^\dagger$ gives the Cauchy--Schwarz bound
\begin{equation}
|\widetilde\eta_{\mu\tau}|^2\le \widetilde\eta_{\mu\mu}\widetilde\eta_{\tau\tau}.
\label{eq:nonunitarity_CS_main}
\end{equation}
At $95\%$ C.L. the limits used here are
\begin{equation}
\widetilde\eta_{ee}<1.4\times10^{-3},\qquad
\widetilde\eta_{\mu\mu}<1.4\times10^{-4},
\label{eq:NU_bounds_diag1_main}
\end{equation}
\begin{equation}
\widetilde\eta_{\tau\tau}<8.9\times10^{-4},\qquad
|\widetilde\eta_{\mu\tau}|<1.8\times10^{-4}.
\label{eq:NU_bounds_diag2_main}
\end{equation}
The light-neutrino mass matrix of the present model contains, in addition to the Type-I contribution, the independent Weinberg-like term $M_w$. We therefore do not impose the extra light-mass correlations of a pure three-RH-neutrino Type-I fit.

Without inserting numerical values for the texture parameters $a,b,t,d$, Eqs.~\eqref{eq:eta_matrix_main}--\eqref{eq:NU_bounds_diag2_main} give the simple lower bounds
\begin{equation}
|P|>18.90\,|a-b|,
\qquad
|R|>29.88\,|b+2t|,
\label{eq:NU_PR_bounds_main}
\end{equation}
and, from the $\mu\tau$ entry,
\begin{equation}
|R|>37.27\sqrt{|b+2t|\,|2b+d|}.
\label{eq:NU_Rmutau_bound_main}
\end{equation}
The $\tau\tau$ constraint is intrinsically correlated and should not be split into two independent bounds. It reads
\begin{equation}
\frac{|b+2t|^2}{8|Q|^2}
+\frac{|2b+d|^2}{2|R|^2}
<8.9\times10^{-4},
\label{eq:NU_tau_corr_main}
\end{equation}
which, for a specified $R$, implies
\begin{equation}
|Q|^2>
\frac{|b+2t|^2}
{8\left[8.9\times10^{-4}-\dfrac{|2b+d|^2}{2|R|^2}\right]},
\label{eq:NU_QgivenR_main}
\end{equation}
provided the denominator is positive. The translation of these inequalities into the original Yukawa--VEV combinations entering $P,Q,R$ is given in Appendix~\ref{app:nonunitarity_details}.

The bounds in Eqs.~\eqref{eq:NU_PR_bounds_main}--\eqref{eq:NU_QgivenR_main} are complementary to the CLFV constraints rather than replacements for them. In particular, the exact gauge-mediated CLFV amplitudes of Subsection~\ref{subsubsec:gauge_mediator} contain heavy-mass-dependent loop functions and cannot in general be reduced to a single number proportional to $\widetilde\eta_{\alpha\beta}$ \cite{r47,r91}. For the comparison with the heavy-neutrino global fit we also require the heavy states relevant to that fit to lie above the electroweak production threshold; in the present analysis we impose $M_0,M_+,M_->M_Z$ as an applicability condition, not as an additional CLFV bound.

Thus the non-unitarity analysis constrains correlated combinations of $P,Q,R$ and the texture parameters while preserving the characteristic leading zeros in the electron off-diagonal entries. The present bounds mainly restrict magnitudes and do not remove the remaining CP-phase freedom.

\subsubsection{Numerical implications}
\label{subsubsec:NU_numerical}

To show the size of these constraints, we now insert the retained texture combinations at the best-fit points obtained in the model analysis. The combinations entering Eq.~\eqref{eq:eta_matrix_main} are
\begin{equation}
\begin{array}{c|cc}
&{\rm NO}&{\rm IO}\\ \hline
|a-b|&0.01690~{\rm eV}&0.06863~{\rm eV}\\
|b+2t|&0.07386~{\rm eV}&0.10703~{\rm eV}\\
|2b+d|&0.00840~{\rm eV}&0.00508~{\rm eV}
\end{array}
\label{eq:texture_combinations_numeric_main}
\end{equation}
Using Eqs.~\eqref{eq:NU_PR_bounds_main}--\eqref{eq:NU_QgivenR_main}, one obtains
\begin{equation}
\begin{array}{c|cc}
&{\rm NO}&{\rm IO}\\ \hline
|P|_{\min}&0.319~{\rm eV}&1.30~{\rm eV}\\
|Q|_{\min}^{\rm proj}&0.875~{\rm eV}&1.268~{\rm eV}\\
|R|_{\min}&2.21~{\rm eV}&3.20~{\rm eV}
\end{array}
\label{eq:PQR_numeric_main}
\end{equation}
The three entries in each ordering are separate lower projections obtained without fixing additional benchmark values and are not a simultaneously saturated parameter point. In particular, the $Q$ projection follows conservatively from the positive first term in $\widetilde\eta_{\tau\tau}$; the full $Q$--$R$ constraint remains the correlated inequality in Eq.~\eqref{eq:NU_tau_corr_main}. The detailed numerical substitution is given in Appendix~\ref{app:numerical_constraints}.

It is also useful to translate these limits into the combinations appearing in the original higher-dimensional Dirac operators,
\begin{equation}
{\cal Y}_1=\frac{|y_1|v_\Phi}{\Lambda},\qquad
{\cal Y}_2=\frac{|y_2|v_\Phi v_\chi}{\Lambda^2},\qquad
{\cal Y}_3=\frac{|y_3|v_\Phi v_\psi}{\Lambda^2}.
\label{eq:Y123_numeric_main}
\end{equation}
For $v_H=246$ GeV the corresponding lower limits are
\begin{equation}
\begin{array}{c|cc}
&{\rm NO}&{\rm IO}\\ \hline
{\cal Y}_1&4.11\times10^{-12}&5.95\times10^{-12}\\
{\cal Y}_2&2.12\times10^{-12}&8.61\times10^{-12}\\
{\cal Y}_3&1.46\times10^{-11}&2.12\times10^{-11}
\end{array}
\label{eq:Y123_numeric_bounds_main}
\end{equation}
These are constraints on the products of Yukawa couplings, flavon VEVs and powers of $\Lambda^{-1}$ appearing in Eq.~\eqref{eq:PQR_def_main}; they do not separately determine $y_i$, the flavon VEVs or the cutoff. Likewise, the eV-scale numbers in Eq.~\eqref{eq:PQR_numeric_main} are lower limits on the Dirac entries $P,Q,R$, not masses of the RH neutrinos. The comparison with the non-unitarity fit still requires the physical heavy eigenstates to satisfy the heavy-state applicability condition discussed above.

In the common-heavy limit, $G_\gamma^{\tau\mu}\simeq\widetilde\eta_{\mu\tau}$. The non-unitarity bound $|\widetilde\eta_{\mu\tau}|<1.8\times10^{-4}$ is then stronger than the direct radiative constraint in Eq.~\eqref{eq:Ggamma_numeric_main}. Away from this limit the full mass-dependent loop functions must be retained. Thus CLFV and non-unitarity probe related but not identical combinations of the same heavy sector.

\subsection{Baryon Asymmetry of the Universe}
\label{subsec:leptogenesis}

The observed excess of matter over antimatter in the Universe remains one of the outstanding problems in particle physics and cosmology. The baryon-to-entropy ratio inferred from cosmological observations is of order $Y_B\simeq8.7\times10^{-11}$ \cite{r18,r74}. A well-known dynamical possibility is leptogenesis \cite{r72,r73}, where the out-of-equilibrium decay of heavy Majorana neutrinos generates a lepton asymmetry which is subsequently converted into a baryon asymmetry by electroweak sphalerons. In the present work we restrict ourselves to the conventional Type-I two-body channels
\begin{equation}
N_i\rightarrow L_\alpha H,
\qquad
N_i\rightarrow \overline{L}_\alpha H^\dagger .
\label{eq:conventional_lepto_channel_main}
\end{equation}

Using the Dirac matrix in the charged-lepton mass basis established above and the same positive-mass Takagi basis used for the heavy-neutrino sector, the decay Yukawa matrix can be written as
\begin{equation}
{\cal Y}=\frac{\sqrt2}{v_H}
\begin{pmatrix}
0&0&P^*\\
c_NQ^*&i s_NQ^*&0\\
s_NR^*&-i c_NR^*&0
\end{pmatrix},
\label{eq:Ylepto_main}
\end{equation}
where the columns are ordered as $(N_+,N_-,N_0)$, the rows are $(e,\mu,\tau)$, and $c_N=\cos\theta_N$, $s_N=\sin\theta_N$. No assumption that $P,Q,R$ are real is required. The Hermitian combination entering the conventional one-loop source is then
\begin{equation}
H\equiv{\cal Y}^\dagger{\cal Y}
=\frac{2}{v_H^2}
\begin{pmatrix}
|Q|^2c_N^2+|R|^2s_N^2 & i c_Ns_N(|Q|^2-|R|^2)&0\\
-i c_Ns_N(|Q|^2-|R|^2)&|Q|^2s_N^2+|R|^2c_N^2&0\\
0&0&|P|^2
\end{pmatrix}.
\label{eq:Hlepto_main}
\end{equation}
Thus $H_{+-}$ is purely imaginary, while $H_{+0}=H_{-0}=0$. Consequently,
\begin{equation}
\operatorname{Im}\!\left[H_{ij}^{\,2}\right]=0,
\qquad i\neq j.
\label{eq:unflav_invariant_zero_main}
\end{equation}
For non-resonant conventional Type-I leptogenesis the unflavoured asymmetry has the form \cite{r72,r73}
\begin{equation}
\epsilon_i^{I}=
\frac{1}{8\pi H_{ii}}
\sum_{j\ne i}
\operatorname{Im}\!\left[H_{ij}^{\,2}\right]
 f\!\left(\frac{M_j^2}{M_i^2}\right),
\label{eq:eps_unfl_main}
\end{equation}
where $f$ denotes the usual real vertex-plus-self-energy loop function. Equation~\eqref{eq:unflav_invariant_zero_main} therefore gives
\begin{equation}
\epsilon_i^{I}=0
\qquad\text{for all }i.
\label{eq:eps_zero_main}
\end{equation}
This is a structural consequence of the heavy-basis Yukawa matrix and does not rely on a numerical benchmark.

The same conclusion holds for the leading flavour-resolved decay asymmetries. For the only potentially non-zero pair $(+,-)$,
\begin{equation}
{\cal Y}_{\mu+}^*{\cal Y}_{\mu-}
=\frac{2i|Q|^2c_Ns_N}{v_H^2},
\qquad
{\cal Y}_{\tau+}^*{\cal Y}_{\tau-}
=-\frac{2i|R|^2c_Ns_N}{v_H^2},
\label{eq:flav_products_main}
\end{equation}
while the corresponding electron products vanish. Since $H_{+-}$ is also purely imaginary, the vertex- and self-energy-type flavoured invariants are real and hence
\begin{equation}
\epsilon_{i\alpha}^{I}=0
\qquad\text{for all }i,\alpha.
\label{eq:flav_zero_main}
\end{equation}

Within the conventional Type-I mechanism considered here, the leading CP asymmetries therefore vanish as a consequence of the heavy-basis Yukawa structure. Contributions associated with the independent Weinberg-like operator, scalar-assisted channels, and a full flavour-density treatment are beyond the scope of the present manuscript. The stability of this conventional result under the controlled soft deformation introduced below is discussed in Section~\ref{sec:soft_breaking}.

The phenomenological discussion in this section has used the scalar spectrum obtained from the symmetry-preserving potential of Section~\ref{sec:vacuum_scalar}. The same analysis, however, also showed that the viable M--R vacuum contains seven physical CP-odd zero modes. Four of them are Goldstone bosons of accidental global $U(1)$ symmetries that remain after the Yukawa restrictions are imposed, while the other three are phase-flat directions of the renormalisable scalar potential, as summarised in Eq.~\eqref{eq:physical_zero_origin_main}. A realistic scalar spectrum therefore requires these directions to be lifted. In the next section we introduce a controlled dimension-two deformation of the scalar potential and examine whether this can be done without disturbing the M--R vacuum, the triadic texture, or the phenomenological conclusions obtained above.

\section{Controlled soft symmetry breaking and its phenomenological consequences}
\label{sec:soft_breaking}

\subsection{Motivation and form of the soft terms}
\label{subsec:soft_motivation}

The viable M--R vacuum obtained in Section~\ref{sec:vacuum_scalar} contains seven physical CP-odd zero modes in addition to the neutral electroweak Goldstone. Four of them originate from accidental global $U(1)$ symmetries that survive the Yukawa restrictions, while the remaining three are phase-flat directions of the renormalisable scalar potential. Exactly massless physical pseudoscalars can be phenomenologically restrictive through stellar, astrophysical or early-Universe effects, although the corresponding bounds depend on their couplings and production history. We therefore do not treat masslessness alone as a model-independent exclusion; instead, we seek a controlled deformation that lifts all seven modes without changing the real M--R alignment or the triadic texture established in Section~\ref{sec:phenomenology}.

A convenient choice is a set of real dimension-two phase-sensitive bilinears. For the two $A_4$ triplets we take
\begin{equation}
\Delta V_X=s_X\left[(X^\dagger X)_1-
\frac12\left\{(XX)_1+(X^\dagger X^\dagger)_1\right\}\right],
\qquad X=\Phi,\xi,
\label{eq:soft_triplet_main}
\end{equation}
and for the five singlet flavons
\begin{equation}
\Delta V_S=s_S\left[S^\dagger S-
\frac12\left(S^2+S^{\dagger2}\right)\right],
\qquad S=\psi,\chi,\eta,\rho,\kappa.
\label{eq:soft_singlet_main}
\end{equation}
The complete soft deformation is
\begin{equation}
\Delta V_{\rm soft}=\Delta V_\Phi+\Delta V_\xi+
\sum_{S=\psi,\chi,\eta,\rho,\kappa}\Delta V_S,
\qquad [s_X]={\rm mass}^2,
\label{eq:soft_total_main}
\end{equation}
with real coefficients. In the M--R basis, $(XX)_1=X_1^2+X_2^2+X_3^2$, so,  on the selected real branch
\begin{align}
\Delta V_{\rm soft}={}&s_\Phi\sum_{i=1}^3\phi_{Ii}^2
+s_\xi\sum_{i=1}^3\xi_{Ii}^2
+s_\psi\psi_I^2+s_\chi\chi_I^2+s_\eta\eta_I^2
+s_\rho\rho_I^2+s_\kappa\kappa_I^2 .
\label{eq:soft_phaseonly_main}
\end{align}
The terms are therefore soft in the technical sense: they break flavour/shaping symmetries only through operators of mass dimension two and introduce no hard symmetry-breaking quartic coupling. They are gauge invariant and, for real $s_X$, preserve the standard CP transformation used in the scalar analysis.

The breaking pattern is not identical for all seven terms. Table~\ref{tab:soft_symmetry_audit_main} summarises the declared discrete symmetries and a convenient basis $U(1)_{A,B,C,D}$ for the four accidental global rephasings. The derivation of the latter is given in Appendix~\ref{app:soft_scalar_details}.

\begin{table}[!htbp]
\centering
\scriptsize
\setlength{\tabcolsep}{4.2pt}
\caption{Symmetry content of the dimension-two soft terms. \YesMark\ means that the symmetry is preserved and \NoMark\ that it is explicitly broken. The last column lists the accidental global $U(1)$ generators, in the basis defined in Appendix~\ref{app:soft_scalar_details}, broken by each term. All seven terms preserve the electroweak gauge group, the displayed $Z_3$, and CP for real coefficients.}
\label{tab:soft_symmetry_audit_main}
\renewcommand{\arraystretch}{1.12}
\begin{tabularx}{\textwidth}{Yccccc>{\raggedright\arraybackslash}p{0.28\textwidth}}
\toprule
Term & $A_4$ & $Z_{10}$ & $Z_7$ & $Z_5$ & $Z_3$ & accidental $U(1)$ broken \\
\midrule
$s_\Phi$   & \YesMark & \YesMark & \YesMark & \NoMark  & \YesMark & $U(1)_A,\ U(1)_C$ \\
$s_\xi$    & \YesMark & \YesMark & \NoMark  & \YesMark & \YesMark & $U(1)_A$ \\
$s_\psi$   & \YesMark & \NoMark  & \YesMark & \NoMark  & \YesMark & $U(1)_C,\ U(1)_D$ \\
$s_\chi$   & \YesMark & \NoMark  & \YesMark & \NoMark  & \YesMark & $U(1)_B,\ U(1)_C$ \\
$s_\eta$   & \YesMark & \NoMark  & \NoMark  & \YesMark & \YesMark & $U(1)_B$ \\
$s_\rho$   & \NoMark  & \YesMark & \NoMark  & \NoMark  & \YesMark & $U(1)_C$ \\
$s_\kappa$ & \YesMark & \NoMark  & \NoMark  & \NoMark  & \YesMark & $U(1)_D$ \\
\bottomrule
\end{tabularx}
\end{table}

For the triplets, $(\Phi\Phi)_1$ and $(\xi\xi)_1$ are $A_4$ singlets. The exceptional $A_4$ breaking comes from $\rho$: since $\rho\sim1''$, one has $\rho^2\sim1'$ rather than $1$. The cyclic entries follow from the condition $2q_X=0\pmod N$ for a phase-sensitive quadratic. Collectively the seven terms explicitly break all four accidental continuous rephasings and give positive curvature to the remaining three phase-flat directions. The residual discrete symmetry left by these operators is relevant for the domain-wall question and will be discussed separately in Section~\ref{sec:domain_walls}.

\subsection{Vacuum and scalar spectrum after soft breaking}
\label{subsec:soft_scalar_spectrum}

Equation~\eqref{eq:soft_phaseonly_main} shows that the new terms vanish on the real M--R vacuum together with all of their first derivatives,
\begin{equation}
\left.\Delta V_{\rm soft}\right|_0=0,
\qquad
\left.\frac{\partial\Delta V_{\rm soft}}{\partial\varphi_a}\right|_0=0,
\label{eq:soft_tadpoles_zero_main}
\end{equation}
for every real or imaginary scalar fluctuation $\varphi_a$. Hence the original tadpole equations are unchanged and the VEVs do not shift. Moreover,
\begin{equation}
\Delta M_{\rm even}^2=0,
\label{eq:soft_even_zero_main}
\end{equation}
whereas the CP-odd interaction-basis correction is
\begin{equation}
\Delta M_{\rm odd}^2=
{\rm diag}\left(0,
2s_\Phi,2s_\Phi,2s_\Phi,
2s_\xi,2s_\xi,2s_\xi,
2s_\psi,2s_\chi,2s_\eta,2s_\rho,2s_\kappa\right).
\label{eq:soft_odd_shift_main}
\end{equation}
The first entry is the SM neutral Goldstone and remains exactly massless. The seven physical zero modes acquire
\begin{equation}
\begin{aligned}
m_{\Phi_{I0}}^2&=2s_\Phi,\qquad
m_{\xi_{I0}}^2=2s_\xi,\qquad
m_{\psi_I}^2=2s_\psi,\qquad
m_{\chi_I}^2=2s_\chi,\\
m_{\eta_I}^2&=2s_\eta,\qquad
m_{\rho_I}^2=2s_\rho,\qquad
m_{\kappa_I}^2=2s_\kappa.
\end{aligned}
\label{eq:soft_seven_masses_main}
\end{equation}
Thus all seven are lifted for positive $s_X$.

We introduce the soft terms as a controlled way to lift the exactly massless CP-odd states. Whether the massless states are phenomenologically excluded depends on their couplings and production history, so no universal lower mass bound is imposed here. Accordingly, the coefficients $s_X$ are kept independent and positive, without introducing an arbitrary common pseudoscalar mass scale.

The same diagonal shifts act on the two massive transverse CP-odd blocks. With
\begin{equation}
A_s=A+2s_\Phi,
\qquad
B_s=B+2s_\xi,
\qquad
D_s=D,
\label{eq:soft_ABD_main}
\end{equation}
the four transverse masses become
\begin{align}
m_{1\pm,s}^2&=\frac12\left[A_s+B_s\pm
\sqrt{(A_s-B_s)^2+4D^2}\right],\notag\\
m_{2\pm,s}^2&=\frac12\left[A_s+B_s\pm
\sqrt{(A_s-B_s)^2+36D^2}\right].
\label{eq:soft_transverse_masses_main}
\end{align}
The detailed Hessian reduction is collected in Appendix~\ref{app:soft_scalar_details}.

The quartic potential is unchanged, so the BFB conditions of Section~\ref{sec:vacuum_scalar} are unaffected. For $s_X\ge0$, $\Delta V_{\rm soft}$ is positive semidefinite and vanishes at the chosen real M--R vacuum. Thus a point that was already a global minimum in the sufficient no-soft region remains a global minimum after this deformation, although uniqueness among symmetry-related real vacua does not follow from these seven terms alone.

\subsection{Consequences for FCNC, CLFV and non-unitarity}
\label{subsec:soft_flavour}

The stationarity-preserving deformation does not change the fermion operators or the VEVs entering them. At the tree level considered here,
\begin{equation}
\Delta M_\ell=0,
\qquad
\Delta M_D=0,
\qquad
\Delta M_R=0,
\qquad
\Delta M_w=0.
\label{eq:soft_fermion_matrices_main}
\end{equation}
Consequently,
\begin{equation}
\Delta\Theta=0,
\qquad
\Delta\widetilde\eta=0,
\qquad
\Delta B_{\alpha I}=0,
\qquad
\Delta C_{IJ}=0.
\label{eq:soft_theta_eta_main}
\end{equation}
The gauge/HNL CLFV amplitudes and the projections on $P,Q,R$ derived in Section~\ref{sec:phenomenological_constraints}, in particular Subsections~\ref{subsec:CLFV}--\ref{subsec:nonunitarity}, are therefore unchanged. In particular,
\begin{equation}
\begin{array}{c|cc}
&{\rm NO}&{\rm IO}\\ \hline
|P|_{\min}&0.319~{\rm eV}&1.30~{\rm eV}\\
|Q|_{\min}^{\rm proj}&0.875~{\rm eV}&1.268~{\rm eV}\\
|R|_{\min}&2.21~{\rm eV}&3.20~{\rm eV}
\end{array}
\label{eq:soft_PQR_unchanged_main}
\end{equation}
remain the separate no-soft lower projections. No newly lifted zero-mode direction introduces a new leading charged-lepton FCNC Yukawa tensor at this order.

The scalar-mediated FCNC and CLFV amplitudes do change because the flavour-active CP-odd propagators are shifted. The charged-lepton Yukawa matrices and the CP-even projected propagator $g_E$ remain unchanged, whereas
\begin{equation}
X_O\longrightarrow X_O^s=X_O+2s_\Phi,
\qquad
Y_O\longrightarrow Y_O^s=Y_O+2s_\xi,
\qquad
Z_O\longrightarrow Z_O.
\label{eq:soft_XYZ_main}
\end{equation}
Thus
\begin{equation}
g_{O1}^s=\frac{Y_O^s}{X_O^sY_O^s-Z_O^2},
\qquad
g_{O2}^s=\frac{Y_O^s}{X_O^sY_O^s-9Z_O^2},
\label{eq:soft_gO_main}
\end{equation}
and we define
\begin{equation}
\delta g_s=g_{O1}^s-g_{O2}^s,
\qquad
F_-^s=2g_E-g_{O1}^s-g_{O2}^s,
\qquad
F_+^s=2g_E+g_{O1}^s+g_{O2}^s.
\label{eq:soft_Fpm_main}
\end{equation}
Equivalently,
\begin{equation}
\delta g_s=-\frac{8Y_O^sZ_O^2}
{(X_O^sY_O^s-Z_O^2)(X_O^sY_O^s-9Z_O^2)}.
\label{eq:soft_dgs_main}
\end{equation}
For positive $s_\Phi,s_\xi$ the projected CP-odd propagators and their sum decrease, $|\delta g_s|$ decreases, $F_+^s$ decreases, and $F_-^s$ increases algebraically. The scalar three-body rates are obtained from the no-soft expressions by $(\delta g,F_-,F_+)\to(\delta g_s,F_-^s,F_+^s)$. The complete scalar $\tau^-\to\mu^-\mu^-e^+$ rate decreases for positive $s_\Phi,s_\xi$, while the scalar $\tau^-\to e^-e^-\mu^+$ rate is not universally monotonic because the correlated $F_-^s$ and $F_+^s$ terms compete.

Using the same experimental inputs and the scalar--HNL interference treatment described above, the constraints obtained without fixing additional benchmark values become
\begin{equation}
\frac{|\delta g_s|}{v_\Phi^2}<5.6903\times10^{-7}\ {\rm GeV}^{-4},
\label{eq:soft_dgs_bound_main}
\end{equation}
\begin{equation}
0<\frac{F_+^s}{v_\Phi^2}<5.9052\times10^{-7}\ {\rm GeV}^{-4},
\qquad
\frac{|F_-^s|}{v_\Phi^2}<5.9052\times10^{-7}\ {\rm GeV}^{-4}.
\label{eq:soft_Fpm_bound_main}
\end{equation}
For $\tau^-\to e^-e^-\mu^+$ the scalar--Majorana-box interference vanishes in the retained massless-daughter approximation, and the HNL condition remains
\begin{equation}
|F_{\rm Box}^{\tau ee\mu}|<0.2601.
\label{eq:soft_box_bound_main}
\end{equation}
The five singlet soft coefficients lift their corresponding zero modes but do not enter the leading transverse charged-lepton propagators in Eq.~\eqref{eq:soft_gO_main}.

The CP-even scalar dipole cancellation found in the no-soft analysis is unchanged. The CP-odd dipole remains proportional to the same $Z_O^2$ breaking structure, with the no-soft spectral weights replaced by the shifted ones. For $\tau\to\mu\gamma$ the physical amplitude is still the coherent sum
\begin{equation}
A_D^{\rm tot,s}=A_D^{S,s}+A_D^N,
\label{eq:soft_dipole_total_main}
\end{equation}
where the HNL term is unchanged. No model-independent lower bound on an individual scalar mass or soft coefficient follows from this dipole because it depends on a difference of spectral loop weights. Further details are given in Appendix~\ref{app:soft_phenomenology}. The discussion in this subsection is at tree level in the dimension-two soft terms. Loop corrections induced by the soft breaking are not included.

\subsection{Conventional Type-I leptogenesis}
\label{subsec:soft_leptogenesis}

The restricted leptogenesis discussion of Subsection~\ref{subsec:leptogenesis} depends only on the conventional two-body Yukawa matrix ${\cal Y}$ and the RH Majorana matrix $M_R$. Both are unchanged by Eq.~\eqref{eq:soft_fermion_matrices_main}. Hence the matrix ${\cal H}={\cal Y}^\dagger{\cal Y}$ retains the heavy-basis structure derived in Section~\ref{subsec:leptogenesis}, and
\begin{equation}
\operatorname{Im}\!\left[({\cal Y}^\dagger{\cal Y})_{ij}^{\,2}\right]=0,
\qquad
\epsilon_i^{I}=0
\label{eq:soft_lepto_zero_main}
\end{equation}
continues to hold at the same leading order. Thus the soft deformation required to lift the physical CP-odd zero modes does not alter the conventional Type-I conclusion. Independent Weinberg-like, scalar-assisted, and full flavour-density leptogenesis sources remain outside the restricted leptogenesis analysis of the present manuscript.

\section{Domain walls and the residual discrete vacuum degeneracy}
\label{sec:domain_walls}

The spontaneous breaking of an exact discrete symmetry can produce disconnected degenerate vacua and hence cosmological domain walls if the transition takes place after inflation \cite{r92,r93,r94}. This question is relevant here because the flavour construction contains several discrete symmetries. However, the number of physically distinct vacua cannot be obtained by simply multiplying the orders of all the groups: before soft breaking, four accidental continuous rephasings connect many phase-related configurations.

After the seven phase-sensitive soft terms of Section~\ref{sec:soft_breaking} are included, the exact discrete group acting on the scalar vacuum is
\begin{equation}
G_{\rm DW}=V_4\times (Z_2)^4,
\label{eq:GDW_main}
\end{equation}
where $V_4\simeq Z_2\times Z_2$ is the Klein subgroup of $A_4$. The chosen real M--R vacuum has a trivial stabiliser under this group. Its exact symmetry orbit therefore contains
\begin{equation}
|G_{\rm DW}|=4\times16=64
\label{eq:DW_orbit_main}
\end{equation}
real symmetry-related vacua. The seven soft terms that lift the physical CP-odd zero modes do not split these vacua, because they vanish on every all-real member of the orbit. The derivation of the residual group and the orbit is given in Appendix~\ref{app:domain_wall_details}.

A small explicit bias between the degenerate vacua is the standard way of removing a stable wall network \cite{r95,r96}; similar issues have also been discussed in discrete flavour models \cite{r97}. In the present case one may choose a positive-semidefinite quadratic bias of the form
\begin{equation}
\Delta V_{\rm DW}=\sum_i\mu_i^2 F_i^\dagger F_i,
\qquad
F_i(v_{\rm MR})=0,
\label{eq:DW_bias_schematic_main}
\end{equation}
so that
\begin{equation}
\Delta V_{\rm DW}(v_{\rm MR})=0,
\qquad
\left.\nabla\Delta V_{\rm DW}\right|_{v_{\rm MR}}=0.
\label{eq:DW_stationary_main}
\end{equation}
The desired vacuum therefore remains stationary, while the symmetry-related vacua can be lifted. Since the bias is quadratic, the quartic boundedness conditions are unchanged. Since it is non-negative and vanishes at the chosen vacuum, the sufficient global-minimum result obtained above is also retained wherever it applied. The leading tree-level triadic texture is unchanged. The CP-even scalar masses, on the other hand, can receive shifts from this additional bias; we do not assume otherwise.

The other familiar possibility is that the discrete flavour symmetry is broken before or during inflation and is not restored afterwards. A quantitative calculation of the wall tension, annihilation temperature, big-bang nucleosynthesis (BBN) limit or gravitational-wave signal would require the finite-temperature multi-field potential and is beyond the purpose of the present work. Here our aim is only to show that the domain-wall problem is identifiable and that the exact residual degeneracy can be removed without moving the desired vacuum.

\section{Summary and Discussion}
\label{sec:summary_discussion}

The main results of the present work may be summarised as follows:
\begin{enumerate}
\item We have proposed a minimal texture for Majorana neutrinos, referred to as the triadic texture, characterised by two independent correlation conditions among the neutrino mass matrix elements, namely $(M_\nu)_{22}=-2(M_\nu)_{13}$ and $(M_\nu)_{12}=(M_\nu)_{13}$.

\item The texture leads to several non-trivial correlations among the neutrino observables. With the present correlated oscillation likelihood, both NO and IO solutions are obtained, although the NO solution gives the lower minimum in the present fit. The two orderings occupy clearly different regions of the $\theta_{23}$--$\delta_{\rm CP}$ plane. The stability of these regions has also been checked by controlled $\pm1\sigma$ and $\pm2\sigma$ translations of the atmospheric likelihood. The neutrino masses, Majorana phases and $m_{\beta\beta}$ remain correlated with the oscillation parameters, and the recent DESI bounds on $\Sigma_\nu$ therefore provide an important test of the texture \cite{r88}.

\item A flavour-model realisation of the texture is constructed using two distinct vacuum-alignment configurations in two different bases of the non-Abelian group $A_4$, namely the Altarelli--Feruglio and Ma--Rajasekaran bases. Both constructions reproduce the same leading triadic neutrino mass structure. In the M--R basis the required charged-lepton permutation is included before the neutrino mass matrix is compared with the physical flavour ordering; the Majorana matrix transforms by the transpose congruence $M_\nu\to U_L^{\prime T}M_\nu U'_L$.

\item The scalar sector is analysed with the complete renormalisable potential without any \textit{ad hoc} assumptions on the quartic couplings. The A--F alignment contains a physical CP-even zero mode in the tree-level Hessian. The M--R alignment, on the other hand, admits a finite region with a strictly positive physical CP-even spectrum. The CP-even and CP-odd spectra, boundedness-from-below conditions and sufficient conditions for the chosen M--R vacuum to be a global minimum are obtained explicitly.

\item The M--R scalar spectrum contains seven physical CP-odd zero modes in addition to the electroweak Goldstone boson. Four follow from accidental continuous rephasings, while the remaining three are phase-flat directions. A set of dimension-two soft terms lifts all seven physical zero modes. These terms leave the chosen real M--R vacuum stationary and do not change the CP-even Hessian or the leading fermion mass matrices at tree level. They therefore remove the unwanted massless physical scalars without undoing the leading texture construction.

\item The resulting Yukawa structure and scalar interactions restrict the allowed charged lepton flavour violating channels. The heavy-neutrino sector gives a complementary set of loop contributions, and the same active--heavy mixing induces non-unitarity of the effective PMNS matrix. Present experimental bounds constrain combinations of the scalar propagators and the Dirac entries $P,Q,R$. After the above soft terms are added, the gauge/heavy-neutrino contribution and the non-unitarity matrix are unchanged at the retained tree-level order, whereas the scalar amplitudes contain the shifted CP-odd propagators.

\item Conventional Type-I leptogenesis has been considered. In the positive-mass heavy-neutrino basis, the CP-odd invariants entering both the unflavoured and flavoured two-body asymmetries vanish at leading order. Hence the vanishing of the conventional Type-I source follows directly from the Yukawa and RH-neutrino structure used in the model. The independent Weinberg-like contribution, scalar-assisted decay channels and a full flavour-density treatment are separate extensions and are not included in the restricted leptogenesis discussion of the present manuscript.

\item The spontaneous breaking of the remaining exact discrete symmetry also leads to a domain-wall question. After the seven pseudoscalar soft terms are included, the scalar vacuum is acted on by $V_4\times(Z_2)^4$ and the chosen real M--R vacuum has a $64$-element exact symmetry orbit. These seven soft terms do not split the all-real orbit. We have therefore shown that an additional stationary positive-semidefinite quadratic bias can lift the residual degeneracy without shifting the desired vacuum or the leading tree-level texture. The usual pre-inflationary breaking possibility remains an alternative \cite{r92,r93,r94,r95,r96,r97}.
\end{enumerate}

Beyond the results summarised above, several comments are in order. The triadic texture proposed in this work is phenomenologically interesting because it is economical at the level of the neutrino mass matrix and, at the same time, correlates several quantities that will continue to be tested by oscillation, cosmology and neutrinoless double beta decay experiments. The explicit flavour realisation, however, should not be confused with a minimal ultraviolet theory. It contains several flavon fields and shaping symmetries whose role is to organise the effective operators. Our purpose here is to show that the texture can be embedded in a consistent flavour construction and that the same construction can be tested through its vacuum structure and low-energy phenomenology.

Although the present work realises the triadic texture through an effective Type-I seesaw plus a Weinberg-like dimension-six operator framework, the same mass structure may also be generated through other mechanisms, such as a Type-I plus Type-II hybrid seesaw framework. A Type-II realisation would introduce additional electroweak scalar multiplets with charged components, and the corresponding vacuum and scalar-spectrum analysis would be enlarged considerably. We leave such alternatives for future study.

The use of the non-Abelian group $A_4$ is motivated by its ability to organise fields into singlet and triplet representations and to produce correlated lepton-mass structures through suitable vacuum alignments. The comparison between the A--F and M--R realisations is useful in this respect. The same leading neutrino texture may be obtained in the two bases, while the scalar potential provides an independent test of the corresponding spectra and vacuum stability. The auxiliary cyclic symmetries play a different role: they organise the allowed effective operators and suppress unwanted lower-order terms. Since the construction is effective, the expansion is understood in the regime $|v_X|/\Lambda<1$ discussed in Section~\ref{subsec:model_setup_eft}; higher-dimensional operators are then suppressed by additional powers of the flavon-to-cutoff ratios.

A complete renormalisation-group analysis is beyond the scope of the present work. Once explicit soft breaking is present, radiative communication of the breaking and threshold corrections can in principle modify the scalar and effective-operator parameters. Such effects should be studied by evolving the complete set of couplings between the flavour scale, the heavy-neutrino thresholds and the electroweak scale.

The conventional Type-I leptogenesis conclusion also has a definite scope. What vanishes is the leading two-body Type-I source generated by the RH Majorana matrix and Yukawa couplings used here. This does not imply that every possible baryogenesis source in the complete effective theory vanishes. The independent Weinberg-like operator, physical-flavon channels and flavour effects require a separate analysis.

In neutrino phenomenology, the search for viable mass textures will remain important until the structure of the neutrino mass matrix is fixed unambiguously by experiment. The present work shows, in particular, that reproducing a texture through a flavour symmetry is not by itself sufficient. The scalar potential, the vacuum alignment and the resulting phenomenology have to be considered together. In the triadic construction studied here, this combined analysis identifies the M--R realisation as the branch for which a strictly positive physical CP-even spectrum is explicitly established, and it gives several independent ways in which the model can be tested.

\appendix

\section{Representations of $A_4$ in Different Bases}
\label{app:A4_rules}

All elements of $A_4$ can be written in terms of the generators $S$ and $T$, satisfying
\begin{equation}
S^2=T^3=(ST)^3=e.
\end{equation}
The two bases used in Section~\ref{sec:symmetry_realisation} are fixed by which generator is diagonal.

\subsection{Altarelli--Feruglio basis}

For the triplet representation,
\begin{equation}
S_1=\frac13\begin{pmatrix}-1&2&2\\2&-1&2\\2&2&-1\end{pmatrix},
\qquad
T_1=\begin{pmatrix}1&0&0\\0&\omega^2&0\\0&0&\omega\end{pmatrix}.
\end{equation}
For two triplets $a=(a_1,a_2,a_3)^T$ and $b=(b_1,b_2,b_3)^T$,
\begin{align}
(ab)_1&=a_1b_1+a_2b_3+a_3b_2,\notag\\
(ab)_{1'}&=a_3b_3+a_1b_2+a_2b_1,\notag\\
(ab)_{1''}&=a_2b_2+a_1b_3+a_3b_1,\notag\\
(ab)_{3_s}&=\frac13\begin{pmatrix}
2a_1b_1-a_2b_3-a_3b_2\\
2a_3b_3-a_1b_2-a_2b_1\\
2a_2b_2-a_1b_3-a_3b_1
\end{pmatrix},\notag\\
(ab)_{3_a}&=\frac12\begin{pmatrix}
a_2b_3-a_3b_2\\
a_1b_2-a_2b_1\\
a_1b_3-a_3b_1
\end{pmatrix}.
\end{align}

\subsection{Ma--Rajasekaran basis}

The triplet generators are
\begin{equation}
S_2=\begin{pmatrix}1&0&0\\0&-1&0\\0&0&-1\end{pmatrix},
\qquad
T_2=\begin{pmatrix}0&0&1\\1&0&0\\0&1&0\end{pmatrix}.
\end{equation}
The triplet products are
\begin{align}
(ab)_1&=a_1b_1+a_2b_2+a_3b_3,\notag\\
(ab)_{1'}&=a_1b_1+\omega a_2b_2+\omega^2a_3b_3,\notag\\
(ab)_{1''}&=a_1b_1+\omega^2 a_2b_2+\omega a_3b_3,\notag\\
(ab)_{3_s}&=\begin{pmatrix}
a_2b_3+a_3b_2\\
a_3b_1+a_1b_3\\
a_1b_2+a_2b_1
\end{pmatrix},\notag\\
(ab)_{3_a}&=\begin{pmatrix}
a_2b_3-a_3b_2\\
a_3b_1-a_1b_3\\
a_1b_2-a_2b_1
\end{pmatrix}.
\end{align}
The two triplet bases are related by
\begin{equation}
U_\omega=\frac1{\sqrt3}
\begin{pmatrix}
1&1&1\\
1&\omega&\omega^2\\
1&\omega^2&\omega
\end{pmatrix}.
\end{equation}

\section{Neutral-scalar Hessian checks, spectra and diagonalising transformations}
\label{app:complete_scalar_diagonalisation}

This appendix gives the details of the neutral-scalar analysis discussed in Section~\ref{sec:vacuum_scalar}. No additional restriction on the scalar couplings is imposed here. The interaction and mass bases are defined before the corresponding transformations are written, and the larger matrices are given in block or factorised form.

\subsection{A--F CP-even zero mode}
\label{app:AF_CPeven_flat}

In the interaction basis of Eq.~\eqref{eq:AF_even_basis}, impose the tadpole solutions in Eqs.~\eqref{eq:AF_residual_tadpoles}--\eqref{eq:AF_mu2_solutions} and split the basis as the first seven fields $(h,\phi_{R1},\phi_{R2},\phi_{R3},\xi_{R1},\xi_{R2},\xi_{R3})$ plus the five singlet radial fields. The complete $12\times12$ CP-even Hessian is then
\begin{equation}
(M_{\rm even}^2)_{\rm AF}=\begin{pmatrix}A_{\rm AF}&Y_{\rm AF}\\Y_{\rm AF}^T&S_{\rm AF}\end{pmatrix},
\label{eq:app_AF_hessian_block}
\end{equation}
with
\begin{equation}
A_{\rm AF}=\begin{pmatrix}
\alpha_H&p_{H\Phi}&0&0&0&-p_{H\xi}&-p_{H\xi}\\
p_{H\Phi}&\alpha_\Phi&0&0&0&u&u\\
0&0&K_{\Phi2}&C&D_1&0&0\\
0&0&C&B&D_2&0&0\\
0&0&D_1&D_2&E&0&0\\
-p_{H\xi}&u&0&0&0&F&F\\
-p_{H\xi}&u&0&0&0&F&F
\end{pmatrix},\qquad
Y_{\rm AF}=\begin{pmatrix}
\mathbf p_H^T\\ \mathbf p_\Phi^T\\ \mathbf0^T\\ \mathbf0^T\\ \mathbf0^T\\-\mathbf p_\xi^T\\-\mathbf p_\xi^T
\end{pmatrix}.
\label{eq:app_AF_hessian_explicit}
\end{equation}
The entries are
\begingroup
\allowdisplaybreaks[4]
\begin{align}
\alpha_H&=2\lambda_Hv_H^2,\qquad
p_{H\Phi}=\lambda_{H\Phi}v_Hv_\Phi,\qquad
p_{H\xi}=\lambda_{H\xi}v_Hv_\xi,\notag\\
\alpha_\Phi&=\frac{2}{9}(9\lambda_{\Phi1}+4\lambda_{\Phi3})v_\Phi^2,\qquad
F=\frac{2}{9}(9\lambda_{\xi1}+4\lambda_{\xi3})v_\xi^2,\notag\\
u&=-\frac{v_\Phi v_\xi}{12}
\begin{aligned}[t]
\Bigl[{}&12\lambda_{\Phi\xi1}-3\lambda_{\Phi\xi12}+2\lambda_{\Phi\xi14}+6\lambda_{\Phi\xi2}-4\lambda_{\Phi\xi4}+18\lambda_{\Phi\xi9}\Bigr],
\end{aligned}\notag\\
K_{\Phi2}&=\frac{v_\xi^2}{24}
\begin{aligned}[t]
\Bigl[{}&-3\lambda_{\Phi\xi12}+2\lambda_{\Phi\xi14}+18\lambda_{\Phi\xi2}+4\lambda_{\Phi\xi4}+12\lambda_{\Phi\xi8}+6\lambda_{\Phi\xi9}\Bigr],
\end{aligned}\notag\\
B&=\frac{v_\xi^2}{24}
\begin{aligned}[t]
\Bigl[{}&-3\lambda_{\Phi\xi12}-6\lambda_{\Phi\xi14}+18\lambda_{\Phi\xi2}+4\lambda_{\Phi\xi4}+12\lambda_{\Phi\xi8}+6\lambda_{\Phi\xi9}\Bigr],
\end{aligned}\notag\\
C&=\frac1{72}
\begin{aligned}[t]
\Bigl[{}&(72\lambda_{\Phi2}-32\lambda_{\Phi3})v_\Phi^2+\bigl(27\lambda_{\Phi\xi12}-42\lambda_{\Phi\xi14}+54\lambda_{\Phi\xi2}+12\lambda_{\Phi\xi4}\phantom{+\bigl(}{}\\
&+36\lambda_{\Phi\xi8}+18\lambda_{\Phi\xi9}\bigr)v_\xi^2\Bigr],
\end{aligned}\notag\\
D_1&=-\frac{v_\Phi v_\xi}{24}
\begin{aligned}[t]
\Bigl[{}&3\lambda_{\Phi\xi12}-2\lambda_{\Phi\xi14}+18\lambda_{\Phi\xi2}+4\lambda_{\Phi\xi4}+12\lambda_{\Phi\xi8}+6\lambda_{\Phi\xi9}\Bigr],
\end{aligned}\notag\\
D_2&=-\frac{v_\Phi v_\xi}{24}
\begin{aligned}[t]
\Bigl[{}&3\lambda_{\Phi\xi12}+6\lambda_{\Phi\xi14}+18\lambda_{\Phi\xi2}+4\lambda_{\Phi\xi4}+12\lambda_{\Phi\xi8}+6\lambda_{\Phi\xi9}\Bigr],
\end{aligned}\notag\\
E&=\frac{v_\Phi^2}{24}
\begin{aligned}[t]
\Bigl[{}&3\lambda_{\Phi\xi12}-10\lambda_{\Phi\xi14}+18\lambda_{\Phi\xi2}+4\lambda_{\Phi\xi4}+12\lambda_{\Phi\xi8}+6\lambda_{\Phi\xi9}\Bigr].
\end{aligned}
\label{eq:app_AF_hessian_entries}
\end{align}
\endgroup
For $s,t\in\mathcal S$, the five-component vectors and singlet block are
\begin{equation}
(\mathbf p_H)_s=\lambda_{Hs}v_Hv_s,\qquad
(\mathbf p_\Phi)_s=\lambda_{\Phi s}v_\Phi v_s,\qquad
(\mathbf p_\xi)_s=\lambda_{\xi s}v_\xi v_s,
\end{equation}
\begin{equation}
(S_{\rm AF})_{ss}=2\lambda_sv_s^2,\qquad
(S_{\rm AF})_{st}=\lambda_{st}v_sv_t\quad(s\ne t).
\label{eq:app_AF_singlet_block}
\end{equation}
Equations~\eqref{eq:app_AF_hessian_block}--\eqref{eq:app_AF_singlet_block} display every entry of the Hessian. In particular, its sixth and seventh rows (and columns) are identical. Therefore, for generic values of the remaining quartic couplings,
\begin{equation}
\operatorname{rank}\!\left[(M_{\rm even}^2)_{\rm AF}\right]=11,
\qquad
(M_{\rm even}^2)_{\rm AF}\,u_{\rm AF}^{(0)}=0,
\label{eq:app_AF_rank_null}
\end{equation}
with
\begin{equation}
u_{\rm AF}^{(0)}=(0,0,0,0,0,-1/\sqrt2,1/\sqrt2,0,0,0,0,0)^T.
\label{eq:app_AF_nullvector}
\end{equation}
Thus $S_{\rm AF}^{(0)}=(\xi_{R3}-\xi_{R2})/\sqrt2$ has $m^2=0$, confirming explicitly the physical tree-level CP-even zero mode quoted in Sub-subsection~\ref{subsubsec:AF_CPeven}.

\subsection{M--R CP-even sector}
\label{app:CPeven_complete}

\subsubsection{Radial basis, mass matrix and the meaning of $r_{ai}$}

The eight-dimensional radial interaction basis is denoted by
\begin{equation}
\mathbf R=(R_1,\ldots,R_8)^T
=(h,\Phi_R,\xi_R,\psi_R,\chi_R,\eta_R,\rho_R,\kappa_R)^T.
\label{eq:app_radial_basis_defined}
\end{equation}
The corresponding radial mass basis is
\begin{equation}
\mathbf H_R=(H_{R1},\ldots,H_{R8})^T,
\qquad
\mathbf R=O_R\mathbf H_R.
\label{eq:app_radial_transform_defined}
\end{equation}
We write
\begin{equation}
O_R=(r_{ai}),\qquad a=1,\ldots,8,\quad i=1,\ldots,8,
\label{eq:app_rai_definition}
\end{equation}
so that the symbol $r_{ai}$ has the unambiguous meaning
\begin{equation}
R_a=\sum_{i=1}^{8}r_{ai}H_{Ri}.
\label{eq:app_rai_component_meaning}
\end{equation}
Equivalently, the $i$th column
$\mathbf u_i=(r_{1i},\ldots,r_{8i})^T$ is the normalised eigenvector belonging to
$m_{Ri}^2$:
\begin{equation}
(M_R^2-m_{Ri}^2 I_8)\mathbf u_i=0,
\qquad
\mathbf u_i^T\mathbf u_j=\delta_{ij}.
\label{eq:app_OR_definition}
\end{equation}
Therefore
\begin{equation}
O_R^TO_R=I_8,
\qquad
O_R^TM_R^2O_R=\operatorname{diag}(m_{R1}^2,\ldots,m_{R8}^2).
\label{eq:app_OR_check}
\end{equation}

The same $8\times8$ matrix used in the main text can be written as three readable blocks,
\begin{equation}
M_R^2=
\begin{pmatrix}
M_{3R}^2&X_R\\
X_R^T&M_S^2
\end{pmatrix},
\label{eq:app_MR_block_form}
\end{equation}
where
\begin{equation}
M_{3R}^2=
\begin{pmatrix}
2\lambda_Hv_H^2&\sqrt3\lambda_{H\Phi}v_Hv_\Phi&\sqrt3\lambda_{H\xi}v_Hv_\xi\\
\sqrt3\lambda_{H\Phi}v_Hv_\Phi&2A_\Phi v_\Phi^2&C_Rv_\Phi v_\xi\\
\sqrt3\lambda_{H\xi}v_Hv_\xi&C_Rv_\Phi v_\xi&2A_\xi v_\xi^2
\end{pmatrix},
\label{eq:app_M3R_readable}
\end{equation}
with $A_\Phi$, $A_\xi$ and $C_R$ defined in
Eqs.~\eqref{eq:Aphi_Axi_main} and \eqref{eq:CR_CT_main}. The $3\times5$ portal block is
\begin{equation}
X_R=
\begin{pmatrix}
\lambda_{H\psi}v_Hv_\psi&\lambda_{H\chi}v_Hv_\chi&\lambda_{H\eta}v_Hv_\eta&\lambda_{H\rho}v_Hv_\rho&\lambda_{H\kappa}v_Hv_\kappa\\
\sqrt3\lambda_{\Phi\psi}v_\Phi v_\psi&\sqrt3\lambda_{\Phi\chi}v_\Phi v_\chi&\sqrt3\lambda_{\Phi\eta}v_\Phi v_\eta&\sqrt3\lambda_{\Phi\rho}v_\Phi v_\rho&\sqrt3\lambda_{\Phi\kappa}v_\Phi v_\kappa\\
\sqrt3\lambda_{\xi\psi}v_\xi v_\psi&\sqrt3\lambda_{\xi\chi}v_\xi v_\chi&\sqrt3\lambda_{\xi\eta}v_\xi v_\eta&\sqrt3\lambda_{\xi\rho}v_\xi v_\rho&\sqrt3\lambda_{\xi\kappa}v_\xi v_\kappa
\end{pmatrix},
\label{eq:app_XR_readable}
\end{equation}
and, in the singlet ordering $(\psi_R,\chi_R,\eta_R,\rho_R,\kappa_R)$,
\begin{equation}
M_S^2=
\begin{pmatrix}
2\lambda_\psi v_\psi^2&\lambda_{\psi\chi}v_\psi v_\chi&\lambda_{\psi\eta}v_\psi v_\eta&\lambda_{\psi\rho}v_\psi v_\rho&\lambda_{\psi\kappa}v_\psi v_\kappa\\
\lambda_{\psi\chi}v_\psi v_\chi&2\lambda_\chi v_\chi^2&\lambda_{\chi\eta}v_\chi v_\eta&\lambda_{\chi\rho}v_\chi v_\rho&\lambda_{\chi\kappa}v_\chi v_\kappa\\
\lambda_{\psi\eta}v_\psi v_\eta&\lambda_{\chi\eta}v_\chi v_\eta&2\lambda_\eta v_\eta^2&\lambda_{\eta\rho}v_\eta v_\rho&\lambda_{\eta\kappa}v_\eta v_\kappa\\
\lambda_{\psi\rho}v_\psi v_\rho&\lambda_{\chi\rho}v_\chi v_\rho&\lambda_{\eta\rho}v_\eta v_\rho&2\lambda_\rho v_\rho^2&\lambda_{\rho\kappa}v_\rho v_\kappa\\
\lambda_{\psi\kappa}v_\psi v_\kappa&\lambda_{\chi\kappa}v_\chi v_\kappa&\lambda_{\eta\kappa}v_\eta v_\kappa&\lambda_{\rho\kappa}v_\rho v_\kappa&2\lambda_\kappa v_\kappa^2
\end{pmatrix}.
\label{eq:app_MS_readable}
\end{equation}
The eight radial masses are therefore the eight real roots of
\begin{equation}
\det(M_R^2-m_{Ri}^2I_8)=0,
\qquad i=1,\ldots,8.
\label{eq:app_radial_roots}
\end{equation}
No special block-diagonalisation of this general radial matrix is assumed in the present
manuscript.

\subsubsection{Transverse basis and analytic matrix $O_T$}

The four CP-even transverse fields are ordered as
\begin{equation}
\mathbf T_R=(\Phi_{T1},\Phi_{T2},\xi_{T1},\xi_{T2})^T,
\label{eq:app_transverse_basis}
\end{equation}
where the relation to $(\phi_{R1},\phi_{R2},\phi_{R3})$ and
$(\xi_{R1},\xi_{R2},\xi_{R3})$ is fixed by the matrices $U_\Phi$ and $U_\xi$ in
Eq.~\eqref{eq:UPhi_Uxi_main}. The transverse mass matrix is Eq.~\eqref{eq:MT_even_main}.
Its diagonalising angles are defined by Eqs.~\eqref{eq:gamma_even_main} and
\eqref{eq:beta_even_main}. In this appendix we use only the abbreviations
\begin{equation}
c_\beta\equiv\cos\beta,\quad s_\beta\equiv\sin\beta,
\qquad
c_\gamma\equiv\cos\gamma,\quad s_\gamma\equiv\sin\gamma.
\label{eq:app_beta_gamma_abbrev}
\end{equation}
With the transverse mass basis
\begin{equation}
\mathbf T_m=(T_{+1},T_{+2},T_{-1},T_{-2})^T,
\qquad
\mathbf T_R=O_T\mathbf T_m,
\end{equation}
we have
\begin{equation}
O_T=
\begin{pmatrix}
c_\beta c_\gamma&-c_\beta s_\gamma&-s_\beta c_\gamma&-s_\beta s_\gamma\\
c_\beta s_\gamma& c_\beta c_\gamma&-s_\beta s_\gamma& s_\beta c_\gamma\\
s_\beta c_\gamma& s_\beta s_\gamma& c_\beta c_\gamma&-c_\beta s_\gamma\\
s_\beta s_\gamma&-s_\beta c_\gamma& c_\beta s_\gamma& c_\beta c_\gamma
\end{pmatrix},
\label{eq:app_OT_explicit}
\end{equation}
and
\begin{equation}
O_T^TM_T^2O_T=
\operatorname{diag}(m_{T+}^2,m_{T+}^2,m_{T-}^2,m_{T-}^2).
\label{eq:app_OT_check}
\end{equation}
The analytic expressions for $m_{T\pm}^2$ are given in
Eq.~\eqref{eq:transverse_even_masses_main}. Hence
\begin{equation}
\operatorname{Spec}(M_{\rm even}^2)=
\{m_{R1}^2,\ldots,m_{R8}^2,m_{T+}^2,m_{T+}^2,m_{T-}^2,m_{T-}^2\}.
\label{eq:app_even_spectrum}
\end{equation}

\subsubsection{Complete CP-even transformation}

It is useful to state the complete transformation without introducing an additional matrix
of shorthand entries. We define the radial-first adapted vector
\begin{equation}
\mathcal B_{\rm even}^{\rm(ad)}=
(h,\Phi_R,\xi_R,\psi_R,\chi_R,\eta_R,\rho_R,\kappa_R,
\Phi_{T1},\Phi_{T2},\xi_{T1},\xi_{T2})^T.
\label{eq:app_even_adapted_order}
\end{equation}
The permutation $P_{8+4}$ is defined completely by its action
\begin{align}
P_{8+4}\,\mathcal B_{\rm even}^{\rm(ad)}={}&
(h,\Phi_{T1},\Phi_{T2},\Phi_R,
\xi_{T1},\xi_{T2},\xi_R,\notag\\
&\psi_R,\chi_R,\eta_R,\rho_R,\kappa_R)^T.
\label{eq:app_P84_definition}
\end{align}
Thus $P_{8+4}$ contains only zeros and ones and introduces no new parameter. With
\begin{equation}
U_{\rm trip}=\operatorname{diag}(1,U_\Phi,U_\xi,I_5),
\end{equation}
the original interaction basis of Eq.~\eqref{eq:CPeven_basis_main} is related to the mass
basis
\begin{equation}
\mathcal B_{\rm even,m}=
(H_{R1},\ldots,H_{R8},T_{+1},T_{+2},T_{-1},T_{-2})^T
\end{equation}
by the exact $12\times12$ transformation
\begin{equation}
\mathcal B_{\rm even}=U_{\rm even}\mathcal B_{\rm even,m},
\qquad
U_{\rm even}=U_{\rm trip}P_{8+4}
\begin{pmatrix}
O_R&0_{8\times4}\\
0_{4\times8}&O_T
\end{pmatrix}.
\label{eq:app_Ueven_factorised}
\end{equation}
Using the definitions above, Eq.~\eqref{eq:app_Ueven_factorised} gives the complete CP-even transformation. Here $O_R$ diagonalises the radial sector, $O_T$ diagonalises the transverse sector, $P_{8+4}$ restores the triplet ordering, and $U_{\rm trip}$ returns to the original component basis. It obeys
\begin{equation}
U_{\rm even}^TU_{\rm even}=I_{12},
\qquad
U_{\rm even}^TM_{\rm even}^2U_{\rm even}
=\operatorname{diag}(m_{R1}^2,\ldots,m_{R8}^2,
 m_{T+}^2,m_{T+}^2,m_{T-}^2,m_{T-}^2).
\label{eq:app_Ueven_check}
\end{equation}

\subsection{M--R CP-odd sector}
\label{app:CPodd_complete}

The non-trivial CP-odd mixing occurs only in the six triplet-phase components. We first
make the angle notation explicit. The two angles $\theta_1$ and $\theta_2$ are those of
Eq.~\eqref{eq:odd_angles_main},
\begin{equation}
\tan 2\theta_1=\frac{2D}{A-B},
\qquad
\tan 2\theta_2=\frac{6D}{A-B},
\end{equation}
where $A$, $B$ and $D$ are defined in Eq.~\eqref{eq:ABD_odd_main}. We use
\begin{equation}
c_1\equiv\cos\theta_1,\quad s_1\equiv\sin\theta_1,
\qquad
c_2\equiv\cos\theta_2,\quad s_2\equiv\sin\theta_2.
\label{eq:app_c1c2_definitions}
\end{equation}
The original triplet-phase interaction basis is ordered as
\begin{equation}
\mathbf I_6=(\phi_{I1},\phi_{I2},\phi_{I3},\xi_{I1},\xi_{I2},\xi_{I3})^T,
\end{equation}
and the corresponding ordered mass/adapted basis is
\begin{equation}
\mathbf A_6=(A_{1+},A_{2+},\Phi_{I0},A_{1-},A_{2-},\xi_{I0})^T.
\end{equation}
Their relation is
\begin{equation}
\mathbf I_6=O_6\mathbf A_6,
\end{equation}
with
\begin{equation}
O_6=
\begin{pmatrix}
-\frac{c_1}{\sqrt2}&-\frac{c_2}{\sqrt6}&\frac1{\sqrt3}&\frac{s_1}{\sqrt2}&\frac{s_2}{\sqrt6}&0\\
0&\frac{2c_2}{\sqrt6}&\frac1{\sqrt3}&0&-\frac{2s_2}{\sqrt6}&0\\
\frac{c_1}{\sqrt2}&-\frac{c_2}{\sqrt6}&\frac1{\sqrt3}&-\frac{s_1}{\sqrt2}&\frac{s_2}{\sqrt6}&0\\
-\frac{s_1}{\sqrt2}&\frac{s_2}{\sqrt6}&0&-\frac{c_1}{\sqrt2}&\frac{c_2}{\sqrt6}&\frac1{\sqrt3}\\
0&\frac{2s_2}{\sqrt6}&0&0&\frac{2c_2}{\sqrt6}&-\frac1{\sqrt3}\\
\frac{s_1}{\sqrt2}&\frac{s_2}{\sqrt6}&0&\frac{c_1}{\sqrt2}&\frac{c_2}{\sqrt6}&\frac1{\sqrt3}
\end{pmatrix}.
\label{eq:app_O6_odd_explicit}
\end{equation}
This is equivalently $O_6=\operatorname{diag}(U_\Phi,U_\xi)O_{\rm pair}$, with
$O_{\rm pair}$ given in Eq.~\eqref{eq:Opair_odd_main}.

The complete neutral CP-odd interaction and mass bases are
\begin{align}
\mathcal B_{\rm odd}={}&
(\Omega,\phi_{I1},\phi_{I2},\phi_{I3},\xi_{I1},\xi_{I2},\xi_{I3},
\psi_I,\chi_I,\eta_I,\rho_I,\kappa_I)^T,
\notag\\
\mathcal B_{\rm odd,m}={}&
(\Omega,A_{1+},A_{2+},\Phi_{I0},A_{1-},A_{2-},\xi_{I0},
\psi_I,\chi_I,\eta_I,\rho_I,\kappa_I)^T.
\end{align}
The complete CP-odd transformation is therefore given by the block matrix
\begin{equation}
\mathcal B_{\rm odd}=U_{\rm odd}\mathcal B_{\rm odd,m},
\qquad
U_{\rm odd}=\operatorname{diag}(1,O_6,I_5).
\label{eq:app_Uodd_block}
\end{equation}
In Eq.~\eqref{eq:app_Uodd_block}, the first entry acts on $\Omega$, $O_6$ acts on the six triplet phases, and $I_5$ acts on $(\psi_I,\chi_I,\eta_I,\rho_I,\kappa_I)$. It follows that
\begin{equation}
U_{\rm odd}^TU_{\rm odd}=I_{12},
\end{equation}
\begin{equation}
U_{\rm odd}^TM_{\rm odd}^2U_{\rm odd}
=\operatorname{diag}(0,m_{1+}^2,m_{2+}^2,0,m_{1-}^2,m_{2-}^2,0,0,0,0,0,0).
\label{eq:app_Uodd_check}
\end{equation}
The four non-zero masses $m_{1\pm}^2$ and $m_{2\pm}^2$ are given in
Eq.~\eqref{eq:odd_four_masses_main}. The complete spectrum is therefore
\begin{equation}
\operatorname{Spec}(M_{\rm odd}^2)
=\{m_{1+}^2,m_{2+}^2,m_{1-}^2,m_{2-}^2,0,0,0,0,0,0,0,0\}.
\label{eq:app_odd_spectrum}
\end{equation}
The first zero eigenstate is $G^0\equiv\Omega$. The remaining seven zero eigenstates are the
physical CP-odd modes discussed in
Subsubsection~\ref{subsubsec:accidental_U1_zero_modes}.

\section{Radial quartic matrix in the original couplings}
\label{app:raw_radial_quartic}

The BFB and global-minimum discussion uses the non-negative variables in the fixed ordering
\begin{equation}
\mathbf z=(z_H,z_\Phi,z_\xi,z_\psi,z_\chi,z_\eta,z_\rho,z_\kappa)^T
\end{equation}
of Eq.~\eqref{eq:BFB_zvector_main}. The radial quartic polynomial is
$V_{4,{\rm rad}}=\mathbf z^TQ_{\rm MR}\mathbf z$. To keep the matrix readable while retaining
all of its entries, we use the same $3+5$ partition as for the radial mass matrix:
\begin{equation}
Q_{\rm MR}=
\begin{pmatrix}
Q_3&Q_P\\
Q_P^T&Q_S
\end{pmatrix}.
\label{eq:app_QMR_block}
\end{equation}
No physical assumption is associated with this partition; it is only an ordering of rows and
columns.

We define the following entries directly in terms of the original quartic couplings:
\begin{equation}
\begin{aligned}
q_H&=\lambda_H,\quad
q_\Phi=\frac{3\lambda_{\Phi1}+4\lambda_{\Phi3}}{3},\quad
q_\xi=\frac{3\lambda_{\xi1}+4\lambda_{\xi3}}{3},\quad
q_s=\lambda_s,\\
q_{H\Phi}&=\frac{\lambda_{H\Phi}}2,\quad
q_{H\xi}=\frac{\lambda_{H\xi}}2,\quad
q_{\Phi\xi}=\frac{3\lambda_{\Phi\xi1}+4\lambda_{\Phi\xi11}+4\lambda_{\Phi\xi4}+3\lambda_{\Phi\xi8}}{6}.
\end{aligned}
\label{eq:app_QMR_qdefs_a}
\end{equation}
Here $s\in\{\psi,\chi,\eta,\rho,\kappa\}$, and
\begin{equation}
q_{Hs}=\frac{\lambda_{Hs}}2,
\qquad
q_{\Phi s}=\frac{\lambda_{\Phi s}}2,
\qquad
q_{\xi s}=\frac{\lambda_{\xi s}}2,
\qquad
q_{st}=\frac{\lambda_{st}}2\quad(s\ne t).
\label{eq:app_QMR_qdefs_b}
\end{equation}
The $3\times3$ block is
\begin{equation}
Q_3=
\begin{pmatrix}
q_H&q_{H\Phi}&q_{H\xi}\\
q_{H\Phi}&q_\Phi&q_{\Phi\xi}\\
q_{H\xi}&q_{\Phi\xi}&q_\xi
\end{pmatrix}.
\label{eq:app_Q3_explicit}
\end{equation}
The $3\times5$ portal block, with columns ordered as
$(\psi,\chi,\eta,\rho,\kappa)$, is
\begin{equation}
Q_P=
\begin{pmatrix}
q_{H\psi}&q_{H\chi}&q_{H\eta}&q_{H\rho}&q_{H\kappa}\\
q_{\Phi\psi}&q_{\Phi\chi}&q_{\Phi\eta}&q_{\Phi\rho}&q_{\Phi\kappa}\\
q_{\xi\psi}&q_{\xi\chi}&q_{\xi\eta}&q_{\xi\rho}&q_{\xi\kappa}
\end{pmatrix},
\label{eq:app_QP_explicit}
\end{equation}
and the symmetric $5\times5$ singlet block is
\begin{equation}
Q_S=
\begin{pmatrix}
q_\psi&q_{\psi\chi}&q_{\psi\eta}&q_{\psi\rho}&q_{\psi\kappa}\\
q_{\psi\chi}&q_\chi&q_{\chi\eta}&q_{\chi\rho}&q_{\chi\kappa}\\
q_{\psi\eta}&q_{\chi\eta}&q_\eta&q_{\eta\rho}&q_{\eta\kappa}\\
q_{\psi\rho}&q_{\chi\rho}&q_{\eta\rho}&q_\rho&q_{\rho\kappa}\\
q_{\psi\kappa}&q_{\chi\kappa}&q_{\eta\kappa}&q_{\rho\kappa}&q_\kappa
\end{pmatrix}.
\label{eq:app_QS_explicit}
\end{equation}
Equations~\eqref{eq:app_QMR_block}--\eqref{eq:app_QS_explicit} specify every one of the
64 entries of the symmetric $8\times8$ matrix (36 independent entries) without introducing any
new coupling. In particular, the symbols $q$ are only typographical abbreviations for the original
quartics in Eqs.~\eqref{eq:app_QMR_qdefs_a} and \eqref{eq:app_QMR_qdefs_b}.

For the stronger positive-definite region used in the sufficient global-minimum argument,
Sylvester's criterion may be applied directly to the assembled matrix:
\begin{equation}
\det\!\left[(Q_{\rm MR})_{1:n,1:n}\right]>0,
\qquad n=1,\ldots,8.
\label{eq:app_QMR_sylvester}
\end{equation}
For BFB alone, copositivity on the non-negative domain is the weaker requirement, as explained
in the main text. No strict-diagonal-dominance condition and no absolute-value table is imposed.

\subsection{Angular identities and the radial completed square}
\label{app:global_minimum_derivation}

The two algebraic steps used in the global-minimum discussion are given below.
For a triplet $X=\Phi,\xi$, we define
\begin{equation}
\upsilon_X=\frac{(X^\dagger X)_{1'}}{z_X},
\qquad
\mathbf a_X=\frac{(X^\dagger X)_{3_a}}{z_X}.
\label{eq:app_angular_auxiliary}
\end{equation}
Because $X^\dagger X$ is Hermitian, $(X^\dagger X)_{1''}/z_X=\upsilon_X^*$ and
$\mathbf a_X$ is purely imaginary. The product rules of Appendix~\ref{app:A4_rules} therefore give
\begin{equation}
U_X=|\upsilon_X|^2,
\qquad
j_2=2\,\mathrm{Re}(\upsilon_\Phi\upsilon_\xi^*),
\qquad
j_{14}=\frac{4}{\sqrt3}\,\mathrm{Im}(\upsilon_\Phi\upsilon_\xi^*).
\label{eq:app_j2_j14_reduction}
\end{equation}
Consequently,
\begin{equation}
j_2^2+\frac34j_{14}^2
=4|\upsilon_\Phi|^2|\upsilon_\xi|^2
=4U_\Phi U_\xi,
\label{eq:app_first_angular_identity}
\end{equation}
which is the first identity in Eq.~\eqref{eq:mixed_angular_identities_main}. Writing
$\mathbf a_X=i\mathbf w_X$ with real $\mathbf w_X$, one also has
\begin{equation}
W_X=\mathbf w_X\!\cdot\!\mathbf w_X,
\qquad
j_5=-\mathbf w_\Phi\!\cdot\!\mathbf w_\xi.
\label{eq:app_j5_reduction}
\end{equation}
The ordinary Cauchy--Schwarz inequality then yields
\begin{equation}
|j_5|\le\sqrt{W_\Phi W_\xi},
\label{eq:app_second_angular_identity}
\end{equation}
which is the second identity used in the main text.

Along the fixed M--R angular directions the potential is a quadratic polynomial in the radial
variables $\mathbf z$,
\begin{equation}
V_{\rm MR}(\mathbf z)=\mathbf z^TQ_{\rm MR}\mathbf z+\mathbf b^T\mathbf z+V_0.
\label{eq:app_radial_polynomial}
\end{equation}
Stationarity at $\mathbf z_0$ gives
\begin{equation}
2Q_{\rm MR}\mathbf z_0+\mathbf b=0.
\label{eq:app_radial_stationarity_vector}
\end{equation}
Eliminating $\mathbf b$ with Eq.~\eqref{eq:app_radial_stationarity_vector} and subtracting the
vacuum value gives, without any further assumption,
\begin{align}
V_{\rm MR}(\mathbf z)-V_{\rm MR}(\mathbf z_0)
&=\mathbf z^TQ_{\rm MR}\mathbf z
-2\mathbf z_0^TQ_{\rm MR}\mathbf z
+\mathbf z_0^TQ_{\rm MR}\mathbf z_0\notag\\
&=(\mathbf z-\mathbf z_0)^TQ_{\rm MR}(\mathbf z-\mathbf z_0),
\label{eq:app_radial_completed_square}
\end{align}
which reproduces Eq.~\eqref{eq:radial_completed_square_main}. Hence $Q_{\rm MR}>0$ makes the
radial difference strictly positive away from $\mathbf z_0$, while the independent angular
inequalities of the main text ensure that changing the triplet directions cannot lower the energy.

\section{Details of scalar-mediated FCNC and CLFV}
\label{app:scalar_clfv_details}

This appendix gives the technical steps behind Subsubsection~\ref{subsubsec:scalar_mediators} for the complete scalar potential used in the main text.

\subsection{Transverse charged-lepton Yukawa matrices}

For the M--R alignment the orthonormal directions in the $\Phi$ triplet may be chosen as
\begin{equation}
u_R=\frac1{\sqrt3}(1,1,1),\qquad
u_A=\frac1{\sqrt2}(0,1,-1),\qquad
u_B=\frac1{\sqrt6}(2,-1,-1).
\end{equation}
The first vector is parallel to the VEV. Substitution in the charged-lepton operator of Eq.~\eqref{eq:yukawa_effective} gives a diagonal flavour matrix for the radial direction. For the two transverse directions, the charged-lepton rotation already established in Section~\ref{sec:symmetry_realisation} gives
\begin{equation}
\mathcal Y_A=\frac1{\sqrt2}
\begin{pmatrix}
0&i y_{L2}&-i y_{L3}\\
-i y_{L1}&0&i y_{L3}\\
i y_{L1}&-i y_{L2}&0
\end{pmatrix},
\qquad
\mathcal Y_B=\frac1{\sqrt2}
\begin{pmatrix}
0&y_{L2}&y_{L3}\\
y_{L1}&0&y_{L3}\\
y_{L1}&y_{L2}&0
\end{pmatrix}.
\label{eq:YA_YB_app}
\end{equation}
The PCLC ordering implies the physical charged-lepton identifications
\[
(m_e,m_\mu,m_\tau)=\frac{\sqrt3\,v_Hv_\Phi}{2\Lambda}
(y_{L1},y_{L3},y_{L2}).
\]
Together with the scalar-field normalisation, this reduces Eq.~\eqref{eq:YA_YB_app} to the matrices $\Gamma_A$ and $\Gamma_B$ quoted in Eq.~\eqref{eq:GammaAB_main}. This also makes clear why the radial direction is flavour diagonal while the two transverse directions carry all tree-level scalar FCNC.

\subsection{Equivalence of mass-eigenstate and propagator descriptions}

We denote the interaction-basis neutral scalars by $\varphi_r$ and the corresponding mass eigenstates by $S_n$,
\begin{equation}
\varphi_r=\sum_n U_{rn}S_n,
\qquad
\Gamma_n=\sum_r U_{rn}\Gamma_r.
\end{equation}
Orthogonality of $U$ gives
\begin{align}
\sum_n\frac{\Gamma_n^{(f)}\Gamma_n^{(i)}}{q^2-m_n^2}
&=\sum_{r,s}\Gamma_r^{(f)}
\left[U(q^2I-M_{\rm diag}^2)^{-1}U^T\right]_{rs}
\Gamma_s^{(i)}\notag\\
&=\sum_{r,s}\Gamma_r^{(f)}
\left[(q^2I-M^2)^{-1}\right]_{rs}
\Gamma_s^{(i)},
\end{align}
which proves Eq.~\eqref{eq:yukawa_propagator_bridge}.

\subsection{Projected propagators for the complete scalar potential}

The CP-even transverse mass matrix has the block form
\begin{equation}
M_E^2=
\begin{pmatrix}
X_EI_2&\mathcal X\\
\mathcal X^T&Y_EI_2
\end{pmatrix},
\qquad
\mathcal X\mathcal X^T=Z_E^2I_2.
\end{equation}
Taking the Schur complement of the lower $2\times2$ block gives
\begin{align}
S_\Phi
&=(X_E-q^2)I_2-\mathcal X[(Y_E-q^2)I_2]^{-1}\mathcal X^T\notag\\
&=\left[(X_E-q^2)-\frac{Z_E^2}{Y_E-q^2}\right]I_2,
\end{align}
and therefore
\begin{equation}
S_\Phi^{-1}=\frac{Y_E-q^2}{(X_E-q^2)(Y_E-q^2)-Z_E^2}I_2,
\end{equation}
which is Eq.~\eqref{eq:gE_maintext}.

The two massive CP-odd transverse blocks are
\begin{equation}
M_{O1}^2=\begin{pmatrix}X_O&Z_O\\Z_O&Y_O\end{pmatrix},
\qquad
M_{O2}^2=\begin{pmatrix}X_O&3Z_O\\3Z_O&Y_O\end{pmatrix}.
\end{equation}
Direct inversion gives Eq.~\eqref{eq:gO12_main}. Their difference is
\begin{equation}
\delta g
=Y_O\frac{(X_OY_O-9Z_O^2)-(X_OY_O-Z_O^2)}
{(X_OY_O-Z_O^2)(X_OY_O-9Z_O^2)},
\end{equation}
which immediately yields Eq.~\eqref{eq:deltag_maintext}. Since $Z_O=\Delta v_\Phi v_\xi$, no term linear in $\Delta$ can appear in the projected CP-odd difference.

\subsection{Contact-limit three-body widths}

The four-lepton effective operator is obtained by contracting two Yukawa vertices with the projected scalar propagator. For the two same-sign modes, after the identical-fermion exchange term is included, the closed contact-limit widths are
\begin{align}
\Gamma(\tau^-\to e^-e^-\mu^+)={}&
\frac{m_\tau^5}{442368\pi^3v_\Phi^4}
\Big[(m_\mu^2m_\tau^2+m_e^4)|F_-|^2\notag\\
&\hspace{14mm}+2m_e^2(m_\tau^2+m_\mu^2)|F_+|^2\Big],
\label{eq:GeeMu_app}
\end{align}
and
\begin{align}
\Gamma(\tau^-\to\mu^-\mu^-e^+)={}&
\frac{m_\tau^5}{442368\pi^3v_\Phi^4}
\Big[(m_e^2m_\tau^2+m_\mu^4)|F_-|^2\notag\\
&\hspace{14mm}+2m_\mu^2(m_\tau^2+m_e^2)|F_+|^2\Big].
\label{eq:GmumuE_app}
\end{align}
The opposite-sign mixed channels are proportional to $|\delta g|^2$ and therefore begin at ${\cal O}(\Delta^4)$ in the rate. The branching ratio is $\mathrm{BR}=\Gamma/\Gamma_\tau$. These contact expressions require $m_S^2\gg q^2\lesssim m_\tau^2$; if a scalar is light or resonant, the full momentum dependence must be retained. Standard treatments of three-body and radiative CLFV amplitudes may be found in Refs.~\cite{r75,r76,r77,r78}.

\subsection{Radiative scalar amplitudes}

For a neutral scalar and an internal charged lepton, the most general one-loop dipole is a linear combination of four Yukawa bilinears. We use the convention of Ref.~\cite{r77} and translate its chiral couplings into the present $\Gamma_A,\Gamma_B$ notation. Because every $A$-direction bilinear is equal and opposite to the corresponding $B$-direction bilinear, while the two CP-even directions carry exactly the same spectral weight, one obtains
\begin{equation}
\sigma_L^{E}=\sigma_R^{E}=0\qquad(i\ne j)
\end{equation}
for the CP-even sector. For the CP-odd blocks the two spectral weights differ by an amount proportional to $Z_O^2$, and hence
\begin{equation}
\mathcal A_O(\ell_i\to\ell_j\gamma)={\cal O}(\Delta^2),
\qquad
\mathrm{BR}_O={\cal O}(\Delta^4).
\end{equation}
The numerical use of these amplitudes will be combined with the gauge/HNL contribution in the phenomenological constraint discussion after Subsubsection~5.1.2.

\section{Details of gauge-boson/heavy-neutrino CLFV}
\label{app:gauge_clfv_details}

This appendix gives the details of the gauge/HNL CLFV calculation used in Subsubsection~\ref{subsubsec:gauge_mediator}, including the heavy-neutrino mixing and the loop amplitudes.

\subsection{Leading active-heavy mixing}

In the physical charged-lepton ordering already established in Section~\ref{sec:symmetry_realisation}, the Dirac and leading RH Majorana matrices relevant for the gauge/HNL calculation are
\begin{equation}
M_D=\begin{pmatrix}0&P&0\\Q&0&0\\0&0&R\end{pmatrix},\qquad
M_R=\begin{pmatrix}
\mathcal M_1&0&\mathcal M_2\\
0&\mathcal M_3&0\\
\mathcal M_2&0&0
\end{pmatrix}.
\end{equation}
The inverse of the Majorana matrix is
\begin{equation}
\big(M_R\big)^{-1}=
\begin{pmatrix}
0&0&1/\mathcal M_2\\
0&1/\mathcal M_3&0\\
1/\mathcal M_2&0&-\mathcal M_1/\mathcal M_2^2
\end{pmatrix}.
\label{eq:MRinv_appE}
\end{equation}
Direct multiplication gives
\begin{equation}
\Xi=M_D\big(M_R\big)^{-1}=
\begin{pmatrix}
0&P/\mathcal M_3&0\\
0&0&Q/\mathcal M_2\\
R/\mathcal M_2&0&-\mathcal M_1R/\mathcal M_2^2
\end{pmatrix},
\qquad \Theta=\Xi^*,
\end{equation}
which is Eq.~\eqref{eq:Theta_explicit_main}. The approximation is the leading seesaw result; terms of schematic order $M_wM_D/M_R^2$ and $M_D^3/M_R^3$ are not retained in the present phenomenological analysis.

The Type-I part of the light-neutrino mass matrix is
\begin{equation}
-M_D\big(M_R\big)^{-1}(M_D)^T=
\begin{pmatrix}
-P^2/\mathcal M_3&0&0\\
0&0&-QR/\mathcal M_2\\
0&-QR/\mathcal M_2&\mathcal M_1R^2/\mathcal M_2^2
\end{pmatrix}.
\label{eq:typeI_appE}
\end{equation}
Comparison with the triadic matrix and the Weinberg-like contribution isolates the combinations $a-b$, $b+2t$ and $2b+d$, leading to Eq.~\eqref{eq:Mi_texture_main}.

\subsection{Takagi diagonalisation and heavy-state support}

After phase redefinitions the displayed $(1,3)$ block may be taken real for the purpose of obtaining the physical masses. Defining
\begin{equation}
\Delta_N=\sqrt{\mathcal M_1^2+4\mathcal M_2^2},\qquad
\tan2\theta_N=\frac{2\mathcal M_2}{\mathcal M_1},
\end{equation}
the positive Takagi masses are
\begin{equation}
M_+=\frac{\Delta_N+\mathcal M_1}{2},\qquad
M_-=\frac{\Delta_N-\mathcal M_1}{2},\qquad
M_0=|\mathcal M_3|.
\end{equation}
The negative signed eigenvector of the real symmetric $(1,3)$ block carries the usual phase $i$ in the Takagi matrix. Transforming the physical charged-current coefficient $\Theta$ to the heavy mass basis then gives Eq.~\eqref{eq:Bmatrix_main}. The resulting support is
\begin{equation}
N_0:\ e\ \text{only},\qquad N_\pm:\ \mu,\tau\ \text{only}.
\end{equation}
A useful algebraic cross-check is
\begin{equation}
s_Nc_N\left(\frac1{M_+^2}-\frac1{M_-^2}\right)
=-\frac{\mathcal M_1}{\mathcal M_2^3},
\end{equation}
which reconstructs the weak-basis element $(\Theta\Theta^\dagger)_{\mu\tau}$ from the heavy-mass-basis couplings.

\subsection{Radiative form factors}

We use the convention of Ilakovac and Pilaftsis \cite{r89}. With $x=M^2/M_W^2$,
\begin{align}
F_\gamma(x)={}&\frac{7x^3-x^2-12x}{12(1-x)^3}
-\frac{x^4-10x^3+12x^2}{6(1-x)^4}\ln x,\\
G_\gamma(x)={}&-\frac{2x^3+5x^2-x}{4(1-x)^3}
-\frac{3x^3}{2(1-x)^4}\ln x.
\end{align}
The charged-current sums are
\begin{equation}
F_\gamma^{\alpha\beta}=\sum_I B_{\alpha I}^*B_{\beta I}F_\gamma(x_I),\qquad
G_\gamma^{\alpha\beta}=\sum_I B_{\alpha I}^*B_{\beta I}G_\gamma(x_I).
\end{equation}
The support in Eq.~\eqref{eq:HNL_support_main} gives the two exact zeros in Eq.~\eqref{eq:Ggamma_support_main}. For the surviving transition,
\begin{equation}
G_\gamma^{\tau\mu}=Q^*Rs_Nc_N
\left[\frac{G_\gamma(x_+)}{M_+^2}-\frac{G_\gamma(x_-)}{M_-^2}\right].
\end{equation}
In the common-heavy limit $G_\gamma(x)\to1/2$ this becomes
\begin{equation}
G_\gamma^{\tau\mu}\simeq\frac12(\Theta\Theta^\dagger)_{\mu\tau},
\end{equation}
and Eq.~\eqref{eq:BRrad_HNL_main} reduces to
\begin{equation}
\mathrm{BR}(\tau\to\mu\gamma)\simeq
\frac{3\alpha_{\rm em}}{8\pi}\frac{\Gamma_0}{\Gamma_\tau}
|(\Theta\Theta^\dagger)_{\mu\tau}|^2,
\qquad
\Gamma_0=\frac{G_F^2m_\tau^5}{192\pi^3}.
\end{equation}
This limiting expression is used only as a cross-check; the mass-dependent form factor in the main text is the phenomenological result.

\subsection{$Z$ penguin and box structures}

We define the neutral-current heavy-state combination
\begin{equation}
C_{IJ}=\sum_{\ell=e,\mu,\tau}B_{\ell I}^*B_{\ell J}.
\end{equation}
In the same convention \cite{r89}, the $Z$-penguin form factor is
\begin{align}
F_Z^{\ell\ell'}={}&\sum_{IJ}B_{\ell I}^*B_{\ell'J}\Big\{
\delta_{IJ}[F_Z(x_I)+2G_Z(0,x_I)]\notag\\
&+C_{IJ}^*[G_Z(x_I,x_J)-G_Z(0,x_I)-G_Z(0,x_J)]
+C_{IJ}H_Z(x_I,x_J)\Big\}.
\label{eq:FZ_appE}
\end{align}
The explicit kernels are
\begin{equation}
F_Z(x)=-\frac{5x}{2(1-x)}-\frac{5x^2}{2(1-x)^2}\ln x,
\end{equation}
\begin{align}
G_Z(x,y)=-\frac1{2(x-y)}\Bigg[
\frac{x^2(1-y)}{1-x}\ln x
-\frac{y^2(1-x)}{1-y}\ln y\Bigg],
\end{align}
\begin{align}
H_Z(x,y)=\frac{\sqrt{xy}}{4(x-y)}\Bigg[
\frac{x^2-4x}{1-x}\ln x
-\frac{y^2-4y}{1-y}\ln y\Bigg].
\end{align}

For the heavy-only box it is important to retain the unitarity subtraction. Introducing
\begin{equation}
K_{IJ}=F_{\rm Box}(x_I,x_J)-F_{\rm Box}(0,x_J)
-F_{\rm Box}(0,x_I)+1,
\end{equation}
the composite box form factor can be written as
\begin{align}
F_{\rm Box}^{\ell\ell'\ell_1\ell_2}={}&\sum_{IJ}\Big[
B_{\ell I}^*B_{\ell'I}\delta_{\ell_1\ell_2}\delta_{IJ}
\{F_{\rm Box}(0,x_I)-1\}\notag\\
&+B_{\ell I}^*B_{\ell_1I}\delta_{\ell'\ell_2}\delta_{IJ}
\{F_{\rm Box}(0,x_I)-1\}\notag\\
&+B_{\ell I}^*B_{\ell_2J}^*
(B_{\ell'I}B_{\ell_1J}+B_{\ell_1I}B_{\ell'J})K_{IJ}\notag\\
&+B_{\ell I}^*B_{\ell_2I}^*B_{\ell'J}B_{\ell_1J}
G_{\rm Box}(x_I,x_J)\Big].
\label{eq:Fbox_appE}
\end{align}
The last line is the genuinely Majorana contribution. It is the origin of the non-zero leading support for $\tau^-\to e^-e^-\mu^+$ even though there is no leading one-index $\tau\to e$ penguin.

For this channel the support reduces to
\begin{equation}
F_{\rm Box}^{\tau ee\mu}=B_{e0}^{\,2}
\sum_{I=\pm}B_{\tau I}^*B_{\mu I}^*G_{\rm Box}(x_I,x_0).
\label{eq:Majbox_support_appE}
\end{equation}
The remaining box kernels $F_{\rm Box}$ and $G_{\rm Box}$ are those of Ref.~\cite{r89}; no alternative normalisation is mixed into Eqs.~\eqref{eq:BR3mu_HNL_main}--\eqref{eq:BRMajBox_main}.

\subsection{Amplitude-level combination with the scalar sector}

The scalar and gauge/HNL sectors have been derived independently, but a physical final state is unique. Therefore, for channels with overlapping support,
\begin{equation}
\overline{|\mathcal M_{\rm tot}|^2}=
\overline{|\mathcal M_{\rm scalar}|^2}
+\overline{|\mathcal M_{W/N}|^2}
+2\Re\,\overline{\mathcal M_{\rm scalar}\mathcal M_{W/N}^*}.
\end{equation}
For $\tau\to\mu\gamma$ the dipole coefficients should likewise be combined before squaring. The channels with overlap at the order retained are $\tau\to\mu\gamma$, $\tau^-\to\mu^-e^-e^+$ and $\tau^-\to e^-e^-\mu^+$. By contrast, $\tau^-\to e^-\mu^-\mu^+$ and $\tau^-\to\mu^-\mu^-e^+$ have no leading HNL/$W$ support and therefore provide clean scalar-sector probes within the present approximation.

\section{Details of the non-unitarity analysis}
\label{app:nonunitarity_details}

This appendix gives the details of the non-unitarity calculation discussed in Subsection~\ref{subsec:nonunitarity}.

\subsection{Leading active--heavy mixing}

At leading seesaw order we first define the Weyl coefficient
\begin{equation}
\Xi=M_D\big(M_R\big)^{-1},
\qquad \Theta=\Xi^*.
\label{eq:Theta_appF_start}
\end{equation}
For the matrices of Eq.~\eqref{eq:MDc_MR_main},
\begin{equation}
\big(M_R\big)^{-1}=
\begin{pmatrix}
0&0&1/\mathcal M_2\\
0&1/\mathcal M_3&0\\
1/\mathcal M_2&0&-\mathcal M_1/\mathcal M_2^2
\end{pmatrix},
\label{eq:MRinverse_appF}
\end{equation}
so direct multiplication gives
\begin{equation}
\Xi=
\begin{pmatrix}
0&P/\mathcal M_3&0\\
0&0&Q/\mathcal M_2\\
R/\mathcal M_2&0&-\mathcal M_1R/\mathcal M_2^2
\end{pmatrix},
\qquad \Theta=\Xi^*.
\label{eq:Theta_appF_M}
\end{equation}
Using Eq.~\eqref{eq:Mi_texture_main},
\begin{align}
\frac{P}{\mathcal M_3}&=-\frac{a-b}{P},
&\frac{Q}{\mathcal M_2}&=-\frac{b+2t}{2R},\notag\\
\frac{R}{\mathcal M_2}&=-\frac{b+2t}{2Q},
&-\frac{\mathcal M_1R}{\mathcal M_2^2}&=-\frac{2b+d}{R},
\end{align}
which reproduces Eq.~\eqref{eq:Theta_texture_main} after complex conjugation for the physical charged-current coefficient.

\subsection{Explicit multiplication of $\Theta\Theta^\dagger$}

The three row vectors of the physical charged-current coefficient are
\begin{equation}
(\Theta)_e=\left(0,-\frac{(a-b)^*}{P^*},0\right),
\qquad
(\Theta)_\mu=\left(0,0,-\frac{(b+2t)^*}{2R^*}\right),
\end{equation}
\begin{equation}
(\Theta)_\tau=\left(-\frac{(b+2t)^*}{2Q^*},0,-\frac{(2b+d)^*}{R^*}\right).
\end{equation}
The nonzero scalar products are therefore
\begin{align}
(\Theta\Theta^\dagger)_{ee}&=\frac{|a-b|^2}{|P|^2},\notag\\
(\Theta\Theta^\dagger)_{\mu\mu}&=\frac{|b+2t|^2}{4|R|^2},\notag\\
(\Theta\Theta^\dagger)_{\tau\tau}&=
\frac{|b+2t|^2}{4|Q|^2}+\frac{|2b+d|^2}{|R|^2},\notag\\
(\Theta\Theta^\dagger)_{\mu\tau}&=
\frac{(b+2t)^*(2b+d)}{2|R|^2}.
\end{align}
The electron row has no common nonzero heavy column with either the muon or tau row, giving
\begin{equation}
(\Theta\Theta^\dagger)_{e\mu}=(\Theta\Theta^\dagger)_{e\tau}=0.
\end{equation}
Dividing by two gives Eq.~\eqref{eq:eta_matrix_main}. Hermiticity is manifest, and positive semidefiniteness gives the general check
\begin{equation}
|\widetilde\eta_{\mu\tau}|^2\le
\widetilde\eta_{\mu\mu}\widetilde\eta_{\tau\tau}.
\end{equation}
For the present texture one also obtains the simple ratio in Eq.~\eqref{eq:eta_ratio_main}.

\subsection{Translation of the global bounds}

The $ee$, $\mu\mu$ and $\mu\tau$ limits in Eqs.~\eqref{eq:NU_bounds_diag1_main}--\eqref{eq:NU_bounds_diag2_main} give, respectively,
\begin{align}
|P|&>\frac{|a-b|}{\sqrt{2(1.4\times10^{-3})}}
=18.90|a-b|,\notag\\
|R|&>\frac{|b+2t|}{\sqrt{8(1.4\times10^{-4})}}
=29.88|b+2t|,\notag\\
|R|&>\sqrt{\frac{|b+2t|\,|2b+d|}{4(1.8\times10^{-4})}}
=37.27\sqrt{|b+2t|\,|2b+d|}.
\end{align}
For the tau diagonal entry the two positive terms must be kept together, leading to Eqs.~\eqref{eq:NU_tau_corr_main} and \eqref{eq:NU_QgivenR_main}. Positivity of the bracket in Eq.~\eqref{eq:NU_QgivenR_main} alone implies
\begin{equation}
|R|>23.70|2b+d|.
\end{equation}
This last inequality is only the condition that a finite allowed $Q$ can exist; it is not an independent experimental measurement.

\subsection{Original Yukawa--VEV combinations}

It is useful to display how the bounds act on the combinations entering the Dirac mass matrix. For this translation, we define
\begin{equation}
\mathcal Y_1=\frac{|y_1|v_\Phi}{\Lambda},\qquad
\mathcal Y_2=\frac{|y_2|v_\Phi v_\chi}{\Lambda^2},\qquad
\mathcal Y_3=\frac{|y_3|v_\Phi v_\psi}{\Lambda^2}.
\label{eq:Y123_appF}
\end{equation}
From Eq.~\eqref{eq:PQR_def_main},
\begin{equation}
|Q|=\frac{\sqrt3}{2}v_H\mathcal Y_1,
\qquad
|P|=\frac{\sqrt3}{2\sqrt2}v_H\mathcal Y_2,
\qquad
|R|=\frac{\sqrt3}{2\sqrt2}v_H\mathcal Y_3.
\end{equation}
The non-unitarity entries can then be written as
\begin{align}
\widetilde\eta_{ee}&=\frac{4|a-b|^2}{3v_H^2\mathcal Y_2^2},\notag\\
\widetilde\eta_{\mu\mu}&=\frac{|b+2t|^2}{3v_H^2\mathcal Y_3^2},\notag\\
\widetilde\eta_{\tau\tau}&=
\frac{|b+2t|^2}{6v_H^2\mathcal Y_1^2}
+\frac{4|2b+d|^2}{3v_H^2\mathcal Y_3^2},\notag\\
|\widetilde\eta_{\mu\tau}|&=
\frac{2|b+2t|\,|2b+d|}{3v_H^2\mathcal Y_3^2}.
\end{align}
Thus the clean one-dimensional lower limits become
\begin{equation}
\mathcal Y_2>30.86\frac{|a-b|}{v_H},
\qquad
\mathcal Y_3>48.80\frac{|b+2t|}{v_H},
\end{equation}
\begin{equation}
\mathcal Y_3>60.86\frac{\sqrt{|b+2t|\,|2b+d|}}{v_H},
\end{equation}
while the $\tau\tau$ condition remains correlated in $(\mathcal Y_1,\mathcal Y_3)$.

\subsection{Relation to the heavy-neutrino mass basis and applicability}

The charged-current matrix $B_{\ell I}$ in Eq.~\eqref{eq:Bmatrix_main} obeys
\begin{equation}
\frac12\sum_I B_{\alpha I}B_{\beta I}^*
=\widetilde\eta_{\alpha\beta}
\end{equation}
at the same leading seesaw order. This provides an independent check that the weak-basis and heavy-mass-basis calculations use the same physical flavour ordering. The exact CLFV loop amplitudes nevertheless contain the mass-dependent functions of Subsection~\ref{subsubsec:gauge_mediator}; only when those loop weights are effectively common can they reduce to a constant times a single $\widetilde\eta_{\alpha\beta}$ entry \cite{r47,r91}.

The global-fit expressions used here assume that the heavy neutrinos are not produced as extra light states in the electroweak precision observables being fitted. We therefore retain
\begin{equation}
M_0,M_+,M_->M_Z
\end{equation}
as an applicability condition for the comparison with Ref.~\cite{r47}. It is not an additional flavour bound.

\section{Numerical translation of the phenomenological constraints}
\label{app:numerical_constraints}

The numerical substitutions used in Subsubsections~\ref{subsubsec:clfv_constraints} and \ref{subsubsec:NU_numerical} are given here, together with the input normalisations and assumptions used in obtaining the quoted bounds.

\subsection{Experimental inputs}
For the charge-resolved three-lepton tau modes we use the HFLAV end-2023 combined upper limits at $90\%$ C.L. \cite{r90},
\begin{align}
{\rm BR}(\tau^-\to e^-\mu^-\mu^+)&<2.0\times10^{-8},&
{\rm BR}(\tau^-\to\mu^-e^-e^+)&<1.3\times10^{-8},\notag\\
{\rm BR}(\tau^-\to e^-e^-\mu^+)&<1.1\times10^{-8},&
{\rm BR}(\tau^-\to\mu^-\mu^-e^+)&<1.4\times10^{-8}.
\label{eq:HFLAV_inputs_appG}
\end{align}
The same compilation gives ${\rm BR}(\tau\to\mu\gamma)<2.7\times10^{-8}$ at $90\%$ C.L. The MEG~II limit is ${\rm BR}(\mu\to e\gamma)<1.5\times10^{-13}$ at $90\%$ C.L. \cite{r91}. The non-unitarity limits in Eqs.~\eqref{eq:NU_bounds_diag1_main} and \eqref{eq:NU_bounds_diag2_main} are kept at their published $95\%$ C.L. \cite{r47}; no rescaling between confidence levels is made.

\subsection{Scalar contact coefficients}
Using the measured charged-lepton masses and tau width in Eqs.~\eqref{eq:BRtaueMuMu_scalar_main}--\eqref{eq:BRtaumumuE_scalar_main}, the two opposite-sign modes become
\begin{align}
{\rm BR}(\tau^-\to e^-\mu^-\mu^+)&=4.0290\times10^4
\left|\frac{\delta g}{v_\Phi^2}\right|^2,\notag\\
{\rm BR}(\tau^-\to\mu^-e^-e^+)&=4.0148\times10^4
\left|\frac{\delta g}{v_\Phi^2}\right|^2,
\label{eq:opp_numeric_appG}
\end{align}
where the contact combinations are in ${\rm GeV}^{-4}$. The first channel is a clean scalar mode at the order retained, and the first limit in Eq.~\eqref{eq:HFLAV_inputs_appG} therefore gives
\begin{equation}
\left|\frac{\delta g}{v_\Phi^2}\right|<7.05\times10^{-7}\ {\rm GeV}^{-4}.
\end{equation}
For the two same-sign modes one obtains
\begin{align}
{\rm BR}(\tau^-\to e^-e^-\mu^+)&=
2.0074\times10^4\left|\frac{F_-}{v_\Phi^2}\right|^2
+0.9424\left|\frac{F_+}{v_\Phi^2}\right|^2,\notag\\
{\rm BR}(\tau^-\to\mu^-\mu^-e^+)&=
71.45\left|\frac{F_-}{v_\Phi^2}\right|^2
+4.0147\times10^4\left|\frac{F_+}{v_\Phi^2}\right|^2.
\label{eq:samesign_numeric_appG}
\end{align}
The second of these is a clean scalar constraint. Since every term is non-negative,
\begin{equation}
4.0147\times10^4\left|\frac{F_+}{v_\Phi^2}\right|^2<1.4\times10^{-8},
\end{equation}
and hence
\begin{equation}
0<\frac{F_+}{v_\Phi^2}<5.905\times10^{-7}\ {\rm GeV}^{-4}.
\label{eq:Fplus_numeric_appG}
\end{equation}
In the stable massive scalar region the inverse propagator eigenvalues are positive. This gives $|F_-|<F_+$ and $|\delta g|<F_+$ and yields the common clean scale quoted in Eq.~\eqref{eq:CS_master_main}. The direct clean opposite-sign mode gives $|\delta g|/v_\Phi^2<7.05\times10^{-7}$ GeV$^{-4}$, while the amplitude-level scalar--HNL treatment of $\tau^-\to\mu^-e^-e^+$ sharpens this to
\begin{equation}
\frac{|\delta g|}{v_\Phi^2}<5.6903\times10^{-7}\ {\rm GeV}^{-4}.
\end{equation}
For $\tau^-\to e^-e^-\mu^+$ the only leading HNL contribution is the genuine Majorana box. In the retained massless-daughter approximation its interference with the scalar term vanishes, and the experimental limit gives
\begin{equation}
|F_{\rm Box}^{\tau ee\mu}|<0.2601.
\end{equation}

\subsection{Numerical non-unitarity bounds}
The best-fit texture combinations used in the numerical projection are
\begin{equation}
\begin{array}{c|cc}
&{\rm NO}&{\rm IO}\\ \hline
|a-b|&0.01690&0.06863\\
|b+2t|&0.07386&0.10703\\
|2b+d|&0.00840&0.00508
\end{array}\quad {\rm eV}.
\label{eq:texture_inputs_appG}
\end{equation}
The direct $ee$ and $\mu\mu$ bounds give
\begin{align}
|P|&>18.898|a-b|,\notag\\
|R|&>29.881|b+2t|,
\end{align}
while the $\mu\tau$ entry gives $|R|>37.27\sqrt{|b+2t|\,|2b+d|}$. Substitution of Eq.~\eqref{eq:texture_inputs_appG} gives
\begin{align}
|P|_{\rm NO}&>0.319~{\rm eV},& |P|_{\rm IO}&>1.30~{\rm eV},\notag\\
|R|_{\rm NO}&>2.21~{\rm eV},& |R|_{\rm IO}&>3.20~{\rm eV}.
\label{eq:PR_numeric_appG}
\end{align}
Because both terms in $\widetilde\eta_{\tau\tau}$ are positive, the first term alone gives the conservative one-dimensional projections
\begin{equation}
|Q|_{\rm NO}>0.875~{\rm eV},\qquad
|Q|_{\rm IO}>1.268~{\rm eV}.
\label{eq:Q_numeric_appG}
\end{equation}
These are separate lower projections, not a simultaneously saturated point with the $P$ and $R$ limits. The exact $Q$--$R$ allowed region is still governed by Eq.~\eqref{eq:NU_tau_corr_main}.

Using Eq.~\eqref{eq:PQR_def_main},
\begin{equation}
|Q|=\frac{\sqrt3}{2}v_H{\cal Y}_1,\qquad
|P|=\frac{\sqrt3}{2\sqrt2}v_H{\cal Y}_2,\qquad
|R|=\frac{\sqrt3}{2\sqrt2}v_H{\cal Y}_3,
\end{equation}
with the definitions in Eq.~\eqref{eq:Y123_numeric_main}. For $v_H=246$ GeV, Eqs.~\eqref{eq:PR_numeric_appG} and \eqref{eq:Q_numeric_appG} give the values in Eq.~\eqref{eq:Y123_numeric_bounds_main}. The conversion uses $1~{\rm eV}=10^{-9}$ GeV. These are bounds on the products ${\cal Y}_i$, not on the individual Yukawa couplings, VEVs or cutoff.

\subsection{Radiative HNL limit}
For Eq.~\eqref{eq:BRrad_HNL_main} we use the on-shell electroweak definitions
\begin{equation}
\alpha_w=\frac{\sqrt2G_FM_W^2}{\pi},\qquad
s_w^2=1-\frac{M_W^2}{M_Z^2},
\end{equation}
with $G_F=1.1663787\times10^{-5}$ GeV$^{-2}$, $M_W=80.377$ GeV and $M_Z=91.1876$ GeV. The resulting coefficient is approximately
\begin{equation}
{\rm BR}(\tau\to\mu\gamma)\simeq6.45\times10^{-4}|G_\gamma^{\tau\mu}|^2.
\end{equation}
The HFLAV limit then gives Eq.~\eqref{eq:Ggamma_numeric_main}. The last digit varies slightly with the standard electroweak input scheme; the relevant amplitude scale is of order $10^{-3}$. The exact phenomenology retains the mass-dependent loop functions of Ref.~\cite{r89}.

\section{Soft-breaking potential, symmetry structure and scalar Hessian}
\label{app:soft_scalar_details}

This appendix gives the details of the soft-breaking analysis of Section~\ref{sec:soft_breaking}, including the symmetry-breaking pattern and the corresponding Hessian shifts.

\subsection{Component form and stationarity}

Using the M--R product rule $(XX)_1=X_1^2+X_2^2+X_3^2$ and the field parametrisation of Eq.~\eqref{eq:scalar_field_expansion},
\begin{equation}
(X^\dagger X)_1-\frac12\big[(XX)_1+(X^\dagger X^\dagger)_1\big]
=\sum_{i=1}^3 X_{Ii}^2,
\qquad X=\Phi,\xi.
\label{eq:soft_triplet_identity_appH}
\end{equation}
For a singlet $S=(v_S+S_R+iS_I)/\sqrt2$ one similarly has
\begin{equation}
S^\dagger S-\frac12(S^2+S^{\dagger2})=S_I^2.
\label{eq:soft_singlet_identity_appH}
\end{equation}
Equations~\eqref{eq:soft_triplet_identity_appH} and \eqref{eq:soft_singlet_identity_appH} imply that the soft contribution contains no CP-even fluctuation. Hence all real tadpoles and the complete CP-even Hessian receive zero correction. Every imaginary tadpole is proportional to the corresponding CP-odd field and therefore also vanishes on a real vacuum. This establishes Eq.~\eqref{eq:soft_tadpoles_zero_main} without a perturbative VEV shift.

Taking two derivatives gives Eq.~\eqref{eq:soft_odd_shift_main}. Since the triplet corrections are proportional to the $3\times3$ identity in each triplet subspace, the same shift is obtained after applying the vacuum-adapted transformations $U_\Phi$ and $U_\xi$. The phase-parallel states receive $2s_\Phi$ and $2s_\xi$, while the two transverse $\Phi$ entries and two transverse $\xi$ entries receive the same respective shifts. The two massive $2\times2$ blocks therefore become
\begin{equation}
M_{1,s}^2=
\begin{pmatrix}A+2s_\Phi&D\\D&B+2s_\xi\end{pmatrix},
\qquad
M_{2,s}^2=
\begin{pmatrix}A+2s_\Phi&3D\\3D&B+2s_\xi\end{pmatrix},
\label{eq:soft_blocks_appH}
\end{equation}
which gives Eq.~\eqref{eq:soft_transverse_masses_main}. Positive $s_X$ add a positive-semidefinite matrix to a CP-odd sector that was already positive in its four massive directions, and therefore preserve local positivity while lifting all seven physical zeros.

\subsection{Discrete symmetry breaking pattern}

For a field with additive $Z_N$ charge $q_X$, the modulus $X^\dagger X$ is invariant, whereas the phase-sensitive quadratic $X^2$ is invariant only if
\begin{equation}
2q_X=0\pmod N.
\label{eq:soft_ZN_condition_appH}
\end{equation}
Applying Eq.~\eqref{eq:soft_ZN_condition_appH} to Table~\ref{tab:field_assignments} gives the $Z_{10}$, $Z_7$ and $Z_5$ entries in Table~\ref{tab:soft_symmetry_audit_main}. All flavons are neutral under the displayed $Z_3$ assignment, and hence $Z_3$ remains exact under the present deformation.

For $A_4$, $(\Phi\Phi)_1$ and $(\xi\xi)_1$ are singlets, while $\psi,\chi,\eta$ and $\kappa$ transform as $1$. Their soft terms therefore preserve $A_4$. By contrast, $\rho\sim1''$ and
\begin{equation}
1''\otimes1''=1',
\label{eq:rho_A4_soft_appH}
\end{equation}
so the $\rho^2+\rho^{\dagger2}$ piece explicitly breaks $A_4$. This breaking is nevertheless soft because its coefficient has mass dimension two. The electroweak gauge symmetry is preserved by every term because all seven flavons are gauge singlets. Real $s_X$ also preserve the standard CP transformation.

\subsection{The four accidental global $U(1)$ directions}

The continuous phase freedom of the scalar potential is reduced by the Dirac and Majorana operators in Eq.~\eqref{eq:yukawa_effective}. We denote a continuous rephasing charge by $q_F$. In a convention where barred and charge-conjugated fields are treated explicitly, invariance of the retained operators gives
\begin{equation}
\begin{aligned}
-q_D+q_\Phi+q_H+q_{e,\mu,\tau}&=0, &
-q_D+q_\Phi-q_H+q_{\nu_e}&=0,\\
-q_D+q_\Phi-q_H+q_{\nu_\mu}+q_\chi&=0, &
-q_D+q_\Phi-q_H+q_{\nu_\tau}+q_\psi&=0,\\
2q_{\nu_e}+q_\rho&=0, &
q_{\nu_e}+q_{\nu_\tau}+q_\kappa&=0,\\
2q_{\nu_\mu}+q_\eta&=0, &
-2q_D-2q_H+q_\xi&=0.
\end{aligned}
\label{eq:U1_charge_constraints_appH}
\end{equation}
The solution space of Eq.~\eqref{eq:U1_charge_constraints_appH} is five-dimensional. One direction is precisely the gauged hypercharge assignment. After quotienting by that direction, four independent accidental global symmetries remain, in agreement with the zero-mode discussion of Section~\ref{sec:vacuum_scalar}.

A convenient basis for their scalar charges is
\begin{equation}
\begin{array}{c|rrrrrrrr}
& H&\Phi&\xi&\psi&\chi&\eta&\rho&\kappa\\ \hline
U(1)_A&0&1&2&0&0&0&0&0\\
U(1)_B&0&0&0&0&1&2&0&0\\
U(1)_C&0&1&0&-2&-1&0&2&0\\
U(1)_D&0&0&0&1&0&0&0&1
\end{array}
\label{eq:U1_scalar_basis_appH}
\end{equation}
where the accompanying fermion charges are fixed by Eq.~\eqref{eq:U1_charge_constraints_appH}. The basis is of course not unique; any nonsingular linear recombination describes the same four-dimensional symmetry space. Equation~\eqref{eq:U1_scalar_basis_appH} is chosen because it makes the action of the soft terms transparent. The $\Phi$, $\xi$, $\psi$, $\chi$, $\eta$, $\rho$ and $\kappa$ soft terms break, respectively,
\begin{equation}
(A,C),\qquad A,\qquad(C,D),\qquad(B,C),\qquad B,\qquad C,\qquad D.
\label{eq:soft_U1_breaking_appH}
\end{equation}
Thus every one of the four exact accidental global $U(1)$ directions is explicitly broken by the selected dimension-two deformation. At the same time, Eq.~\eqref{eq:soft_odd_shift_main} gives positive curvature to the remaining three phase-flat directions which were not Goldstones of additional exact continuous symmetries.

\subsection{Boundedness and global character}

The soft operators are quadratic and therefore do not modify the quartic large-field function used in the BFB analysis. For $s_X\ge0$,
\begin{equation}
\Delta V_{\rm soft}\ge0,
\qquad
\left.\Delta V_{\rm soft}\right|_{\rm M-R}=0.
\label{eq:soft_global_appH}
\end{equation}
If the unbroken potential satisfies the sufficient conditions which make the chosen M--R configuration a global minimum, adding Eq.~\eqref{eq:soft_global_appH} cannot lower the potential away from that configuration. The soft deformation removes the continuous/phase-flat zero directions, but it does not lift the complete real residual discrete orbit discussed in Section~\ref{sec:domain_walls}. It nevertheless does not invalidate the sufficient global-minimum region derived above.

\section{Soft-breaking corrections to the phenomenological quantities}
\label{app:soft_phenomenology}

\subsection{Scalar propagators and CLFV combinations}

At vanishing external momentum, the CP-odd $\Phi$-projected inverse propagators are obtained from the two shifted blocks of Appendix~\ref{app:soft_scalar_details}. Writing
\begin{equation}
X_O^s=X_O+2s_\Phi,
\qquad
Y_O^s=Y_O+2s_\xi,
\label{eq:soft_XY_appI}
\end{equation}
with $Z_O$ unchanged, direct inversion gives
\begin{equation}
[(M_{1,s}^2)^{-1}]_{\Phi\Phi}
=\frac{Y_O^s}{X_O^sY_O^s-Z_O^2},
\qquad
[(M_{2,s}^2)^{-1}]_{\Phi\Phi}
=\frac{Y_O^s}{X_O^sY_O^s-9Z_O^2}.
\label{eq:soft_prop_appI}
\end{equation}
These are $g_{O1}^s$ and $g_{O2}^s$ of Eq.~\eqref{eq:soft_gO_main}. The CP-even transverse matrix receives no correction, so $g_E^s=g_E$. The scalar Wilson coefficients and three-body branching ratios therefore follow from the no-soft expressions by
\begin{equation}
\delta g\to\delta g_s,
\qquad
F_-\to F_-^s,
\qquad
F_+\to F_+^s.
\label{eq:soft_substitutions_appI}
\end{equation}
No new charged-lepton flavour matrix appears. The five singlet coefficients $s_{\psi,\chi,\eta,\rho,\kappa}$ lift their corresponding zero modes but do not enter Eq.~\eqref{eq:soft_prop_appI} because those states do not mix with the leading transverse charged-lepton flavour directions in the CP-odd Hessian.

For a positive-definite original CP-odd block and positive $s_\Phi,s_\xi$, increasing the diagonal entries gives
\begin{equation}
g_{O1}^s<g_{O1},\qquad
g_{O2}^s<g_{O2},\qquad
g_{O1}^s+g_{O2}^s<g_{O1}+g_{O2},
\end{equation}
and hence $|\delta g_s|<|\delta g|$, $F_+^s<F_+$ and $F_-^s>F_-$. The last inequality is algebraic and does not imply that every rate containing $F_-^s$ is monotonic. In particular, the scalar $\tau^-\to e^-e^-\mu^+$ rate contains competing $F_-^s$ and $F_+^s$ contributions and is not universally monotonic.

\subsection{Three-body bounds and scalar--HNL interference}

The charge-resolved scalar coefficients are constrained with exactly the same experimental limits used in the no-soft analysis. After the amplitude-level scalar--HNL treatment, the strongest robust opposite-sign result is
\begin{equation}
\frac{|\delta g_s|}{v_\Phi^2}<5.6903\times10^{-7}\ {\rm GeV}^{-4}.
\label{eq:soft_dgs_bound_appI}
\end{equation}
The HNL-clean same-sign channel $\tau^-\to\mu^-\mu^-e^+$ gives
\begin{equation}
0<\frac{F_+^s}{v_\Phi^2}<5.9052\times10^{-7}\ {\rm GeV}^{-4},
\qquad
\frac{|F_-^s|}{v_\Phi^2}<5.9052\times10^{-7}\ {\rm GeV}^{-4}.
\label{eq:soft_Fpm_bound_appI}
\end{equation}
For $\tau^-\to e^-e^-\mu^+$, the only leading HNL piece is the genuine Majorana box. In the massless-daughter approximation specified in Subsubsection~\ref{subsubsec:clfv_constraints}, its interference with the scalar contribution vanishes, and the corresponding condition remains
\begin{equation}
|F_{\rm Box}^{\tau ee\mu}|<0.2601.
\label{eq:soft_box_bound_appI}
\end{equation}
The complete scalar $\tau^-\to\mu^-\mu^-e^+$ rate decreases with positive $s_\Phi,s_\xi$, whereas the scalar $\tau^-\to e^-e^-\mu^+$ rate must be evaluated from the full correlated expression.

\subsection{Radiative modes}

The CP-even scalar dipole cancellation is unchanged because the CP-even Hessian is unchanged. In the CP-odd part, the loop functions are evaluated with the soft-shifted masses and projection weights, but the amplitude remains proportional to the same $Z_O^2$ breaking structure. The physical $\tau\to\mu\gamma$ dipole is therefore
\begin{equation}
A_D^{\rm tot,s}=A_D^{S,s}+A_D^N,
\end{equation}
with the HNL term $A_D^N$ unchanged. Since the scalar dipole involves a difference of spectral loop weights, no model-independent lower limit on a single scalar mass or on one soft coefficient follows from the present radiative bound.

\subsection{Tree-level texture, HNL sector and non-unitarity}

Because Eq.~\eqref{eq:soft_tadpoles_zero_main} leaves all VEVs entering the fermion operators unchanged,
\begin{equation}
\Xi_{\rm soft}=M_D(M_R)^{-1}=\Xi,
\qquad
(\Theta)_{\rm soft}=\Xi_{\rm soft}^*=\Theta,
\qquad
\widetilde\eta_{\rm soft}=\widetilde\eta.
\label{eq:soft_eta_appI}
\end{equation}
The heavy-neutrino charged-current matrix $B$, the neutral-current combination $C$, and the exact HNL loop functions are therefore unchanged at this order. The separate lower projections remain
\begin{equation}
(P,Q,R)_{\rm NO}>(0.319,0.875,2.21)~{\rm eV},
\qquad
(P,Q,R)_{\rm IO}>(1.30,1.268,3.20)~{\rm eV},
\label{eq:soft_PQR_appI}
\end{equation}
with the same qualification as in Subsubsection~\ref{subsubsec:NU_numerical}: the entries are separate lower projections, not a simultaneously saturated point. These equalities hold at tree level for the retained dimension-two soft terms. Possible loop-induced corrections from the soft breaking are not included here.

\subsection{Conventional Type-I invariant}

The conventional leptogenesis Yukawa matrix ${\cal Y}$ depends only on $P,Q,R$ and on the Takagi diagonalisation of $M_R$. Both are unchanged by Eq.~\eqref{eq:soft_total_main}. Consequently
\begin{equation}
({\cal Y}_{\rm soft}^\dagger {\cal Y}_{\rm soft})_{ij}
=({\cal Y}^\dagger {\cal Y})_{ij},
\end{equation}
and the purely imaginary off-diagonal structure of the no-soft result is preserved. Equation~\eqref{eq:soft_lepto_zero_main} therefore follows without an additional loop calculation for the restricted conventional two-body Type-I source.

\section{Residual discrete orbit and a stationary domain-wall bias}
\label{app:domain_wall_details}

This appendix gives the compact group-theoretic check underlying Section~\ref{sec:domain_walls}. The seven phase-sensitive soft terms break $Z_5$ and $Z_7$ completely. Their $Z_{10}$-charged squares leave only the element $g_{10}^5$, while the $\rho^2+\mathrm{h.c.}$ term transforms as $1'$ of $A_4$ and therefore leaves the Klein subgroup
\begin{equation}
V_4=\{1,S,TST^{-1},T^2ST^{-2}\}\simeq Z_2\times Z_2.
\end{equation}
The residual $g_{10}^5$ action is already contained in the sign remnants of the accidental rephasings and must not be counted as an additional factor.

Solving the Yukawa rephasing conditions after the continuous phases have been reduced to signs gives four independent bits. A convenient scalar action is
\begin{equation}
\Phi\to(-1)^p\Phi,
\quad
\chi\to(-1)^c\chi,
\quad
\psi\to(-1)^s\psi,
\quad
(\xi,\eta,\rho)\to(-1)^x(\xi,\eta,\rho),
\quad
\kappa\to(-1)^{x+s}\kappa,
\label{eq:DW_sign_action_appJ}
\end{equation}
with $p,c,s,x\in\{0,1\}$. Thus the scalar sign group is $(Z_2)^4$ and
\begin{equation}
G_{\rm DW}=V_4\times(Z_2)^4.
\label{eq:GDW_appJ}
\end{equation}
For the alignments $\langle\Phi\rangle\propto(1,1,1)$ and $\langle\xi\rangle\propto(1,-1,1)$, no nonidentity element of $V_4$ fixes the full triplet pair, and every independent sign flips at least one field with a nonzero VEV. The stabiliser is therefore trivial and the symmetry orbit contains $4\times16=64$ real vacua.

The original soft potential in Eq.~\eqref{eq:soft_phaseonly_main} is a sum of imaginary-field squares, and hence it vanishes on every real member of this orbit. To demonstrate that the residual wall degeneracy can nevertheless be lifted without moving the desired vacuum, define
\begin{equation}
{\bf a}=(1,1,1)^T,
\qquad
{\bf b}=(1,-1,1)^T,
\qquad
R_0=\frac{{\bf a}{\bf b}^T}{3},
\label{eq:DW_R0_appJ}
\end{equation}
and consider the positive-semidefinite quadratic relations
\begin{align}
\Delta V_1&=\frac{\mu_1^2}{v_\Phi^2+v_\xi^2}
\left\|v_\xi\Phi-v_\Phi R_0\xi\right\|^2,\notag\\
\Delta V_2&=\frac{\mu_2^2}{v_\chi^2+v_\eta^2}
\left|v_\eta\chi-v_\chi\eta\right|^2,\notag\\
\Delta V_3&=\frac{\mu_3^2}{v_\psi^2+v_\eta^2}
\left|v_\eta\psi-v_\psi\eta\right|^2,\notag\\
\Delta V_4&=\frac{\mu_4^2}{v_\kappa^2+v_\eta^2}
\left|v_\eta\kappa-v_\kappa\eta\right|^2,
\qquad \mu_i^2>0.
\label{eq:DW_bias_explicit_appJ}
\end{align}
At the target M--R vacuum all four brackets vanish. Therefore
\begin{equation}
\Delta V_{\rm DW}=\sum_{i=1}^4\Delta V_i\ge0,
\qquad
\Delta V_{\rm DW}(v_{\rm MR})=0,
\qquad
\left.\nabla\Delta V_{\rm DW}\right|_{v_{\rm MR}}=0.
\label{eq:DW_bias_properties_appJ}
\end{equation}
The first relation removes the nontrivial $V_4$ orientations and fixes the relative $\Phi$--$\xi$ sign; the remaining three relations remove the independent singlet/sign images. Together with the residual sign constraint on $\kappa$, the identity is the only element of $G_{\rm DW}$ on which all four terms vanish.

Since Eq.~\eqref{eq:DW_bias_explicit_appJ} is quadratic, the quartic BFB conditions are unchanged. Its Hessian contribution is positive semidefinite. Because it is zero at the target and non-negative elsewhere, any sufficient global-minimum property established for the unbroken potential is inherited. The real VEVs are exactly stationary, so the leading tree-level $M_D$, $M_R$, $M_w$ and $M_\nu$ are unchanged. The CP-even scalar masses, however, can shift because the domain-wall bias acts on real flavon relations; no claim of an unchanged CP-even spectrum is made. The bias in Eq.~\eqref{eq:DW_bias_explicit_appJ} is one possible choice that lifts the residual degeneracy; we do not assume that it is unique or minimal.

\end{document}